\documentclass[aps,twocolumn,10pt,superscriptaddress,tightenlines,nofootinbib,floatfix,preprintnumbers,longbibliography]{revtex4-1}
\usepackage{amsmath,amssymb}
\usepackage{bm}
\usepackage{comment}
\usepackage{graphicx}
\usepackage{color}
\usepackage{cancel}
\usepackage{tikz}
\usepackage{pifont}
\usepackage{MnSymbol}
\usepackage{CJKutf8}
\usepackage{float}
\usepackage{placeins}
\usetikzlibrary{shapes.misc}

\definecolor{kngrey}{HTML}{A6AAA9}
\definecolor{knred}{HTML}{EC5D57}
\definecolor{knorange}{HTML}{F39019}
\definecolor{knyellow}{HTML}{F5D328}
\definecolor{kngreen}{HTML}{70BF41}
\definecolor{knblue}{HTML}{51A7F9}
\definecolor{knpurple}{HTML}{B36AE2}
\definecolor{knbgreen}{HTML}{0B5D12}

\definecolor{evan}{rgb}{0.0, 0.65, 0.58}
\definecolor{enrico}{rgb}{0.65, 0, 0.58}

\usepackage{hyperref}
\hypersetup{
    colorlinks=true,       
    linkcolor=blue,          
    citecolor=blue,        
    filecolor=blue,      
    urlcolor=blue           
}

\def\ga{1.278(21)(26)}

\def\b{{\beta}}
\def\d{{\delta}}
\def\D{{\Delta}}
\def\e{{\epsilon}}

\def\l{{\lambda}}

\def\s{{\sigma}}
\def\t{{\tau}}

\newcommand{\sla}[1]{\slash \!\!\! #1}

\def\eqref#1{{(\ref{#1})}}

\newcommand{\glasgow}{
 School of Physics and Astronomy, 
    University of Glasgow, 
    Glasgow G12 8QQ, UK
 }
\newcommand{\jlab}{
 Theory Center, 
 Thomas Jefferson National Accelerator Facility, 
 Newport News, VA 23606, USA
 }
\newcommand{\jlabcomp}{
 Scientific Computing Group, 
 Thomas Jefferson National Accelerator Facility, 
 Newport News, VA 23606, USA
 }
\newcommand{\julich}{
 Institut f\"{u}r Kernphysik and Institute for Advanced Simulation,
 Forschungszentrum J\"{u}lich, 54245 J\"{u}lich Germany
 }
\newcommand{\liverpool}{
    Theoretical Physics Division, Department of Mathematical Sciences, University of Liverpool, Liverpool L69 3BX, UK
    }
\newcommand{\lblnsd}{
 Nuclear Science Division,
    Lawrence Berkeley National Laboratory,
 Berkeley, CA 94720, USA
 }
\newcommand{\lblnersc}{
 NERSC,
    Lawrence Berkeley National Laboratory,
 Berkeley, CA 94720, USA
 }
\newcommand{\llnl}{
 Nuclear and Chemical Sciences Division,
 Lawrence Livermore National Laboratory,
 Livermore, CA 94550, USA
 }
\newcommand{\nvidia}{
    NVIDIA Corporation,
    2701 San Tomas Expressway, Santa Clara, CA 95050, USA
    }
\newcommand{\riken}{
 RIKEN-BNL Research Center, 
 Brookhaven National Laboratory, 
 Upton, NY 11973, USA
 }
\newcommand{\rutgers}{
 New High Energy Theory Center and Department of Physics and Astronomy,
 Rutgers,
 The State University of New Jersey, Piscataway, NJ 08854, USA
 }
\newcommand{\ucb}{
 Department of Physics,
 University of California,
 Berkeley, CA 94720, USA
 }
\newcommand{\wm}{
 Department of Physics,
 The College of William \& Mary,
 Williamsburg, VA 23187, USA
 }

\newcount\hour \newcount\hourminute \newcount\minute
\hour=\time \divide \hour by 60
\hourminute=\hour \multiply \hourminute by 60
\minute=\time \advance \minute by -\hourminute
\newcommand{\mydate}{\ \today \ - \number\hour :\number\minute}

\begin{document}

\title{An accurate calculation of the nucleon axial charge with lattice QCD}

\author{Evan~Berkowitz}
\affiliation{\julich}
\affiliation{\llnl}

\author{David~Brantley}
\affiliation{\wm}
\affiliation{\lblnsd}

\author{Chris~Bouchard}
\affiliation{\glasgow}
\affiliation{\wm}

\author{Chia~Cheng~Chang \begin{CJK*}{UTF8}{bsmi}(張家丞)\end{CJK*}}
\affiliation{\lblnsd}

\author{M.~A.~Clark}
\affiliation{\nvidia}

\author{Nicolas~Garron}
\affiliation{\liverpool}

\author{B\'{a}lint~Jo\'{o}}
\affiliation{\jlabcomp}

\author{Thorsten~Kurth}
\affiliation{\lblnersc}

\author{Chris~Monahan}
\affiliation{\rutgers}

\author{Henry~Monge-Camacho}
\affiliation{\wm}
\affiliation{\lblnsd}

\author{Amy~Nicholson}
\affiliation{\ucb}
\affiliation{\lblnsd}

\author{Kostas~Orginos}
\affiliation{\wm}
\affiliation{\jlab}

\author{Enrico~Rinaldi}
\affiliation{\riken}
\affiliation{\llnl}

\author{Pavlos~Vranas}
\affiliation{\llnl}
\affiliation{\lblnsd}

\author{Andr\'{e}~Walker-Loud}
\affiliation{\lblnsd}
\affiliation{\llnl}

\date{\mydate}

\begin{abstract}
We report on a lattice QCD calculation of the nucleon axial charge, $g_A$, using M\"{o}bius Domain-Wall fermions solved on the dynamical $N_f=2+1+1$ HISQ ensembles after they are smeared using the gradient-flow algorithm.
The calculation is performed with three pion masses, $m_\pi\sim\{310,220,130\}$~MeV.
Three lattice spacings ($a\sim\{0.15,0.12,0.09\}$~fm) are used with the heaviest pion mass, while the coarsest two spacings are used on the middle pion mass and only the coarsest spacing is used with the near physical pion mass.
On the $m_\pi\sim220$~MeV $a\sim0.12$~fm point, a dedicated volume study is performed with $m_\pi L \sim \{3.22,4.29,5.36\}$.
Using a new strategy motivated by the Feynman-Hellmann Theorem, we achieve a precise determination of $g_A$ with relatively low statistics, and demonstrable control over the excited state, continuum, infinite volume and chiral extrapolation systematic uncertainties, the latter of which remains the dominant uncertainty.
Our final determination at 2.6\% total uncertainty is $g_A = \ga$, with the first uncertainty including statistical and systematic uncertainties from fitting and the second including model selection systematics related to the chiral and continuum extrapolation.
The largest reduction of the second uncertainty will come from a greater number of pion mass points as well as more precise lattice QCD results near the physical pion mass.
\end{abstract}
\maketitle


%
\section{Introduction\label{sec:intro}}

The nucleon axial charge, $g_A$, is one of the most fundamental quantities that characterize the nucleon. This coupling measures the strength with which the weak axial current couples to the nucleon and plays a central role in our theoretical understanding of nuclear physics. The axial charge governs many fundamental nuclear processes, such as nuclear beta decay and pion exchange between nucleons. Furthermore, small changes to its value may have resulted in a profoundly different universe than that which we observe today, as some astrophysical and cosmological processes, such as Big Bang nucleosynthesis, depend very sensitively on $g_A$~\cite{Mathews:2004kc}.

If nature respected chiral symmetry exactly, the axial charge, in conjunction with the vector charge, would be exactly one. However, chiral symmetry is explicitly broken by nonzero quark masses and is spontaneously broken by the QCD vacuum.  The axial charge is measured to be
\begin{equation}
    g^{exp}_A = 1.2723(23)\text{~\cite{Olive:2016xmw}}\, .
\end{equation}
This very precise determination of $g_A$ is the current PDG (Particle Data Group) experimental world average from cold neutron decays~\cite{PhysRevLett.56.919, YEROZOLIMSKY1997240, LIAUD199753, Mostovoi:2001ye, PhysRevLett.100.151801, Mund:2012fq, Mendenhall:2012tz}. Because it can be very well-measured experimentally, the axial charge is a prime candidate for searches for BSM (beyond the Standard Model) physics: if one can determine $g_A$ with similar precision directly from QCD, any deviations from experiment may be interpreted as BSM contributions.

Given the prominence of $g_A$, this quantity should be the first important benchmark calculation for demonstrating that uncertainties associated with LQCD (lattice QCD) calculations relevant to nuclear physics can be appropriately quantified and controlled. However, this quantity has been notoriously difficult to compute in LQCD, largely due to systematics that were not fully appreciated, see for example the review~\cite{Constantinou:2014tga}. Only recently has the first LQCD calculation of $g_A$ been produced with the major sources of systematics addressed, notably the continuum and infinite volume limits, as well as an extrapolation/interpolation to the physical pion mass. This LQCD result, reported in Ref.~\cite{Bhattacharya:2016zcn}, is:
\begin{equation}
g_A^{\text{PNDME}} = 1.195(33)(20)\, ,
\end{equation}
where the first uncertainty is statistical in nature and the second arises from the model extrapolation to the physical point. We note this result differs from the physical PDG value by 2 standard deviations. Given that the LQCD calculation is still an order of magnitude less precise than the experimental determination, it seems premature to assign such a large discrepancy to new physics, and further investigation into the systematics behind $g_A$ is necessary.

In addition to providing a benchmark for nuclear physics from LQCD in general, a precise determination of $g_A$ is an important first step toward calculations of corrections to $g_A$ due to in-medium effects in nuclei~\cite{Arima:1988xa,Towner:1987zz,Brown:1987obh}. This so-called $g_A$ quenching effect is relevant for understanding potential signals for $0\nu\b\b$ (neutrinoless double-beta decay), see for example the recent review~\cite{Menendez:2017chr}, a process which would confirm the Majorana nature of neutrinos as well as provide a potential explanation for the matter-antimatter asymmetry in the universe. Enormous world-wide experimental efforts searching for this process are both planned and underway, and a determination of these axial matrix elements may also be useful for planning future experiments. Recent work has explored the importance of two-body currents for weak matrix elements in light~\cite{Gazit:2008ma,Park:2002yp,Kubodera:2010qx} and medium~\cite{Menendez:2011qq} nuclei using EFT (Effective Field Theory)~\cite{Weinberg:1978kz}. In principle, LQCD can be used to determine these corrections directly from QCD, see for example the exploratory calculation in Refs.~\cite{Shanahan:2017bgi,Tiburzi:2017iux}. However, before having confidence in LQCD calculations of unknown corrections to $g_A$, one must first have full command over the systematics for $g_A$ itself.

The most significant remaining systematic to be addressed, according to Ref.~\cite{Bhattacharya:2016zcn}, concerns the excited state pollution of the ground state matrix elements. In this work, we calculate $g_A$ from a method inspired by the FH (Feynman-Hellmann) Theorem~\cite{Bouchard:2016heu}. This method allows for the removal of excited state contamination through extrapolation in a single time variable, much like the simpler two-point correlation function. This not only gives better control over systematics associated with excited state contamination, but also eliminates the need to vary two time variables independently, greatly saving in computational cost.

Using this method, we present a new determination of $g_A$ with controlled systematics. We use the publicly available HISQ configurations generated by the MILC Collaboration, allowing for an exploration of discretization and finite volume effects, as well as extrapolation to physical pion mass. We use the recently-developed MA (mixed action) from Ref.~\cite{Berkowitz:2017opd}, producing MDWF (M\"obius Domain-Wall fermion) propagators after smearing the gauge fields using the gradient-flow method, which significantly improves the chiral symmetry properties. The aforementioned savings in computational cost from the FH method are compounded through the use of the QUDA library, which significantly accelerates the M\"obius Domain-Wall fermion solutions. These improvements have allowed us to achieve both precision and accuracy in our final determination of
\begin{equation}
g_A = \ga,
\end{equation}
where the first uncertainty is statistical and the second is the extrapolation systematic.  Our result is compatible with the PDG value.

We provide a full explanation of the methods we used to achieve this result. In Sec.~\ref{sec:FH}, we summarize the FH method~\cite{Bouchard:2016heu} for evaluating matrix elements. After that, in Sec.~\ref{sec:MA}, we briefly summarize the mixed action, which is fully described in Ref.~\cite{Berkowitz:2017opd}. Then we present our data analysis methods, in Sec.~\ref{sec:Analysis}, along with results for the correlator analysis. Renormalization is discussed in Sec.~\ref{sec:renorm}, various methods of performing the continuum, infinite volume and chiral extrapolations are summarized in Sec.~\ref{sec:extrap},
and finally our results are presented in Sec.~\ref{sec:results}.

%
\section{\label{sec:FH}The Feynman-Hellmann Method}
\subsection{Correlation functions}
Recently, there has been significant interest and development with methods related to the Feynman-Hellman Theorem~\cite{Bouchard:2016heu, Chambers:2014qaa, Chambers:2015bka, Savage:2016kon}. We employ the method introduced in Ref.~\cite{Bouchard:2016heu} to construct and analyze correlation functions relevant to this work. The Feynman-Hellmann theorem relates matrix elements to linear variations in the spectrum with respect to an external current,
\begin{equation}
\frac{\partial E_n}{\partial\l}=\langle n | H_\l | n \rangle\, ,
\end{equation}
where $H=H_0 + \l H_\l$.
On the lattice, the spectrum is determined from the two-point correlation function,
\begin{equation}
\label{eq:twoptcorrelator}
C_\l(t)=\langle \l | N(t)N^\dagger(0) | \l \rangle\\
\end{equation}
where $N$ is a nucleon interpolating operator~\cite{Basak:2005aq,Basak:2005ir}, and $\l$ denotes the vacuum in the presence of an external source $S_\l=\l \int d^4x j(x)$, where $j(x)$ is some bilinear current density and $J(t)\equiv \int d^3\vec{x} j(t,\vec{x})$. The sourceless zero-temperature vacuum is recovered by taking $\l\rightarrow 0$ and labeled as $\Omega$. In Euclidean calculations, only space-like correlation functions are directly accessible; as a consequence, the effective mass can be constructed from a ratio of two-point correlation functions,
\begin{equation}
\label{eq:meff_source}
    M^{eff}_\l(t,\t)=\frac{1}{\t}\ln\left(\frac{C_\l(t)}{C_\l(t+\t)}\right).
\end{equation}
We invoke the Feynman-Hellmann theorem and take the analytic derivative with respect to $\l$ of Eq.~(\ref{eq:meff_source}) to obtain the \textit{effective derivative},
\begin{equation} \label{eq:derivative_meff}
\frac{\partial M^{eff}(t,\t)}{\partial \l}\bigg|_{\l=0} = \frac{1}{\t} \left[
	\frac{\partial_\l C_\l(t)}{C_\l(t)}-\frac{\partial_\l C_\l(t+\t)}{C_\l(t+\t)}\right]
	\bigg|_{\l=0}
\end{equation}
where the derivative of the two-point correlation function is related to the three-point correlation function,
\begin{align}
\label{eq:twopt_derivative}
    \left. -\frac{\partial C_\l(t)}{\partial \l} \right|_{\l=0} = & -C(t)\int dt^\prime \langle \Omega | J(t^\prime) | \Omega \rangle \nonumber\\
    & + \int dt^\prime \langle \Omega | T\{ N(t) J(t^\prime) N^\dagger(0)\} | \Omega \rangle,
\end{align}
which we define as the \textit{derivative correlation function}.
In subsequent figures and text, we define the quantity 
\begin{equation}
\mathring{g}_\l^{eff} \equiv \partial M_\l^{eff} = \partial_\l M_\l^{eff} |_{\l=0}\, ,
\end{equation}
to denote the effective mass plot of the derivative correlation function constructed with the bare current $J_\l$.

The first term of Eq.~(\ref{eq:twopt_derivative}) has vacuum quantum numbers and is only non-vanishing for the scalar current, and even in this case, exactly cancels in Eq.~\eqref{eq:derivative_meff}. The second term contains the quantity of interest for $t>t^\prime$ as well as contact terms and other unwanted contributions from $t\leq t'$.
The $t$ dependence of the unwanted contributions differs from that of the quantity of interest, allowing it to be removed in the analysis of the derivative correlation function.

Before extracting the spectrum and matrix elements from these correlation functions, we first construct the spin averaged combination for both the two-point and derivative correlation function. Furthermore, for the derivative correlation function, the iso-vector combination is constructed to eliminate contributions from disconnected diagrams.

\subsection{Spectral decomposition}
\label{Sec:spectral_decomp}
For convenience of notation, we define the relevant matrix elements as,
\begin{align}
J_{nm} \equiv & \langle n | J | m \rangle,\\
Z^\dagger_{nj} \equiv & \langle n | N^\dagger | j \rangle,
\end{align}
and overlap factors of various interpolating operators as,
\begin{align}
Z^\dagger_n\equiv &\langle n | N^\dagger | \Omega\rangle,\\
J^\dagger_j \equiv & \langle j | J | \Omega \rangle,\\
Z^\dagger_{J:n} \equiv & \langle n | JN^\dagger | \Omega \rangle.
\end{align}

The spectral decomposition of the two-point correlation function of the nucleon is derived by inserting the resolution of the identity into Eq.~(\ref{eq:twoptcorrelator}), and yields
\begin{equation}
    C(t)=\sum_{n}z_n z^\dagger_n e^{-E_nt}
    \label{eq:twopt_fit_ansatz}
\end{equation}
such that $z_n$ is related to the overlap factor $Z_n$ as
\begin{equation}
    z_n\equiv\frac{Z_n}{\sqrt{2E_n}}
\end{equation}
under relativistic normalization.

The spectral decomposition of the derivative correlation function is similarly derived, assuming a vanishing vacuum expectation value, $\langle \Omega | J | \Omega \rangle=0$, and leads to
\begin{multline}
N_J(t) =  \sum_{n} \left[(t-1)z_{n}g^J_{nn}z^\dagger_{n}+d_n^J\right]e^{-E_nt}
\\
	+\sum_{n,m\neq n}z_n g^J_{nm} z^\dagger_m 
		\frac{e^{-E_nt +\frac{\Delta_{nm}}{2}}-e^{-E_mt +\frac{\Delta_{mn}}{2}}}{e^{\frac{\Delta_{mn}}{2}}-e^{\frac{\Delta_{nm}}{2}}}\, ,
    \label{eq:threept_fit_ansatz}
\end{multline}
where $\D_{mn} = E_m - E_n$ and
\begin{align} 
\label{eq:define_gnm}
g^J_{nm} \equiv & \frac{J_{nm}}{\sqrt{4 E_n E_m}}\, ,
\\
\label{eq:define_dn}
d^J_n \equiv & Z_n Z^\dagger_{J:n} + Z_{J:n}Z^\dagger_n + Z_nZ^\dagger_n\langle \Omega | J | \Omega \rangle 
\nonumber \\& 
	+ \sum_{j} \frac{Z_nZ^\dagger_{nj}J_j^\dagger +J_j Z_{jn} Z_n^\dagger}
		{2E_j\left(e^{E_j}-1\right)}\, .
\end{align}
All contributions arising from the contact terms ($t=0$ and $t=t^\prime$) and unwanted time orderings ($t^\prime > t$) are completely parameterized by $d_n^J$. Furthermore, an estimate of $d_0^J$ may be obtained by observing that
\begin{equation}
N_J(1)=\sum_{n}d^J_n e^{-E_n}
\end{equation}
with the assumption that the ground state dominates the sum for smeared lattice interpolating operators.

Finally, the effective derivative motivated by the Feynman-Hellmann theorem, as expressed by Eq.~(\ref{eq:derivative_meff}), may be analysed from the analogous ratios of the spectral decomposition of the two-point and derivative correlation functions,
\begin{equation}
    G_J(t)\equiv\frac{1}{\t}\left[\frac{N_J(t+\t)}{C(t+\t)}-\frac{N_J(t)}{C(t)}\right].
    \label{eq:dmeff_fit_ansatz}
\end{equation}
In the large time limit, the ground state matrix element of interest is recovered, 
\begin{equation}
\lim_{t\rightarrow\infty} G_J(t) = g^J_{00}\, ,
\end{equation}
as can be verified by plugging in the spectral decomposition of $N_J(t)$ and $C(t)$ into Eq.~\eqref{eq:dmeff_fit_ansatz}.
The numerical implementation of Eq.~\eqref{eq:dmeff_fit_ansatz} serves the same role as the numerical implementation of the traditional effective mass.

\section{\label{sec:MA}Mixed action Lattice QCD}
We perform our calculation on a subset of the publically available HISQ gauge configurations generated by the MILC Collaboration~\cite{Bazavov:2012xda,Bazavov:2014wgs,Bazavov:2015yea} with $N_f=2+1+1$ dynamical sea quarks, consisting of two flavors of dynamical light-quarks with degenerate masses, along with dynamical strange- and charm-quarks with masses near the physical values. Formally the HISQ action has leading discretization errors starting at $\mathrm{O}(\alpha_S a^2, a^4)$, however improved link-smearing greatly suppresses taste-changing interactions leading to numerically smaller discretization errors. The gluons are simulated using the tadpole-improved~\cite{PhysRevD.48.2250}, one-loop Symanzik gauge action~\cite{Alford199587} with leading discretization errors starting at $\mathrm{O}(\alpha_S^2 a^2, a^4)$.

Table~\ref{tab:hisq} lists the HISQ ensembles used in this work. The smallest ensemble consists of approximately 800 configurations and the largest almost 2000 configurations. The list of ensembles includes three different lattice spacings at $\{\sim0.15~\text{fm},\sim0.12~\text{fm},\sim0.09~\text{fm}\}$, allowing for control over the continuum extrapolation. The lightest simulated taste-5 pseudoscalar pion mass, $m_{\pi,5}\sim 130~\text{MeV}$, is slightly smaller than the physical pion mass, and two heavier pion masses are at approximately $220~\text{MeV}$ and $310~\text{MeV}$, allowing for good control over interpolation to the physical pion mass. Most ensembles have large spatial volumes $(m_{\pi,5} L \geq 3.78)$ where finite volume effects are expected to be small~\cite{Durr:2008zz}, with the a15m130 ensemble at a smaller volume $(m_{\pi,5} L = 3.25)$. Control over volume dependence of the matrix element is demonstrated by performing the calculation on three ensembles (a12m220S, a12m220, a12m220L) with separate volumes ($m_{\pi,5} L = 3.22, 4.29, 5.36$ respectively) while holding all other parameters fixed.

For the valence quarks used in this work, we employ the MDWF action~\cite{Brower:2004xi,Brower:2005qw,Brower:2012vk}, and tune the valence pseudoscalar masses to within 2\% of the HISQ taste-5 pseudoscalar masses. 
This action has been used to compute the $\pi^-\rightarrow\pi^+$ matrix element~\cite{Nicholson:2016byl} relevant for $0\nu\b\b$ in the scenario that heavy new physics contributes significantly to the decay~\cite{Prezeau:2003xn}.
A discussion on the salient features of mixed lattice QCD action using MDWF on HISQ is detailed in Ref.~\cite{Berkowitz:2017opd}.
As mentioned above, the gradient-flow is used to smear the HISQ ensembles, highly suppressing the residual chiral symmetry breaking of the MDWF action allowing us to keep the residual mass $m_{res}$ less than 10\% of the valence light quark mass on each ensemble.
After absorbing $m_{res}$ into the quark mass through the PCAC relation, the MDWF action has discretization errors beginning at $\mathrm{O}(a^2)$~\cite{Allton:2008pn}.

\begin{figure}
\includegraphics[width=\columnwidth]{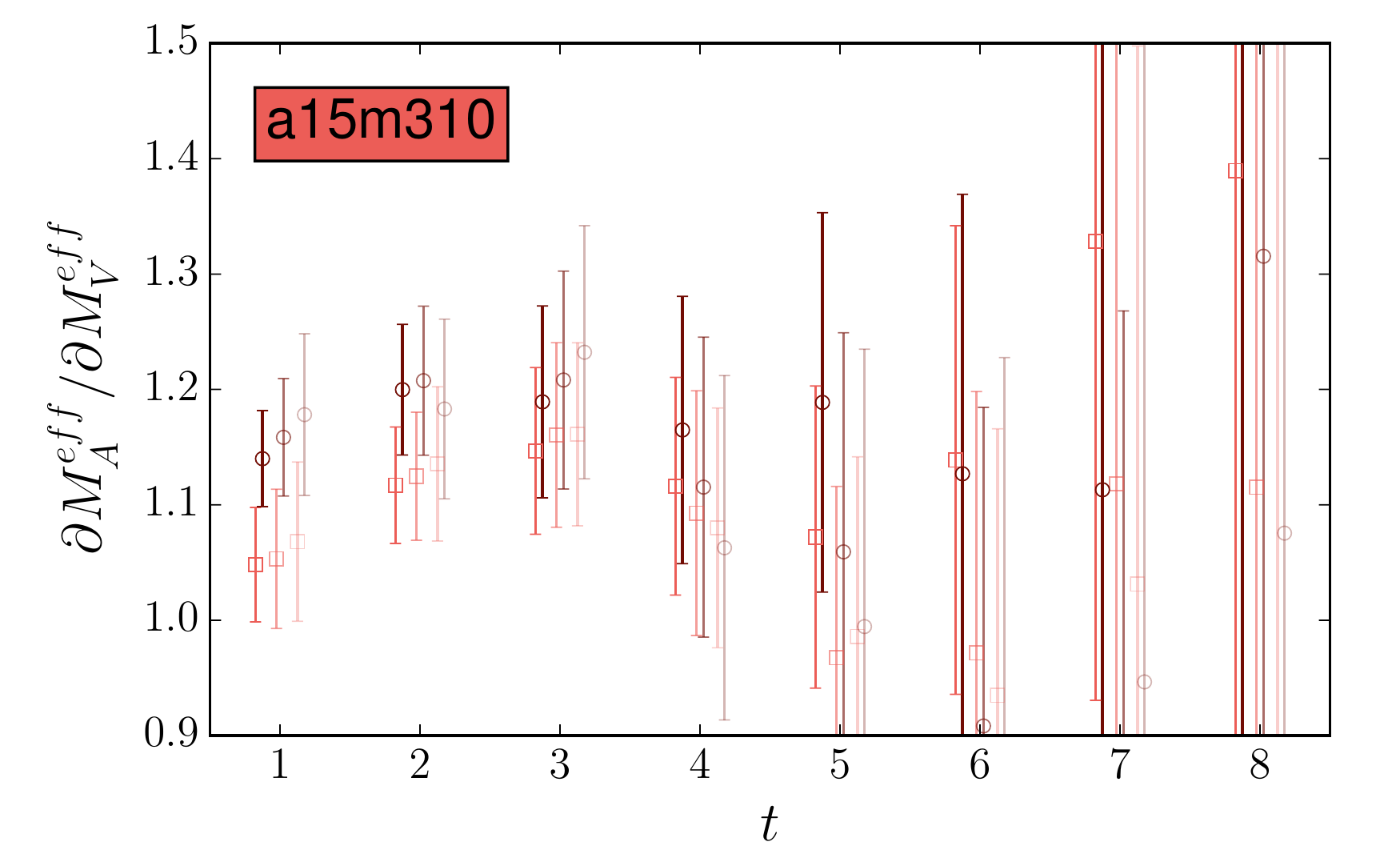}
\includegraphics[width=\columnwidth]{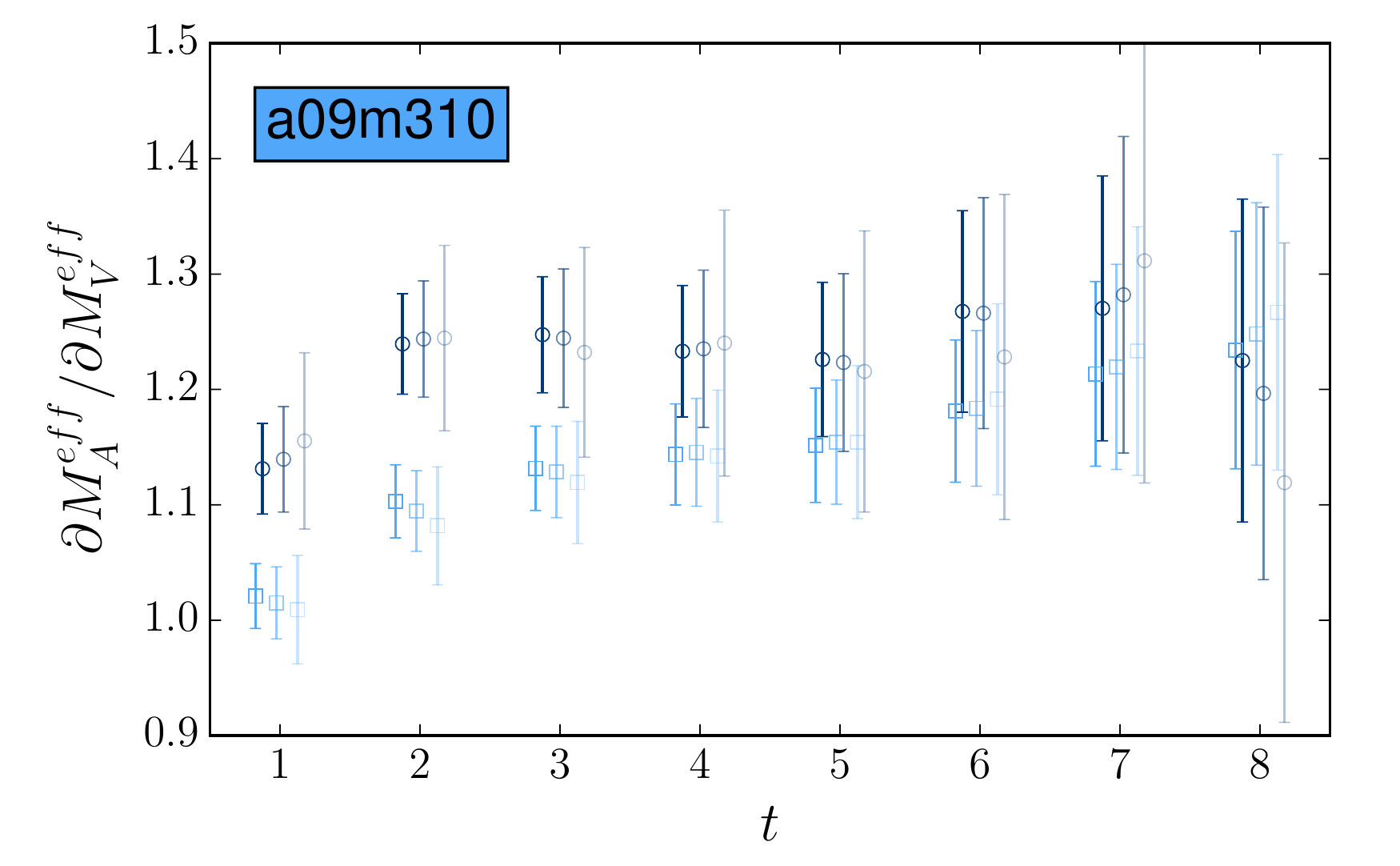}
\caption{\label{fig:ga_flow}
Flow study for the ({\it{top}}) a15m310 and ({\it{bottom}}) a09m310 ensembles. The effective mass of the derivative correlater is shown with point-sink ($\medsquare$) and smeared-sink ($\medcircle$) correlation functions. For each time slice, data for $t_{gf}=\{1.0,~0.6,~0.2\}$ are displayed from left to right with progressively lighter shades.}
\end{figure}

Following Ref.~\cite{Berkowitz:2017opd}, we study the flow time dependence of the gradient-flow smearing used in this calculation. In Fig.~\ref{fig:ga_flow}, we show ratios of the axial over the vector effective derivatives for the a15m310 and a09m310 ensembles with 196 configurations at a single source. The point-sink (squares) and smeared-sink (circles) effective derivatives are plotted with $t_{gf} = 1.0, 0.6, 0.2$ respectively from left to right. In both ensembles, we observe minimal flow time dependence in the ratio of correlators. Additionally the flow time is fixed to $t_{gf}=1.0$ in lattice units on all gauge configurations, ensuring that any quantity extrapolated to the continuum limit will be flow-time independent. For $t_{gf}=1.0$, we find it sufficient to solve the Wilson-flow diffusion equation with 40 integration steps using the Runge-Kutta algorithm. Furthermore, we observe smaller stochastic uncertainty at increasingly larger values of $t_{gf}$. We find these conclusions consistent with what was observed in Ref.~\cite{Berkowitz:2017opd} for other hadronic quantities ({\it{e.g.}} Fig.~3 therein).

Throughout this paper, we use the subscript $5$ (\textit{e.g.} $m_{\pi,5}$) to denote quantities relevant to the HISQ taste-5 pion, and drop the subscript for the pion constructed from Domain-Wall valence quarks (\textit{e.g.} $m_\pi$).

\begin{table*}
\begin{ruledtabular}
\begin{tabular}{llrrcccccc}
abbr. name & ensemble                                           & $N_{cfg}$ & $N_{srcs}$ & volume               & $\sim a$ [fm] & $\sim m_{\pi,5}$ [MeV] & $\sim m_{\pi,5} L$& $\s_{smr}$& $N_{smr}$ \\
\hline
a15m310  &  l1648f211b580m013m065m838a      & 1960        & 24               & $16^3\times48$
& 0.15               & 307                                 & 3.78                    & 4.2            & 60 \\
a12m310  &  l2464f211b600m0102m0509m635a  & 1053        &   4               & $24^3\times64$
& 0.12               & 305                                 & 4.54                    & 4.5            & 60 \\
a09m310  &  l3296f211b630m0074m037m440e    &   784        &   8               & $32^3\times96$
& 0.09               & 313                                 & 4.50                    & 7.5            & 167 \\
a15m220  &  l2448f211b580m0064m0640m828a  & 1000        & 12               & $24^3\times48$
& 0.15               & 215                                 & 3.99                    & 4.5            & 60 \\
a12m220S&  l2464f211b600m00507m0507m628a& 1000        &   4              & $24^3\times64$
& 0.12               & 218                                 & 3.22                    & 6.0            & 90 \\
a12m220  &  l3264f211b600m00507m0507m628a& 1000        &   4              & $32^3\times64$
& 0.12               & 217                                 & 4.29                    & 6.0            & 90 \\
a12m220L&  l4064f211b600m00507m0507m628a& 1000        &   4              & $40^3\times64$
& 0.12               & 217                                 & 5.36                    & 6.0            & 90 \\
a15m130 &  l3248f211b580m00235m0647m831a & 1000        &   5              & $32^3\times48$
& 0.15               & 131                                 & 3.30                    & 4.5            & 60 
\end{tabular}
\end{ruledtabular}
\caption{\label{tab:hisq} The HISQ ensembles used in this work along with the number of configurations $N_{cfg}$, number of sources per configuration $N_{src}$, lattice volume, approximate lattice spacing $a$, approximate HISQ taste-5 pion mass, and approximate value of $m_{\pi,5} L$. Values are obtained from Table I of Ref.~\cite{Bazavov:2012xda} with increased number of configurations.
We also list the values of the gauge invariant Gaussian source smearing algorithm used.}
\end{table*}


\section{\label{sec:Analysis}Correlator analysis}
\subsection{Statistics and autocorrelations}
The large time extent of the HISQ configurations allows us to compute correlation functions with multiple sources, totaling those listed in Tab.~\ref{tab:hisq}.
On each configuration, for a series of evenly spaced time-locations, a seeded random origin is chosen, $(x_0,y_0,z_0,t_0)$.
Then, at this origin, and its antipode, $(x_0,y_0,z_0,t_0) + L/2(1,1,1,0)$ (modular the periodic spatial boundary conditions) a source is generated to solve for the MDWF propagators.
The sources are created with \texttt{Chroma} using a \texttt{SHELL\_SOURCE} with the \texttt{GAUGE\_INV\_GAUSSIAN} routine.  The width ($\s_{smr}$) and number of iterations ($N_{smr}$) are also listed in Table~\ref{tab:hisq}.  These parameters were chosen to reduce the contamination from excited states in the proton while not observing signs of over-smearing.

We shift all time sources to $t_0 \rightarrow 0$ and average over all sources before analyzing the correlation functions, where $t$ is the source-sink separation time. We observe no correlation between different sources, resulting in a reduction of $\sqrt{N_{src}}$ in statistical uncertainty.
We further double the statistical sampling by generating analogous correlation functions for the negative parity nucleon. Under time-reversal, the nucleon reverses parity, allowing us to average the forward propagating nucleon correlation functions with the time-reversed negative-parity nucleon correlation functions.

We perform a simultaneous fit of the two-point correlator and the axial effective derivative together with the vector effective derivative which further constrains the ground state energy of the nucleon, and indirectly leads to a more precise extraction of the axial charge $\mathring{g}_A$, where the ring indicates this as the bare charge. Futhermore, the simultaneous fit naturally gives access to the correlated ratio of $\mathring{g}_A/\mathring{g}_V$ and simplifies the renormalization procedure, which is discussed in Sec.~\ref{sec:renorm}.

The Feynman-Hellmann method used to construct the three-point correlation functions leads to an increase of $\mathrm{O}(t)$ in statistical sampling from equal computation time from the sum over the current insertion. The method also yields the complete source-sink separation time dependence of the correlation function, leading to exponential improvement in the signal-over-noise ratio for the correlator at small source-sink separation times. The improvement in signal quality is demonstrated in detail in Ref.~\cite{Bouchard:2016heu}.

\begin{figure}[h]
\includegraphics[width=\columnwidth]{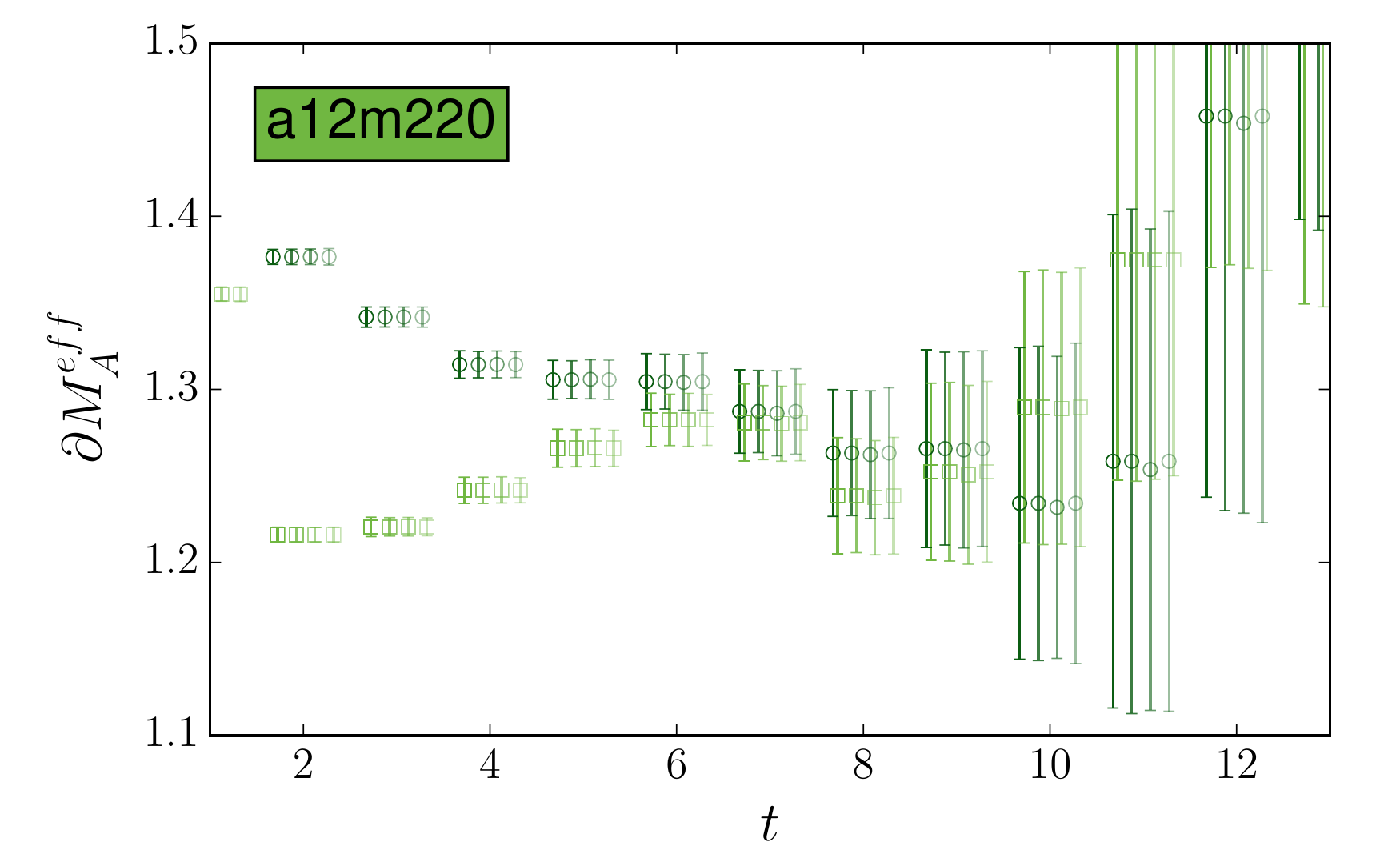}
\caption{Plot of the effective derivative of the axial form factor for the a12m220 ensemble as a function of source-sink separation $t$ for point- ({\color{knbgreen}$\medsquare$}) and smeared-sinks ({\color{kngreen}$\medcircle$}). Bin sizes of 1 (unbinned), 2, 3, and 4 are shown with increasingly lighter shades from left to right for both choices of sink smearing. } \label{fig:autocorrelation}
\end{figure}

We study possible autocorrelations in our data set by binning the derivative correlation functions for every ensemble used in this work. Fig.~\ref{fig:autocorrelation} shows a representative example of a binning study. We observe that both the central value and standard deviation of the raw correlation function is stable under binning for bin sizes up to four, demonstrating that no autocorrelations are present in the data. The complete binning study is presented in Appendix~\ref{app:autocorrelation}. We do not bin any of our data in this work.

\subsection{Method}
We construct the effective derivative from the two-point and derivative correlation function as detailed in Eq.~(\ref{eq:derivative_meff}) for the axial-vector and vector currents. 
A simultaneous fit is then performed to the two-point correlation function in tandem with the effective derivative of both currents, leading to a tripling of the amount of correlated data when determining a large subset of shared fit parameters ({\it{i.e.}} $E_n$ and $z_n$). 
In order to perform the simultaneous fit, we first explore the parameter space with a Bayesian constrained fit. 
The central value of the posterior is then used as an initial guess for the corresponding frequentist fit. 
For our preferred fit, we use the results from the frequentist fit, and motivate our choice of fits under frequentist inference. The correlated uncertainty of the fit parameters is propagated through bootstrap resampling. We perform both the initial Bayesian fit and the final frequentist fit using \texttt{lsqfit}~\cite{lsqfit}.

We use the fit ansatz given by Eqs.~(\ref{eq:twopt_fit_ansatz}) and (\ref{eq:dmeff_fit_ansatz}) respectively for the two-point and effective derivatives respectively. For our preferred fit, we choose to fit the effective derivative, which involves the ratio of the two-point and derivative correlation functions, because the ratio eliminates leading order excited-state contamination. In practice, we often times are able to fit closer to the origin for the effective derivative compared to the two-point correlation function because of this cancellation. However, we find our results statistically consistent compared to fits using Eq.~(\ref{eq:threept_fit_ansatz}) (directly fitting the numerator in the derivative correlation function).

\subsection{Bayesian preconditioning}
\label{sec:bayes_preconditioning}
The initial Bayesian constrained curve fit serves to provide starting values of fit parameters for the preferred unconstrained fit; additionally with sensible priors, convergence to the maximum likelihood estimate of the fit parameters is accelerated, and therefore provides an efficient method for an initial survey of a wide range of fit regions. Since the distribution of expectation values, hence averages, are of interest, the central limit theorem guarantees that the posteriors are Gaussian distributed in the limit of large statistics, therefore we use Gaussian distributions for all the priors.

The two-point correlation function depends on the energies and overlap factors, $E_n$ and $z_n$ respectively. The prior for the ground state energy is chosen by observing the large $t$ limit of the effective mass given by Eq.~(\ref{eq:meff_source}). The prior width of the ground state energy is set to 10\% of the central value, approximately two orders of magnitude larger than the width of the posterior distribution. The excited state energy splitting is set to the two pion splitting with a log-normal distribution, with the lower bound of the 68\% confidence interval set to the one pion splitting. The smeared and point overlap factors are given respectively by,
\begin{align}
z_0^S = & \lim_{t\rightarrow \infty}\sqrt{C_{SS}(t) e^{E_0t}},\\
z_0^P = & \lim_{t\rightarrow \infty}C_{PS}(t)e^{E_0t}/z_0^S.
\end{align}
The prior width of the overlap factors are set to half the central value, constraining the parameter to be non-zero within two standard deviations. The excited state overlap factors are set to a central value of zero, with a width that is five and ten times the ground state central value for the smeared and point overlap factors respectively. The smeared overlap factors are set with a tighter prior with the expectation that smearing enhances the overlap with the ground state.

A simultaneous fit with the axial effective derivative includes $g_{nm}^A$ and $d_n^A$, parameterizing the axial charge and unwanted time-ordering contributions respectively, as described by Eqs.~\eqref{eq:define_gnm} and \eqref{eq:define_dn}. The prior for the ground state axial charge parameter is set by observing the long time limit of the effective derivative given by Eq.~(\ref{eq:derivative_meff}). The prior width is set to half the central value, approximately one order of magnitude larger than the resulting width of the posterior distribution. The excited state axial charge parameters are set to zero, with a width that is one order of magnitude larger than the ground state central value. The time-ordering artifact parameter is set to,
\begin{equation}
d^A_{SS(PS);0} = \frac{1}{2} N_A^{SS(PS)}(1) e^{E_0}
\end{equation}
as discussed in Section~\ref{Sec:spectral_decomp}, where the factor of one-half reflects the expectation that smearing enhances the overlap with the ground state. The width of $d_0^A$ ensures that the parameter is zero within one standard deviation. For excited states, $d_n^A$ is set to zero, and with the same width as the ground state.

The addition of the vector charge correlation function introduces $g_{nm}^V$ and $d_n^V$ for the vector charges, with priors set analogously to the axial charge parameters.

\begin{figure*}[h]
\includegraphics[width=0.49\textwidth]{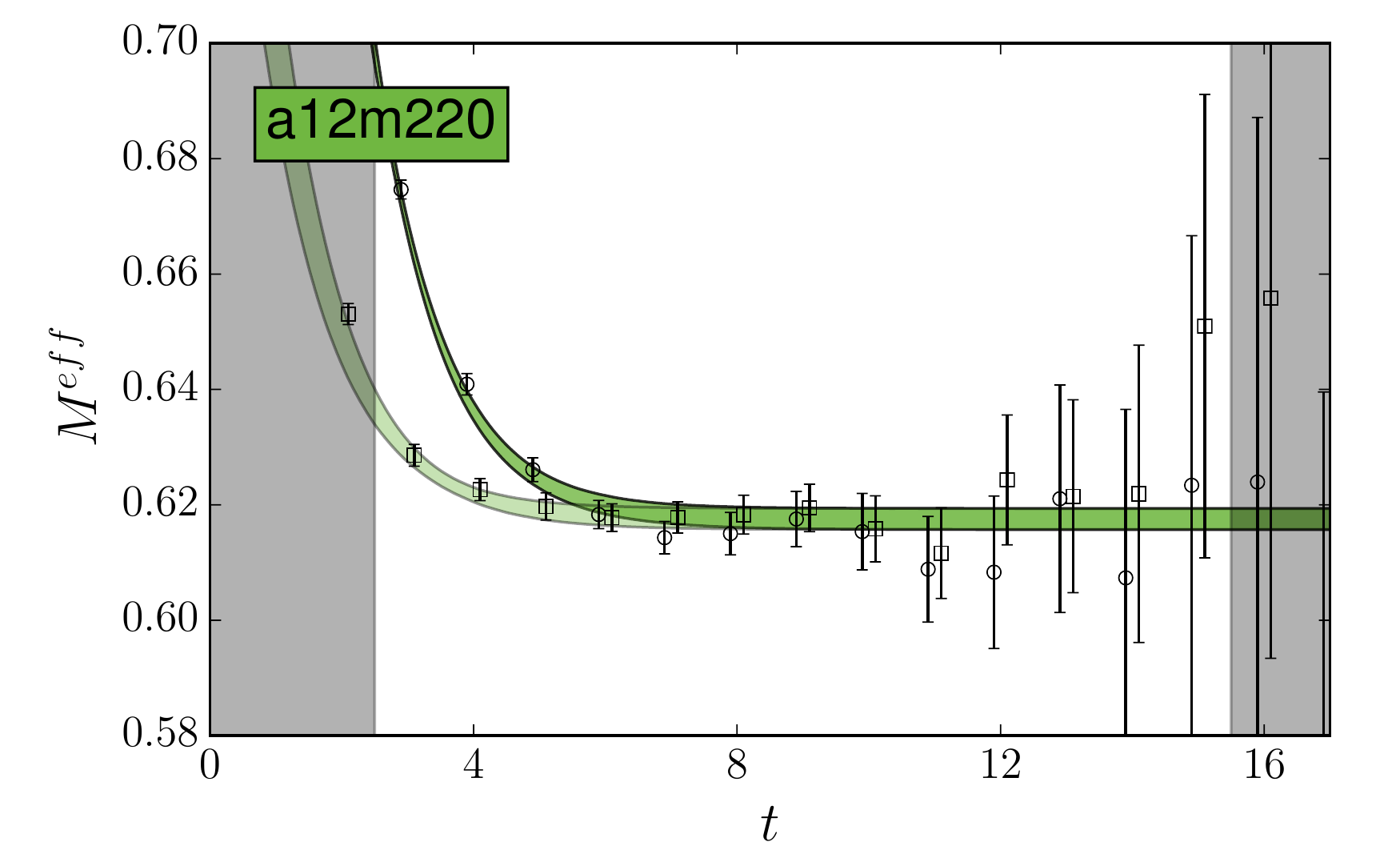}
\includegraphics[width=0.49\textwidth]{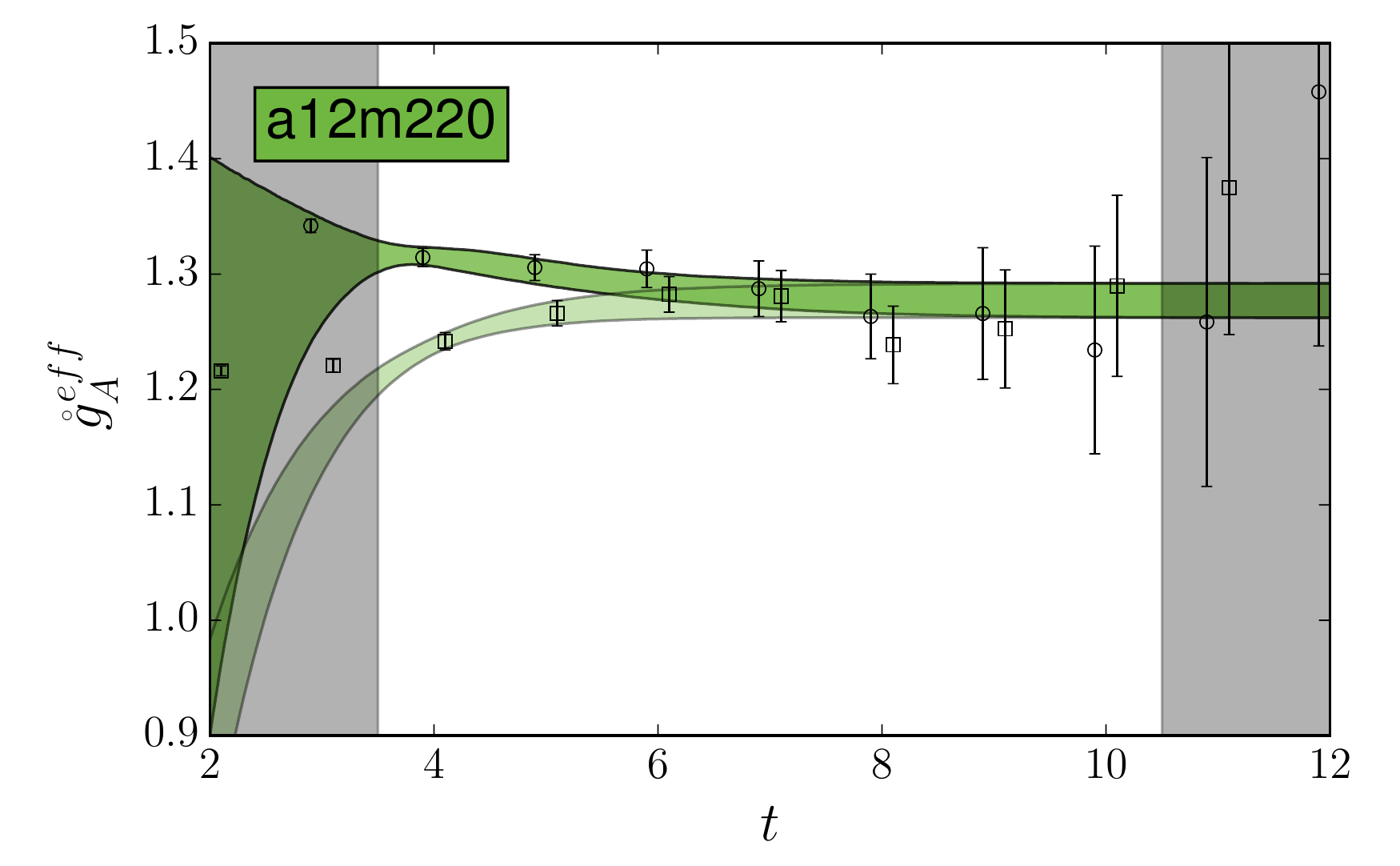}
\includegraphics[width=0.49\textwidth]{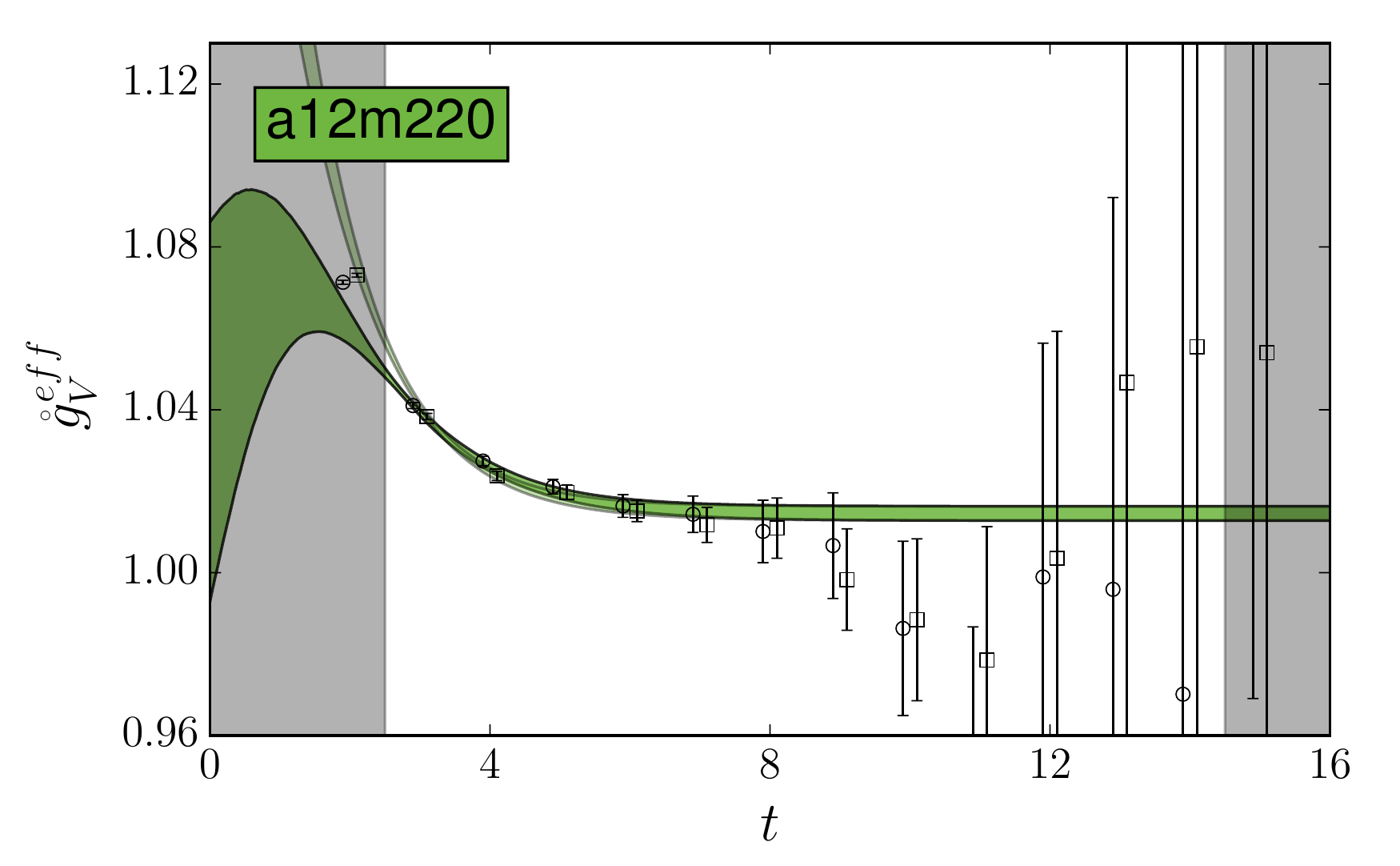}
\includegraphics[width=0.49\textwidth]{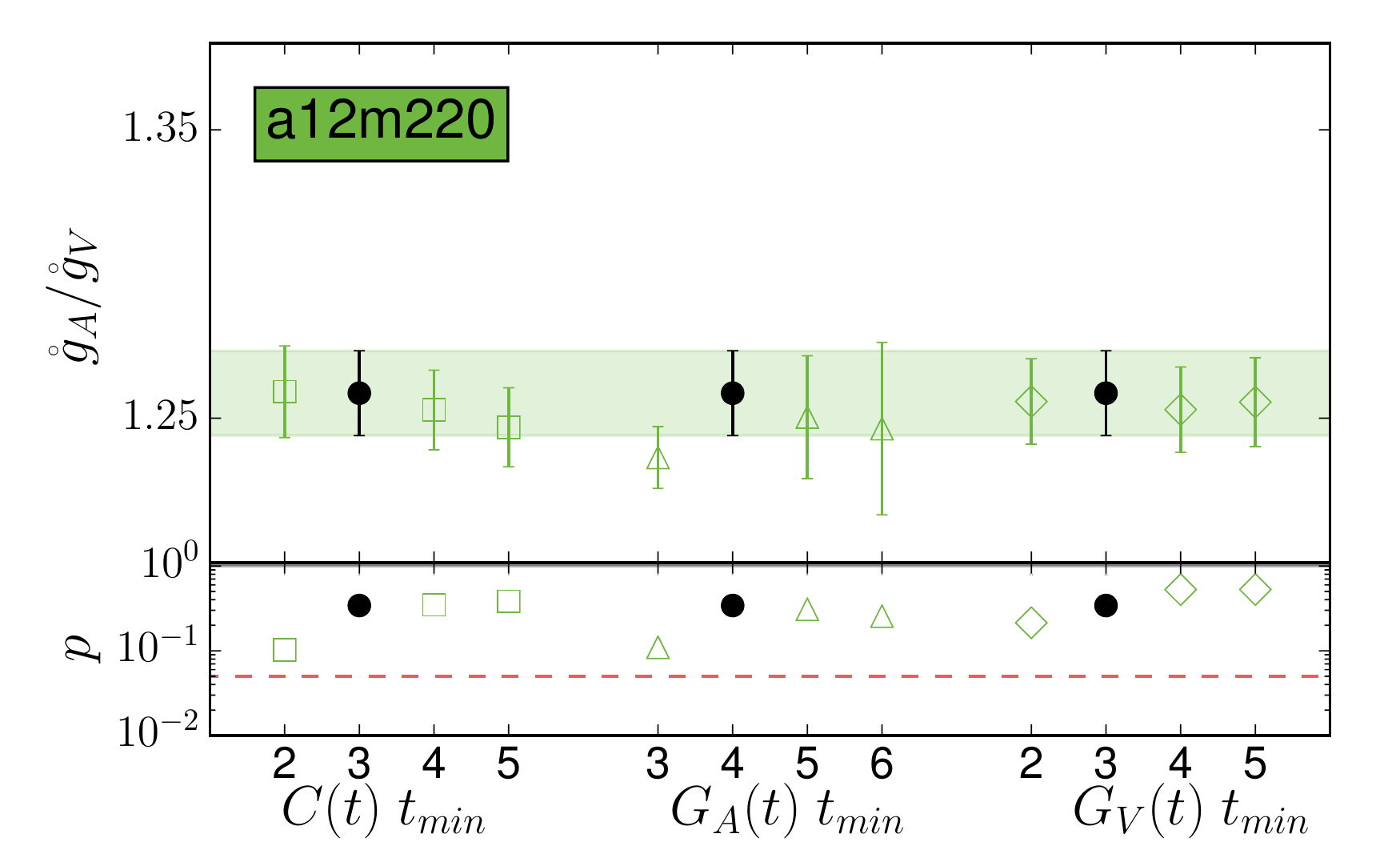}
\includegraphics[width=0.49\textwidth]{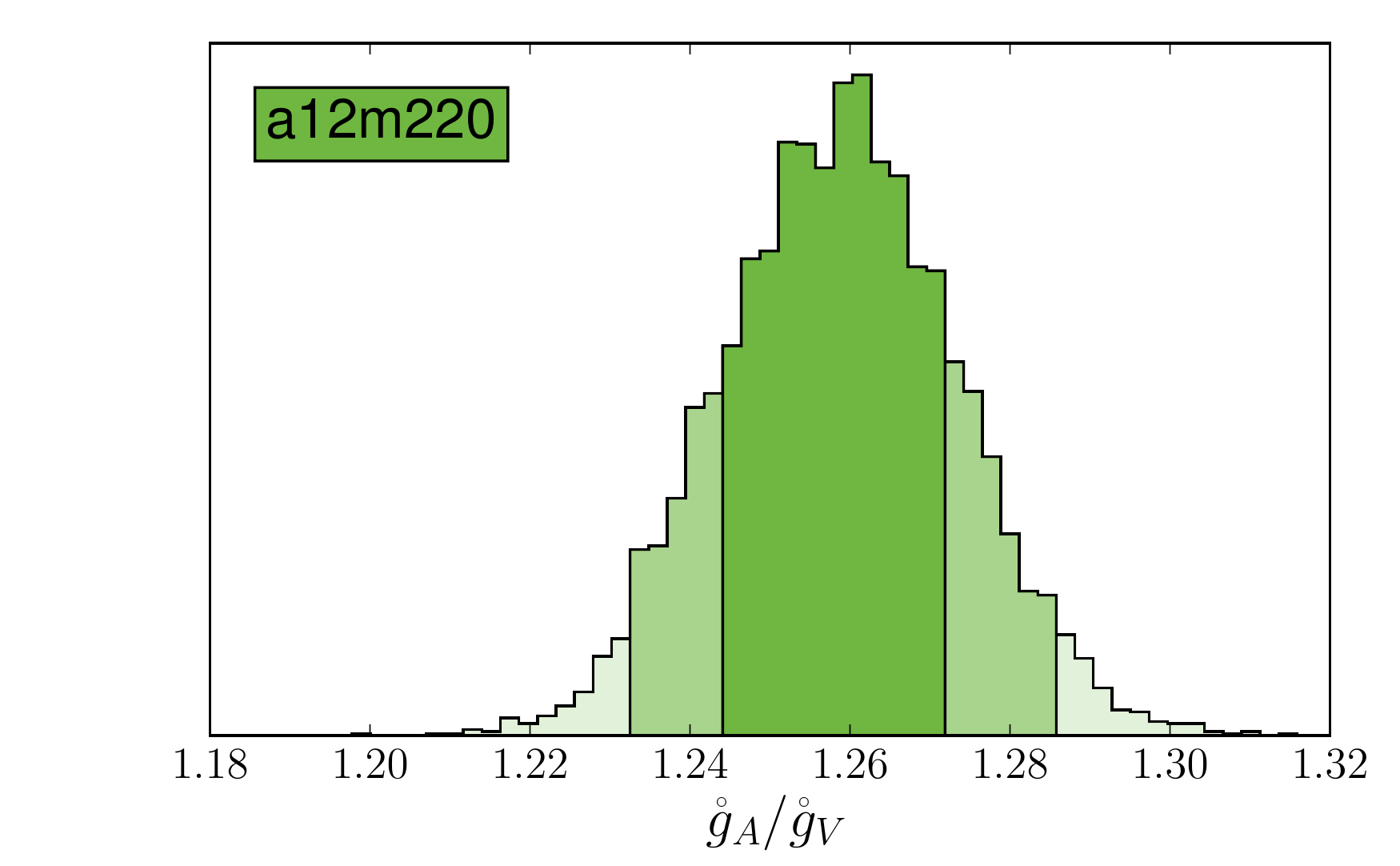}
\includegraphics[width=0.49\textwidth]{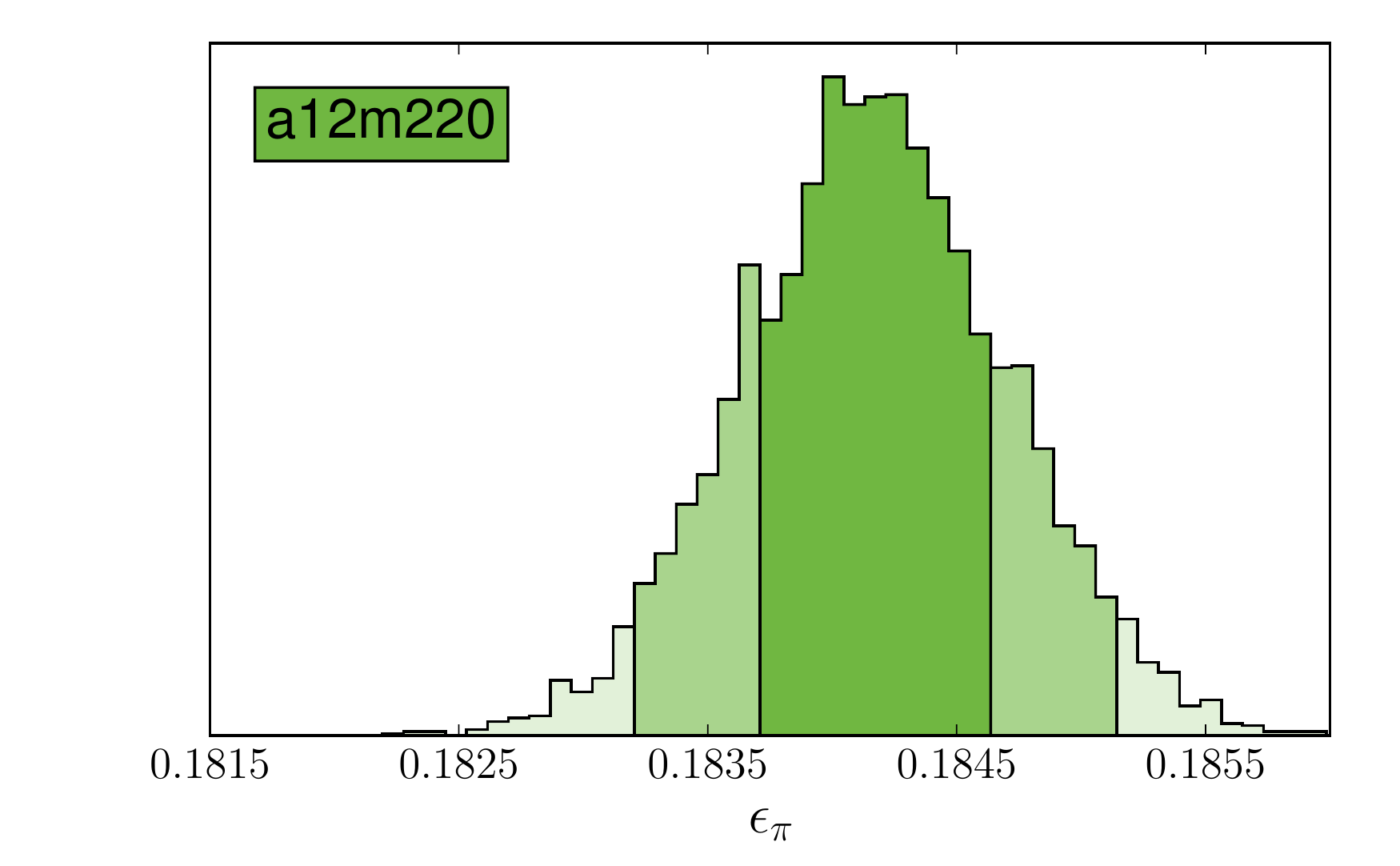}
\caption{Analysis of the a12m220 ensemble with 5000 bootstrap samples. (\textit{top left}) effective mass (\textit{top right}) axial effective derivative (\textit{middle left}) vector effective derivative are shown with smeared-sink ($\medsquare$) and point-sink($\medcircle$) correlation functions, with corresponding reconstructed fit curves plotted in light- and dark-green respectively. The grey regions encompass data not included in the analysis. The data is staggered for clarity. (\textit{middle right}) Stability plot of $\mathring{g}_A/\mathring{g}_V$ for ensemble a12m220. The preferred fit is presented by the solid black symbol, the green band shades the 68\% confidence interval and helps guide the eye. Variations of the fit region of the two-point correlator ({\color{kngreen}$\medsquare$}), $G_A(t)$ ({\color{kngreen}$\medtriangleup$}), and $G_V(t)$ ({\color{kngreen}$\meddiamond$}) are presented. The corresponding frequentist $p$-values are plotted below, with the dashed red line at $p=0.05$ discriminating the statistical significance of the fit results. Uncertainty of the fit variations are determined by 1000 bootstrap samples. (\textit{bottom left}) Bootstrap histogram of $\mathring{g}_A/\mathring{g}_V$. Different shaded regions mark the 68\% and 95\% confidence interval. The central value of $\mathring{g}_A/\mathring{g}_V$ is consistent with the median at the sub-percent level. (\textit{bottom right}) Bootstrap histogram for $\e_\pi$, discussed in Sec.~\ref{sec:e_pi}. The shaded regions are defined similarily to the $\mathring{g}_A/\mathring{g}_V$ histogram.}
 \label{fig:a12m220_curve}
\end{figure*}

\subsection{Frequentist analysis}
With knowledge of the Bayesian analysis, we perform two-state unconstrained simultaneous fits to the two-point correlation function, and axial and vector effective derivatives. Only fit results with a frequentist $p$-value greater than 0.05 are considered as statistically significant. Additionally, excited state systematic uncertainty is controlled by checking for stability of the unconstrained fits under varying regions of source-sink separation in all three correlation functions, with the preferred fit lying in the region of stability. The correlated uncertainty is propagated by bootstrap resampling. For the preferred fit, 5000 bootstraps are sampled, while 1000 bootstraps are sampled for others. In this section the final result of the simultaneous fit for the a12m220 ensemble are presented, and serve as a representative example of all correlator analyses performed in this work. A compilation of all correlator analyses are presented in Appendix~\ref{app:bs_correlator_fits}.

The preferred fit for ensemble a12m220 is presented in Fig.~\ref{fig:a12m220_curve}. A simultaneous two state unconstrained fit is performed to all six correlators shown, with a frequentist $p$-value of 0.34. The correlated ratio of $\mathring{g}_A/\mathring{g}_V$ reconstructed under bootstrap results in a near perfect Gaussian distribution. The $p$-value for all fits are listed in Table~\ref{tab:ME} and accompanying bootstrap histograms in Appendix~\ref{app:bs_correlator_fits}. We further demonstrate the robustness of these fits by demonstrating stability of $\mathring{g}_A/\mathring{g}_V$ under varying values of $t_{min}$. 

Fig.~\ref{fig:a12m220_curve} shows the stability of the preferred fit under varying values of $t_{min}$. In each fit variation, only the fit region of one correlator was varied. We observe statistically insignificant changes under variations of the fit region with the exception of fitting $G_A(t)$ more aggressively to $t_{min}=3$. The approximately one and a half standard deviation shift in the central value observed in the exceptional fit suggests that possibly additional states would be needed to describe excited state systematics of a fit including this time-slice. We conclude from Fig.~\ref{fig:a12m220_curve} that the preferred fit sits in the region of stability under varying fit regions in all three correlators. Similar studies are performed on all ensembles pertinent to this work, with the preferred fits chosen in regions with stability. Fit regions for all ensembles are presented in Tab.~\ref{tab:fit_region}. Stability plots for all ensembles are presented in Appendix~\ref{app:stability}. The complete set of correlator analysis results are tabulated in Tab.~\ref{tab:correlator_results}.

Finally, the reconstructed curve of the preferred fit is shown on top of the effective mass and derivatives. The ability to control the fit at small time separations provides leverage for identifying the plateau region at larger time separations. In particular, we observe agreement in the plateau region of the fit and data for the effective mass and axial effective derivative within one standard deviation, and identify behavior for the vector effective derivative at large time as a statistical fluctuation resulting slight tension. This hypothesis is further supported by the fact that the ratio $\mathring{g}_A/\mathring{g}_V$ is shown to be stable under varying separation time in the axial effective derivative.

In summary, we observe our data to be flow-time independent and free of autocorrelations. We also observe our unconstrained fits to be stable under bootstrap resampling, yielding nearly ideal Gaussian distributions for $\mathring{g}_A/\mathring{g}_V$. Stability of  $\mathring{g}_A/\mathring{g}_V$ under varying fit regions is demonstrated. 
The reconstructed fit curves and error bands are in good simultaneous agreement with the two-point correlation functions and the axial and vector effective derivatives.
With this preponderance of evidence, we demonstrate full control over systematic uncertainty emergent from the fit procedure at unprecedentedly small time separations. As a result, we have a robust, exponentially improved determination of the nucleon axial charge.

\begin{table}
\begin{ruledtabular}
\begin{tabular}{lcccccc}
ensemble& $t^C_{\text{min}}$& $t^C_{\text{max}}$& $t^A_{\text{min}}$& $t^A_{\text{max}}$& $t^V_{\text{min}}$& $t^V_{\text{max}}$ \\
\hline
a15m310  & 4 & 14 & 4 & 10 & 4 & 15 \\
a12m310  & 7 & 12 & 2 & 12 & 5 & 15 \\
a09m310  & 6 & 16 & 3 & 12 & 6 & 17 \\
a15m220  & 2 & 12 & 2 & 8  & 4 & 11 \\
a12m220S & 4 & 10 & 5 & 10 & 3 & 10 \\
a12m220  & 3 & 15 & 4 & 10 & 3 & 14 \\
a12m220L & 4 & 12 & 4 & 12 & 5 & 10 \\
a15m130  & 2 & 10 & 2 & 6  & 2 & 7  \\
\end{tabular}
\end{ruledtabular}
\caption{Fit regions in lattice units for the two-point correlation function ($C$), and the axial ($A$) and vector ($V$) effective deriatives.}
\label{tab:fit_region}
\end{table}

\subsection{$\e_\pi$ and $\e_{ju}$ \label{sec:e_pi}}
The chiral-continuum extrapolation can be reparameterized to depend on the dimensionless quantities
\begin{align}
    \e_\pi\equiv \frac{m_\pi}{4\pi F_\pi} && \text{and} && \e_{ju}\equiv \frac{m_{ju}}{4\pi F_\pi}
\end{align}
circumventing the necessity of performing a scale-setting analysis.
$m_{ju}$ is the mass of the mixed-pion composed of one valence and one sea quark.
Calculation of the pion mass, $m_\pi$, and the pion decay constant, $F_\pi$, are performed on the same lattice actions, gauge configurations, and sources as the main analysis of this paper, while the calculation of $m_{ju}$ is performed with just one source.

A Bayesian constrained fit with a two-state fit ansatz is performed on the pion two-point correlation function in order to extract $m_\pi$ and its overlap factors; in Ref.~\cite{Berkowitz:2017opd}, we show that oscillating states present in the Domain-Wall action are highly suppressed when the gauge fields are smeared with gradient-flow, and therefore are neglected in this analysis. A simultaneous fit to both the point-sink and smeared-sink correlators is performed, and statistics is further doubled by exploiting the property of time-reversal symmetry in the meson correlators. Similarly, a Bayesian constrained fit to a constant is performed to extract $m_\text{res}$ from the $m_{\text{res}}$ correlator, as defined by Eq.~(5) of Ref.~\cite{Berkowitz:2017opd}. The 5D Ward Identity is used to obtain $F_\pi$ from $m_\pi$ and $m_{\text{res}}$ as given in Eq.~(6) of Ref.~\cite{Berkowitz:2017opd}.

A Bayesian constrained fit to 2+2 states is performed on the mixed meson pion correlator due to oscillations present in the HISQ action. A single fit to the point-sink correlator is performed. We increase statistics by folding the correlator.

Similar to Sec.~\ref{sec:bayes_preconditioning}, for both the pion and mixed action pion two-point correlator, the ground state priors for the pion mass and overlap factors are determined by the long-time limit of the effective mass and scaled correlators. The ground state prior widths are set to 10\% of the prior central value, approximately two orders of magnitude larger than the width of the posterior distribution, thus leaving the ground effectively state unconstrained. The prior for the excited state energy splitting is lognormal and is set by approximately the mass splitting between the $a_0$ meson and pion, with a width encompassing the two-pion splitting within one standard deviation. For the mixed pion, we set the splitting to the opposite parity state to $m_\sigma-m_\pi\approx 375~\text{MeV}$. The prior for $m_{\text{res}}$ is set by plotting the $m_{\text{res}}$ correlator, with a width set to 10\% of the central value, and is approximately one order of magnitude larger than its posterior distribution.

The fit regions for the pion two-point correlation functions and the $m_{\text{res}}$ correlators are chosen in the region of stability under varying choices of source-sink separation time to ensure full control over excited state systematic uncertainty.

Uncertainties are propagated by bootstrap resampling. Random bootstrap draws are prepared in advance and saved, guaranteeing that identical bootstrapped configurations are generated for each ensemble across different datasets. For the preferred fits, 5000 bootstrapped configurations are analyzed, allowing for a correlated analysis with the bootstrapped samples of $\mathring{g}_A/\mathring{g}_V$. For each bootstrap, the prior central values for all parameters are set to a value randomly drawn from their corresponding initial prior distributions. The bootstrap histogram for $\e_\pi$ for the a12m220 ensemble is provided in Fig.~\ref{fig:a12m220_curve}. The exhaustive list of histograms for $\e_\pi$ is provided in Appendix~\ref{app:Fpi_bs} and are all demonstrably Gaussian distributed.

\begin{table}
\begin{ruledtabular}
\begin{tabular}{lcccc}
ensemble& $am_\pi$& $am_{res}\times 10^{4}$& $aF_\pi$& $am_{ju}$\\
\hline
a15m310  & 0.2362(2) & 9.56(7) & 0.0753(1) & 0.3061(12)\\
a12m310  & 0.1888(2) & 7.71(6) & 0.0615(1) & 0.2189(08)\\
a09m310  & 0.1409(1) & 2.69(3) & 0.0455(1) &  0.1481(05)\\
a15m220  & 0.1657(2) & 5.75(4) & 0.0727(1) &  0.2554(20)\\
a12m220S & 0.1357(2) & 3.99(4) & 0.0587(1) &  ---\\
a12m220  & 0.1343(1) & 4.04(3) & 0.0587(1) &  0.1770(10)\\
a12m220L & 0.1341(1) & 4.05(2) & 0.0588(1) &  ---\\
a15m130  & 0.1010(2) & 2.55(2) & 0.0708(2) &  0.2252(45)\\
\end{tabular}
\end{ruledtabular}
\caption{\label{tab:spectrum} The pion spectrum and $F_\pi$ are needed for extrapolation.
The quantity $\e_\pi = m_\pi / (4\pi F_\pi)$ is used for the chiral extrapolation. The mixed-pion masses $am_{ju}$ are generated and analyzed on the same ensembles as the rest of this work, except for the a12m220S and a12m220L ensembles, for which we assume the a12m220 value.}
\end{table}

\begin{table}
\begin{ruledtabular}
\begin{tabular}{lllcc}
ensemble   &\multicolumn{1}{c}{$\mathring{g}_A$}&\multicolumn{1}{c}{$\mathring{g}_V$}& $p$ & $Z_A / Z_V -1$ \\
\hline
a15m310   & 1.216(11)           & 1.0009(16)         & 0.32 & $14(01)\times10^{-7}$ \\
a12m310   & 1.261(08)           & 1.0191(11)         & 0.34 & $54(04)\times10^{-7}$ \\
a09m310   & 1.289(14)           & 1.0248(13)         & 0.06 & $69(03)\times10^{-7}$  \\
a15m220   & 1.229(09)           & 0.9959(56)         & 0.07 & $72(46)\times10^{-7}$  \\
a12m220S & 1.294(28)           & 1.0177(34)         & 0.35 & $20(04)\times10^{-6}$ \\
a12m220   & 1.277(15)           & 1.0146(17)         & 0.34 & $20(04)\times10^{-6}$ \\
a12m220L & 1.276(21)           & 1.0198(49)         & 0.09 & $20(04)\times10^{-6}$ \\
a15m130   & 1.262(53)           & 0.994(35)           & 0.05 & $47(53)\times10^{-5}$   \\
\end{tabular}
\end{ruledtabular}
\caption{\label{tab:ME} The resulting bare axial-vector and vector couplings with their corresponding $p$-values from the correlation function fits on ensembles used in this work.  The renormalized axial coupling is determined by $g_A = (Z_A / Z_V) (\mathring{g}_A / \mathring{g}_V)$. The uncertainty is obtained from the bootstrapped standard deviation.}
\label{tab:correlator_results}
\end{table}

\section{\label{sec:renorm}Renormalization}
In the isospin limit, the axial charge $g_A$ is computed from the ratio of bare $\mathring{g}_A / \mathring{g}_V$ and the ratio of the renormalization constants as the vector charge is normalized to $Z_V \mathring{g}_V = 1$,
\begin{equation}
  g_A = (Z_A / Z_V) (\mathring{g}_A / \mathring{g}_V)\, ,
\end{equation}
where $Z_A$ and $Z_V$ are the renormalization factors of the axial-vector and vector current to convert from the bare, local to renormalized currents. Since we use a formulation which preserves chiral symmetry to very good approximation, we expect $Z_A=Z_V$ up to some lattice artifacts and potential IR (infrared) contamination. 
However, since the ratio $Z_A/Z_V$ can be computed very precisely through an NPR (non-perturbative renormalization) procedure, we might observe a deviation from $Z_A/Z_V=1$. We implement the Rome-Southampton renormalization method~\cite{Martinelli:1994ty}, with non-exceptional kinematics~\cite{Sturm:2009kb}. We also implement momentum sources, as proposed in~\cite{Gockeler:1998ye}, leading to very high statistical accuracy (see also~\cite{Arthur:2010ht}). We first solve \begin{equation} \sum_{x} D(y,x) \tilde G_x(p) = e^{ip.y}
\end{equation}
and then multiply by the appropriate phase factor
\begin{equation}
\label{eq:mom_prop}
G_x(p) =  \tilde G_x(p) e^{-ip.x} = \sum_y D^{-1} (x,y) e^{ip.(y-x)}
\end{equation}
to obtain the incoming momentum source propagator with momentum $p$, denoted by $G_x(p)$. As usual, the outgoing propagator (with momentum $-p$) is given by $\gamma_5$-hermiticity:
\begin{equation}
  \bar G_x(p) \equiv
  \gamma_5 G_x(p)^\dagger \gamma_5  = \sum_y D^{-1} (y,x)  {\rm e}^{ -i p\cdot(y-x)}\, .
\end{equation}
Incoming and outgoing refer to the point $x$ where the vertex is located.
For amputation, we introduce the full momentum propagator
\begin{equation}
 G(p) = \sum_x G_x(p) \, .
\end{equation}

Next, we define the bilinear two-point function
\begin{equation}
  V_\Gamma(p_2,p_1)  =
  \sum_x \left[ \bar{ G }_x(p_2) \Gamma G_x(p_1) \right]
\end{equation}
for $\Gamma = \gamma_\mu, \gamma_\mu \gamma_5$.
For the choice of momenta, we follow the SMOM condition:
$p_1 \ne p_2$ with $p_1^2=p_2^2=(p_1-p_2)^2$.
The amputated vertex function reads
\begin{align}
\Pi_\Gamma = &
\langle \bar G(p_2)^{-1} \rangle \langle V_\Gamma(p_2,p_1) \rangle \langle G(p_1)^{-1} \rangle
\;,
\end{align}
where $\langle O \rangle$ denotes the gauge average. The amputated vertex function is a matrix in Dirac-color space ($12\times 12$), which we still have to project onto its tree level value. We implement the so-called $\gamma_\mu$ and $\sla{q}$ schemes, with $q=p_2-p_1$. Explicitly for the vector $\Gamma=\gamma_\mu$, we have
\begin{align}
\Lambda_{\gamma_\mu}^{(\gamma_\mu)} = &{\rm P}^{(\gamma_\mu)}\left[ \Pi_{\gamma_\mu} \right]
= {\rm Tr}[\gamma_\mu \Pi_{\gamma_\mu}]\, ,  
\\
\Lambda_{\gamma_\mu}^{(\sla{q})} = &{\rm P}^{(\sla{q})}\left[ \Pi_{\gamma_\mu}\right]
= \frac{q_\mu}{q^2} {\rm Tr}[\sla{q} \Pi_{\gamma_\mu}]\, ,
\end{align}
where the trace is taken over both color and Dirac indices. Similarly, the projected Green's function for the axial current is obtained by the substitution $\gamma_\mu\to\gamma_\mu \gamma_5$ in the previous equations.
Denoting the scheme (s), and the scale $\mu$ (with $\mu^2 =p_1^2 = p_2^2 = (p_2-p_1)^2$ ) the renormalization conditions read
\begin{equation}
  \label{eq:rcond}
  \frac{ Z_\Gamma}{Z_q^{(s)}(\mu) }\times \Lambda^{(s)}_\Gamma (\mu) = F_\Gamma^{(s)}\;,
    \qquad \Gamma\in \{ \gamma_\mu, \gamma_\mu \gamma_5 \}\;.
\end{equation}
The only scale and scheme dependence in Eq.~(\ref{eq:rcond}) is due to the quark wave function renormalization factor $Z_q$ (as the vector and axial current are protected by Ward Identities). $F_\Gamma^{(s)}$ is the corresponding tree level value. For example, if we want $Z_V=Z_{\gamma_\mu}$, the momentum space propagators can be set to $G = \delta^{ij}\delta_{\alpha\beta}$ (with color indices $i, j$ and Dirac indices $\alpha, \beta$), resulting in
\begin{align}
  \left[ \Pi_{\gamma_\mu} \right]^{ij}_{\alpha\beta}
  \stackrel{\rm tree}{ =} &
  \left[ \gamma_\mu  \right]_{\alpha\beta} \delta^{ij}\, ,
  \\
    {\rm P}^{(\gamma_\mu)}\left[  \Pi_{\gamma_\mu} \right]
    \stackrel{\rm tree}{ =} &
    \left[ \gamma_\mu\right]_{\beta\alpha} \delta^{ji}
    \left[ \gamma_\mu \right]_{\alpha\beta}  \delta^{ij} = 4 \times 12 \;.
\end{align}
As we are only interested in the ratio $Z_A/Z_V$, we do not need to determine $Z_q$. In practice, we compute $Z_A/Z_V$ from
\begin{equation}
\frac{Z_A}{Z_V}  = \frac { \Lambda_{\gamma_\mu}^{(s)} } { \Lambda_{\gamma_\mu \gamma_5}^{(s)} }\, .
\end{equation}
\begin{figure}
\includegraphics[width=\columnwidth]{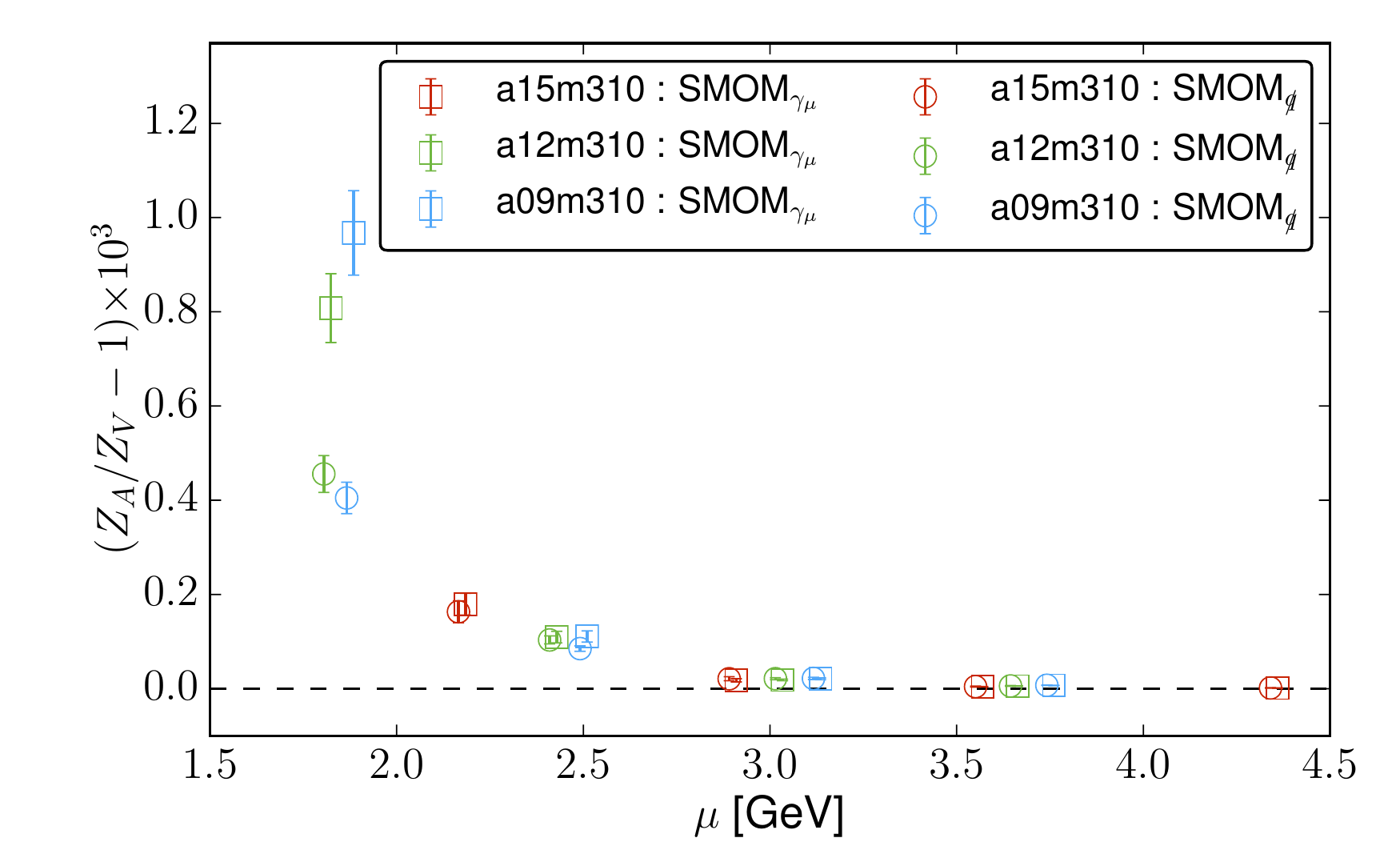}
\caption{\label{fig:za_zv}
Non-perturbative determination of $Z_A/Z_V$ using the SMOM momentum condition for two schemes ($\sla{q}$ and $\gamma_\mu$) on some of the ensembles used in this work.
These are representative of the results found on all ensembles.
}
\end{figure}

The values of the renormalization coefficients calculated from the $\gamma_\mu$ and $\sla{q}$ schemes agree well within uncertainty above $2~\text{GeV}$.  Representative results are shown in Fig.~\ref{fig:za_zv}, with similar results found for all ensembles used in this work. With momenta up to $4.5~\text{GeV}$, there is no evidence for an onset of growing discretization uncertainty for the ratio $Z_{A}/Z_{V}$. 
There is evidence of the IR contamination which all but vanishes for $\mu \gtrsim 3$~GeV.
This demonstrates systematic control over the renormalization coefficients over a wide range of momenta.

In principle, the renormalization conditions Eq.~(\ref{eq:rcond})
are imposed in the chiral limit (for a massless scheme),
so one should take the chiral limit.
In practice, we observe that even at finite mass,
the deviations from $Z_A/Z_V=1$ are tiny---as seen in Table~{\ref{tab:ME}}.
A detailed study of the renormalization of various matrix elements with our action~\cite{Berkowitz:2017opd}, including the vector and axial-vector currents will be provided in a forthcoming publication~\cite{c51_npr}.

\section{\label{sec:extrap}Chiral, continuum and infinite volume extrapolations}

Given the renormalized values of $g_A$, the remaining systematic uncertainties to control are the standard pion (quark) mass%
\footnote{Using $SU(2)$ $\chi$PT (Chiral Perturbation Theory)~\cite{Langacker:1973hh,Gasser:1983yg}, we can freely interchange between a quark mass expansion and pion mass expansion.  Unless specifically noted, we synonymously interchange between describing these chiral corrections as quark mass dependence or pion mass dependence.}
extrapolation and the continuum and infinite volume extrapolations.
EFT methods can be used to parameterize the dependence upon the pion mass, the discretization scale and the finite volume such that these systematic uncertainties can be controlled.

For a static quantity such as $g_A$, standard HB$\chi$PT (Heavy Baryon $\chi$PT)~\cite{Jenkins:1990jv} can be used.
Given the convergence issues of $SU(3)$ HB$\chi$PT~\cite{WalkerLoud:2008bp,Torok:2009dg,Ishikawa:2009vc,Jenkins:2009wv,WalkerLoud:2011ab}, a controlled extrapolation requires the use of the two-flavor theory in which the strange quark is integrated out~\cite{Bernard:1992qa}.
However, even in $SU(2)$ HB$\chi$PT it is not clear if $m_\pi\sim300$~MeV is a sufficiently light pion mass for a converging chiral expansion~\cite{WalkerLoud:2008bp,WalkerLoud:2008pj,Walker-Loud:2013yua}.
The NLO (next-to-leading order) chiral corrections to the nucleon axial charge were first determined in Ref.~\cite{Jenkins:1991es} in $SU(3)$ HB$\chi$PT including explicit decuplet baryons and in Ref.~\cite{Bernard:1992qa}, they were determined in the two-flavor theory without explicit deltas.
If Ref.~\cite{Kambor:1998pi}, the NNLO (next-to-next-to-leading order) chiral corrections of $\mathrm{O}(m_\pi^3)$ were determined in $SU(2)$ HB$\chi$PT.
In Ref.~\cite{Hemmert:2003cb}, explicit delta degrees of freedom were included in the $SU(2)$ corrections using the so-called small scale expansion~\cite{Hemmert:1996xg}.
In Ref.~\cite{Bernard:2006te}, the double logarithm coefficient arising from the two-loop contributions at NNNLO (next-to-next-to-next-to-leading order) was determined using the renormalization group.

We have results at three different pion mass values and should therefore restrict the extrapolation function to have at most two unknown coefficients describing the pion mass dependence, as a third parameter would simply amount to a model that could accomodate all three points.
The same is true for the continuum extrapolation, discussed further below.

At NLO in $SU(2)$ HB$\chi$PT, there are 2 unknown LECs which must be determined, the leading axial coupling in the chiral limit, $g_0$, and a local counter-term associated with a loop divergence proportional to $m_\pi^2$.
At NNLO, $\mathrm{O}(m_\pi^3)$, there is an additional counter-term.  However, this counter-term is not completely arbitrary as the NNLO contribution is purely non-analytic in the quark mass, and therefore it must be proportional to $g_0$.
There are still 3 LECs to be determined at NNLO, and since we have only 3 pion mass points, this extrapolation can be used to estimate the importance of higher order chiral corrections, but not for a robust extrapolation.

$g_A$ is a dimensionless quantity.  If one takes the $\chi$PT dimensional regularization scale to be $\mu=4\pi F_\pi$, then the entire extrapolation function can be expressed in terms of purely dimensionless quantities which we can be determined in the LQCD calculation.
The correction to this formula from using a quark mass dependent scale versus a fixed scale appears at NNNLO, $\mathrm{O}(m_\pi^4)$, as the difference between $4\pi F_\pi$ and $4\pi F_0$ is an $\mathrm{O}(m_\pi^2)$ correction.
Therefore, through NNLO, the pion mass extrapolation function can be simply expressed as
\begin{align}\label{eq:gA_epi}
g_A^{\chi\rm{PT}} &= g_{0}
	- \e_\pi^2 \left[
 		(g_{0} + 2 g_{0}^3) \ln \left( \e_\pi^2 \right)
		- c_2 \right]
\nonumber\\&\phantom{=}
 + g_0 c_3 \e_\pi^3
 +\cdots\, ,
\end{align}
where the $\cdots$ denote terms of higher order in the chiral expansion and all LECs are dimensionless.

Our results indicate very mild pion mass dependence within the range of pion masses used.  This is also consistent with LQCD results for $g_A$ from other groups, which show a nearly flat pion mass dependence up to pion masses on the order of 1~GeV~\cite{Dolgov:2002zm,Edwards:2005ym,Khan:2006de,Yamazaki:2008py,Bratt:2010jn,Alexandrou:2010hf,Capitani:2012gj,Green:2012ud,Horsley:2013ayv,Bhattacharya:2013ehc,Bali:2014nma,Alexandrou:2016xok,Yoon:2016jzj,Liang:2016fgy,Bhattacharya:2016zcn,Alexandrou:2016xok}.
We therefore also consider a simple Taylor expansion around a point $\e_0^2$
\begin{equation}\label{eq:ga_taylor}
g_A^T =
    c_0 + c_2 (\e_\pi^2 - \e_0^2) + c_4 (\e_\pi^2 - \e_0^2)^2 +\cdots\, .
\end{equation}
The Taylor expansion about $\e_\pi^2$ is synonymous with a Taylor expansion in the light quark mass.
One could also consider a Taylor expansion in $\e_\pi$, however this is synonymous with an expansion in $\sqrt{m_l}$, which is not a natural expansion in terms of the input parameters of QCD.
This form can be phenomenologically motivated through the observation of linear pion mass dependence, similar to that observed in the nucleon mass~\cite{WalkerLoud:2008bp,WalkerLoud:2008pj,Walker-Loud:2013yua}.
An extrapolation of our results in $(\e_\pi - \e_0)$ is consistent with that in Eq.~\eqref{eq:ga_taylor}, however, we do not consider it further in this work.

The quadratic term in the Taylor expansion requires the determination of a 3rd unknown parameter, as with the chiral expansion at NNLO, and so one can only use these higher order fits to estimate chiral extrapolation uncertainties.
For the linear extrapolation, the choice of $\e_0$ of course has no impact on the final result.

For sufficiently large volumes, it is trivial to incorporate the corrections arising from a finite periodic volume into $\chi$PT.
In the so-called $p$-regime~\cite{Gasser:1986vb}, one simply replaces the spatial integrals arising in quantum loop corrections with their finite volume sums.
For quantities without kinematic singularities appearing in the momentum integral, the difference between the finite volume and infinite volume quantum corrections can be shown to be suppressed exponentially in the volume, with an asymptotic suppression at least as strong as $e^{-m_\pi L}$.
The leading volume corrections arising at NLO for $g_A$ were first worked out in Ref.~\cite{Beane:2004rf}.
Using the notation of this reference but keeping only the contributions from intermediate nucleons, the volume corrections can be expressed in terms of the two dimensionless quantities $\e_\pi$ and $m_\pi L$,
\begin{align}\label{eq:gA_FV}
\d_L &\equiv g_A(L) - g_A(\infty)
\nonumber\\&
    = \frac{8}{3} \e_\pi^2 \left[
 g_{0}^3 F_1(m_\pi L)
 +g_{0} F_3(m_\pi L)
 \right]
\end{align}
where
\begin{align}
F_1(mL) &= \sum_{\mathbf{n} \neq 0} \left[
 K_0(mL|\mathbf{n}|) - \frac{K_1(mL|\mathbf{n}|)}{mL|\mathbf{n}|}
 \right]\, ,
\nonumber\\
F_3(mL) &= -\frac{3}{2} \sum_{\mathbf{n} \neq 0} \frac{K_1(mL|\mathbf{n}|)}{mL|\mathbf{n}|}\, ,
\end{align}
and $K_\nu(z)$ are modified Bessel functions of the second kind.
The two features of note are that the coefficient of the volume corrections depends upon $g_0$, the LO (leading-order) contribution to $g_A$ and that at fixed $m_\pi L$ the volume corrections scale quadratically in the pion mass.

While the finite volume corrections can be understood through an infrared modification of the low-energy EFT, the discretization corrections can be parameterized through a modification of the ultra-violet behavior of the theory, and similarly mapped into an EFT description.
In order to characterize the finite lattice spacing corrections, one follows a straightforward two-step procedure~\cite{Sharpe:1998xm}.
First, the lattice action is expanded for small lattice spacing into the Symanzik local EFT~\cite{Symanzik:1983dc,Symanzik:1983gh}.
The discretization effects are then encoded in a tower of operators suppressed by higher powers of the discretization scale,
\begin{equation}
S = S_0 + \sum_{n>0} a^n \mathcal{S}_{n}\, ,
\end{equation}
where $S_0=S_{QCD}$ is the QCD action and $S_n$ are the set of effective continuum operators of dimension $4+n$ consistent with all symmetries of the underlying lattice action.
For the mixed lattice action we are using, the leading discretization effects begin at $\mathrm{O}(a^2)$.
The dynamical HISQ action is perturbatively improved such that the leading discretization effects begin at $\mathrm{O}(\alpha_S a^2)$ and the valence MDWF action has leading corrections which begin at $\mathrm{O}(a^2)$.
The MDWF action has an $\mathrm{O}(a)$ correction proportional to the residual chiral symmetry breaking parameter, but this quantity can be absorbed into a redefinition of the quark mass, and in our case, it has been tuned to be less than 10\% of the valence light quark mass~\cite{Berkowitz:2017opd}, see also Table~\ref{tab:spectrum}.

The MA EFT for this action is known~\cite{Bar:2005tu,Tiburzi:2005is}, including the MA EFT expression for $g_A$ at NLO~\cite{Jiang:2007sn,Chen:2007ug},
\begin{align}\label{eq:gA_epi_MA}
g_A^\textrm{MA} &= g_{0}
 - (g_{0} + 2 g_{0}^3)\e_\pi^2 \ln \left( \e_\pi^2 \right)
 + c_{2} \e_\pi^2
\nonumber\\&\phantom{=}
 - \left(g_0 + \frac{24g_0^3 -15g_0^2\bar{g}_0 +14g_0\bar{g}_0^2 +\bar{g}_0^3}{12} \right)
\nonumber\\&\phantom{=}\quad\times
    \left[
     \e_{ju}^2 \ln \left( \e_{ju}^2 \right)
     -\e_\pi^2 \ln \left( \e_\pi^2 \right)
 \right]
\nonumber\\&\phantom{=}
    -g_0 \bar{g}_0^2 \e_{PQ}^2
    \left[1 + \ln \left( \e_\pi^2 \right) \right]
    +c_{2a}^\textrm{MA} \frac{a^2}{w_0^2}\, .
\end{align}
In this expression, there are two additional LECs, $\bar{g}_0$ and $c_{2a}^\text{MA}$, while the other new parameters are all determined in the LQCD calculation.
$\e_{ju}$ is the mixed-meson term
\begin{equation}
    \e_{ju}^2 = \frac{\tilde{m}_{ju}^2}{(4\pi F_\pi)^2}\, ,
\end{equation}
where $\tilde{m}_{ju}$ is the mixed meson mass including discretization corrections~\cite{Orginos:2007tw,Chen:2009su}, which we have determined, see Table~\ref{tab:spectrum}.
With the tuning we have performed ($m_\pi = m_{\pi,5} +\mathrm{O}(2\%)$) the partially quenched parameter is proportional to the taste-Identity HISQ pion splitting
\begin{equation}
    \e_{PQ}^2 = \frac{a^2 \D_\textrm{I}}{(4\pi F_\pi)^2}\, .
\end{equation}
$c_{2a}^\text{MA}$ is the LEC/counter-term for the leading discretization corrections, which we have parameterized with the dimensionless quantity $a/w_0$, where $w_0$ is the gradient flow scale~\cite{Borsanyi:2012zs} which has also been determined with the dynamical HISQ ensembles we are utilizing~\cite{Bazavov:2015yea}.
The new axial coupling $\bar{g}_0$ parameterizes the strength of the singlet axial current coupling to the nucleon.

We find our numerical results are insufficient to constrain all three unknown LECs, $g_0$, $c_2$ and $\bar{g}_0$.
Using partially quenched and $SU(3)$ flavor symmetries, the new coupling can be estimated to be $\bar{g}_0 = D - 3F$ where $D$ and $F$ are the $SU(3)$ axial coupling constants with $g_0 = D+F$ at LO in the $SU(3)$ flavor expansion.
Using phenomenological values of $D\sim0.75$ and $F~\sim0.45$ leads to the estimate $\bar{g}_0 \sim -0.6$ which is consistent with determinations from recent LQCD calculations~\cite{Bhattacharya:2013ehc,Alexandrou:2016xok}.
However, as our numerical results are themselves insufficient to constrain all the LECs, we do not include the MA $\chi$PT fit in the final analysis.

Despite this issue, one observes our results also have very mild discretization corrections.
Further, they are consistent with a pion mass independent continuum extrapolation.
This motivates us to consider a simple form for the extrapolation, with the continuum $\chi$PT expression, Eq.~\eqref{eq:gA_epi}, supplemented explicit discretization corrections.
For example, the leading discretization term is given by
\begin{equation}\label{eq:gA_fa}
    \d_a = c_{2a} \frac{a^2}{w_0^2} +\cdots\, ,
\end{equation}
where the dots represent higher order terms in the continuum extrapolation.
In order to perform the continuum extrapolation with dimensionless LECs, we have chosen to scale the lattice spacing by the gradient flow scale, $w_0$.
The leading discretization corrections for the HISQ action begins at $\mathrm{O}(\alpha_S a^2)$ and so a natural higher order term to include would be such a term.
However, this introduces a 3rd unknown parameter controlling the continuum extrapolation, and therefore can not provide a robust extrapolation as we have only 3 lattice spacings in this work.
One can estimate the systematic uncertainty arising from such a truncated continuum extrapolation by instead extrapolating in terms of $\alpha_S a^2 / w_0^2$ and comparing to an extrapolation with the leading correction.
Alternatively, one could introduce Bayesian priors to prevent the discretization LECs from wandering far away from their natural sizes.
For this work, we find the continuum extrapolated answer is consistent within the 1-sigma level whether one takes $a^2/w_0^2$ as parameterization of the continuum limit versus the term further suppressed by $\alpha_S$ using a standard frequentist minimization.
We conclude this continuum extrapolation systematic is a sub-dominant contribution to the total uncertainty relegating further investigation to future work.

\section{Results and Discussion \label{sec:results}}
In this section, we apply various extrapolations to our results and make a post-diction for $g_A$.
In Table~\ref{tab:ga_extrap}, we provide the various input parameters and the renormalized values of $g_A(\e_\pi, a/w_0, m_\pi L)$.

\begin{table}
\begin{ruledtabular}
\begin{tabular}{lccccc}
ensemble& $\e_\pi$& $m_\pi L$& $a/w_0$\footnote{We use $a/\omega_0$ determined at the physical-mass ensembles for fixed lattice spacing.}& $\alpha_S$& $g_A$ \\
\hline
a15m310   & 0.2495(3)& 3.780(3)& $0.8804(3)$& 0.58801& 1.215(12) \\
a12m310   & 0.2442(5)& 4.531(4)& $0.7036(5)$& 0.53796& 1.237(07) \\
a09m310   & 0.2462(4)& 4.507(4)& $0.5105(3)$& 0.43356& 1.258(14) \\
a15m220   & 0.1814(3)& 3.977(4)& $0.8804(3)$& 0.58801& 1.234(11) \\
a12m220S & 0.1839(5)& 3.257(5)& $0.7036(5)$& 0.53796& 1.272(28) \\
a12m220   & 0.1820(4)& 4.299(4)& $0.7036(5)$&  0.53796& 1.259(15) \\
a12m220L & 0.1814(4)& 5.363(4)& $0.7036(5)$&  0.53796& 1.252(21)\\
a15m130   & 0.1135(5)& 3.233(7)& $0.8804(3)$&  0.58801& 1.270(72)  \\
\end{tabular}
\end{ruledtabular}
\caption{\label{tab:ga_extrap} Renormalized values of $g_A$ and the parameters needed in the various extrapolations determined in this work. The lattice spacing $a/\omega_0$ and strong coupling-constant $\alpha_S$ are obtained from Table IV and III from Ref.~\cite{Bazavov:2015yea} respectively.}
\end{table}

\subsection{Analysis and error budget \label{sec:extrapolation}}
In order to assess the extrapolation uncertainty, we settle on 10 different models for performing the combined continuum, infinite volume and chiral extrapolations.  All 10 of these models have few enough unknown parameters that they provide some measure of predictive power.  We exclude any models with parameters that are in principle determinable with our data set but which have 100\% or greater uncertainty in the resulting fit, as they are clearly not constrained by our results.
This restriction excludes the MA EFT fit, Eq.~\eqref{eq:gA_epi_MA}, as well as the addition of a term proportional to $\e_\pi^2 (a/w_0)^2$ to either the $\chi$PT or Taylor expansion extrapolation functions.
Since we have only 3 lattice spacings, in order to assess the uncertainty in the continuum extrapolation from higher order corrections, we perform extrapolations using Eq.~\eqref{eq:gA_fa} as well as $\alpha_S \d_a$, where $\alpha_S$ is the strong coupling constant.  The discrepancy between the $\mathrm{O}(a^2)$ and $\mathrm{O}(\alpha_S a^2)$ extrapolations should provide a reasonably conservative estimate of the continuum extrapolation uncertainty.
The resulting set of extrapolation functions is given by
\begin{subequations}
\begin{align}
\label{eq:t_esq0_a2}
g_A &= c_0 + \d_a + \d_L\, ,
\\
\label{eq:t_esq0_aSa2}
g_A &= c_0 + \alpha_S \d_a + \d_L\, ,
\\
\label{eq:t_esq1_a0}
g_A &= c_0 + c_2 \e_\pi^2 + \d_L\, ,
\\
\label{eq:t_esq1_a2}
g_A &= c_0 + c_2 \e_\pi^2 + \d_a + \d_L\, ,
\\
\label{eq:t_esq1_aSa2}
g_A &= c_0 + c_2 \e_\pi^2 + \alpha_S \d_a + \d_L\, ,
\\
\label{eq:x_lo_a2}
g_A &= g_0 + \d_a + \d_L\, ,
\\
\label{eq:x_lo_aSa2}
g_A &= g_0 + \alpha_S \d_a + \d_L\, ,
\\
\label{eq:x_nlo_a0}
g_A &= g_0 - (g_0 + 2g_0^3)\e_\pi^2 \ln (\e_\pi^2) + c_2 \e_\pi^2 + \d_L\, ,
\\
\label{eq:x_nlo_a2}
g_A &= g_0 - (g_0 + 2g_0^3)\e_\pi^2 \ln (\e_\pi^2) + c_2 \e_\pi^2 + \d_a + \d_L\, ,
\\
\label{eq:x_nlo_aSa2}
g_A &= g_0 - (g_0 + 2g_0^3)\e_\pi^2 \ln (\e_\pi^2) + c_2 \e_\pi^2 + \alpha_S \d_a + \d_L\, ,
\end{align}
\end{subequations}
where $\d_L$ and $\d_a$ are given by Eqs.~\eqref{eq:gA_FV} and \eqref{eq:gA_fa} respectively.
For the latter five fits from Eqs.(\eqref{eq:x_lo_a2}--\eqref{eq:x_nlo_aSa2}), the unknown coefficient in $\d_L$, $g_0$, is taken to be the same as in the infinite volume $\chi$PT expression while in the Taylor expansion fits, Eqs.~\eqref{eq:t_esq0_a2}-\eqref{eq:t_esq1_aSa2}, the value is left to float as a free parameter, $g_0^{L}$.
To evaluate the finite-volume functions $F_1(m_\pi L)$ and $F_3(m_\pi L)$, we include up to $|\mathbf{n}|=20$ in the summations, well beyond the point of negligible contributions from higher \textit{around-the-world} effects~\cite{Colangelo:2003hf}.
For NLO Taylor expansion fits given by Eqs.~(\ref{eq:t_esq1_a0}--\ref{eq:t_esq1_aSa2}), we expand around $\e_0^2=0$ as the choice of $\e_0$ has no impact in this linear in $\e_\pi^2$ expansion.

We perform a numerical least-squares minimization of the $\chi^2$ constructed from our results and the 10 extrapolation functions listed above.
The input parameters, the independent $x$-variables, have uncertainties which are correlated with the values of $g_A(\e_\pi,a/w0,m_\pi L)$ on each ensemble.
We have not performed calculations with different sets of valence parameters on the same underlying ensemble, so the covariance matrix needed for the $\chi^2$ is diagonal in the ensemble space,
\begin{equation}
\chi^2 = \sum_q \frac{\Big(g_A^q - f(x_q,\theta) \Big)^2}{\s_q^2}\, ,
\end{equation}
where $q$ runs over the ensembles, $g_A^q$ is the mean value of $g_A$, $x_q$ is the set of $x$-variables, $\s_q^2$ is the variance including that from the $x$-variables, all on ensemble $q$ and $\theta$ is the set of unknown parameters in a given extrapolation function.
In order to estimate $\s_q$, we compute the variance
\begin{equation}
\s_q^2 = \textrm{var}\Big( g_A^q[bs] - f(x_q[bs],\theta) \Big)\, ,
\end{equation}
using the bootstrap distributions of $g_A$ and the $x$-variables.
For a function with linear dependence upon $x$, this exactly reproduces the linear least squares variance with uncorrelated uncertainties in $x$ and $y$, $\s^2 = \s_y^2 + \theta^2 \s_x^2$, for a function $f(x,\theta) = \theta x$.
For each minimization, we also perform a fit in which we set the $x$-variance to zero, and find that these uncertainties have no impact on the results within the quoted precision.
We have prepared a set of Python scripts that perform these various minimizations, which we make available with this article.

\begin{figure}
\includegraphics[width=\columnwidth]{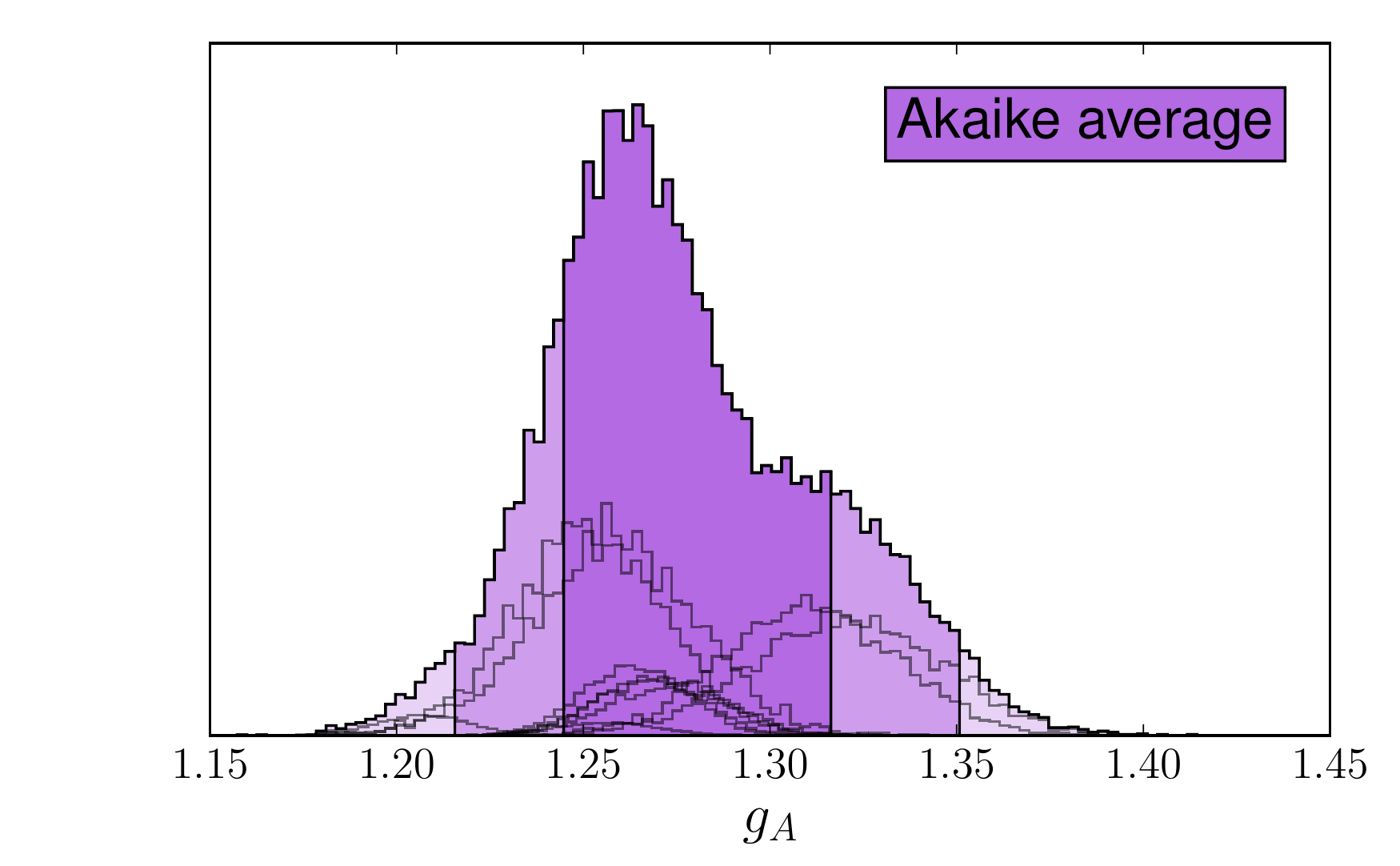}
\caption{\label{fig:ga_hist}
The AIC weighted histogram distribution of our extrapolation results as described in Sec.~\ref{sec:extrapolation}.
The overall magenta distribution is from the weighted bootstrap distributions from each analysis, with the varying shaded regions representing the 1, 2, and 3+ sigma confidence intervals.  The underlying distributions visible are from the weighted distribution from each analysis.}
\end{figure}

To judge the relative quality of these various extrapolations, we utilize the AIC (Akaike information criterion)~\cite{Akaike:1974} defined as,
\begin{equation}
\text{AIC} = 2k-2\ln(\hat{\mathcal{L}})
\end{equation}
for a model with $k$ free parameters and $\hat{\mathcal{L}}$ is the maximum of the likelihood function. The hat on $\mathcal{L}$ indicates that the likelihood is evaluated at its extrema. In the case that the objective function is the $\chi^2$-statistic, $-2\ln(\hat{\mathcal{L}}) = \chi^2_{min}$. The AIC is derived from the Taylor expansion of the Shannon entropy up to quadratic dependence in $\theta$ evaluated at its maximum likelihood estimate. Therefore within a set of models, any model $i$ with AIC$_i$ has the relative probability $P_i$ (where $P_i \leq 1$) of being the ``truth''  compared to the model with AIC$_{\text{min}}$ where,
\begin{equation}
 P_i= \exp\left[-\left(\text{AIC}_i-\text{AIC}_{\text{min}}\right)/2\right].
\end{equation}
It is important to note that AIC does not provide a means for hypothesis testing, but instead only provides a method for model selection. The model-weighted average $\hat{\overline{\theta}}$ and variance $\hat{\bar{\sigma}}_\theta$ of parameter $\theta$ over $N$ models can be calculated from the AIC weights $w_i$,
\begin{align}
\hat{\overline{\theta}} = & \sum_{m=1}^N w_i \hat{\theta}_i,\\
\hat{\overline{\sigma}}_\theta = & \left[\sum_{m=1}^N w_i \sqrt{\hat{\sigma}_{i,\theta} + (\hat{\theta}_i - \hat{\overline{\theta}})^2}\right]^2,\\
w_i = &\frac{P_i}{\sum_{m=1}^N P_m},
\end{align}
where the hat on $\theta_i$ (and its variance) indicates its value determined by the maximum likelihood estimate of model $i$. We employ AIC as means of model selection, and ultimately for estimating the model extrapolation uncertainty.
In Table~\ref{tab:fit_results}, we list quantities from the minimization of the 10 extrapolation functions listed in Eqs.~\eqref{eq:t_esq0_a2}-\eqref{eq:x_nlo_aSa2}.
Our final determination of the axial charge is
\begin{equation}
g_A(\e_\pi^{phys}, a\rightarrow0, m_\pi L\rightarrow\infty) = \ga\, ,
\end{equation}
where the first uncertainty is the statistical/systematic uncertainty arising from the LQCD calculations and the second uncertainty arises from the spread in predictions from the various extrapolation functions considered.
The resulting distribution is displayed in Fig.~\ref{fig:ga_hist}.

\begin{table*}
\begin{ruledtabular}
\begin{tabular}{lclcrcl}
short name& Eq.& extrapolation function& $\chi^2/dof$& \multicolumn{1}{c}{AIC}& weight& \multicolumn{1}{c}{$g_A(\e_\pi^\text{phy})$}\\
\hline
T$\e_\pi^0a^2$& \eqref{eq:t_esq0_a2}& $c_0 + \d_a + \d_L$
	& 4.73/5    & 10.729    & 0.0447    & 1.275(17) \\
T$\e_\pi^0\alpha_Sa^2$& \eqref{eq:t_esq0_aSa2}& $c_0 + \alpha_S \d_a + \d_L$
	& 4.73/5    & 10.728    & 0.0447    & 1.269(14) \\
T$\e_\pi^2a^0$&\eqref{eq:t_esq1_a0}& $c_0 + c_2 \e_\pi^2 + \d_L$
	& 7.82/5    & 13.819    & 0.0095    & 1.257(14) \\
T$\e_\pi^2a^2$&\eqref{eq:t_esq1_a2}& $c_0 + c_2 \e_\pi^2 + \d_a + \d_L$
	& 0.14/4    &  8.138    & 0.1632    & 1.321(26) \\
T$\e_\pi^2\alpha_Sa^2$&\eqref{eq:t_esq1_aSa2}& $c_0 + c_2 \e_\pi^2 + \alpha_S \d_a + \d_L$
	& 0.13/4    &  8.131    & 0.1637    & 1.311(24) \\
$\chi\e_\pi^0a^2$&\eqref{eq:x_lo_a2}& $g_0 + \d_a + \d_L$
	&  6.47/6  &  10.467    & 0.0509    & 1.272(16) \\
$\chi\e_\pi^0\alpha_Sa^2$&\eqref{eq:x_lo_aSa2}& $g_0 + \alpha_S \d_a + \d_L$
	&  6.41/6  &  10.413    & 0.0523    & 1.265(14) \\
$\chi\e_\pi^2a^0$&\eqref{eq:x_nlo_a0}& $g_0 - (g_0 + 2g_0^3)\e_\pi^2 \ln (\e_\pi^2) + c_2 \e_\pi^2 + \d_L$
	& 9.13/6    & 13.126    & 0.0135    & 1.208(12) \\
$\chi\e_\pi^2a^2$&\eqref{eq:x_nlo_a2}& $g_0 - (g_0 + 2g_0^3)\e_\pi^2 \ln (\e_\pi^2) + c_2 \e_\pi^2 + \d_a + \d_L$
	& 1.46/5    &  7.463    & 0.2286    & 1.260(22) \\
$\chi\e_\pi^2\alpha_Sa^2$&\eqref{eq:x_nlo_aSa2}& $g_0 - (g_0 + 2g_0^3)\e_\pi^2 \ln (\e_\pi^2) + c_2 \e_\pi^2 + \alpha_S \d_a + \d_L$
	& 1.46/5    &  7.462    & 0.2289    & 1.253(20) \\
\hline
&&&&& weighted avg. & \ga
\end{tabular}
\end{ruledtabular}
\caption{\label{tab:fit_results} Minimization results from the various extrapolation functions and our weighted average as described in the text.  For the average, the first uncertainty arises from the fitting statistical and systematic uncertainties and the second uncertainty is from the variation due to the model extrapolation.  The resulting distribution is displayed in Fig.~\ref{fig:ga_hist}.  }
\end{table*}

\subsection{Discussion \label{sec:discussion}}

\begin{figure}
\includegraphics[width=\columnwidth]{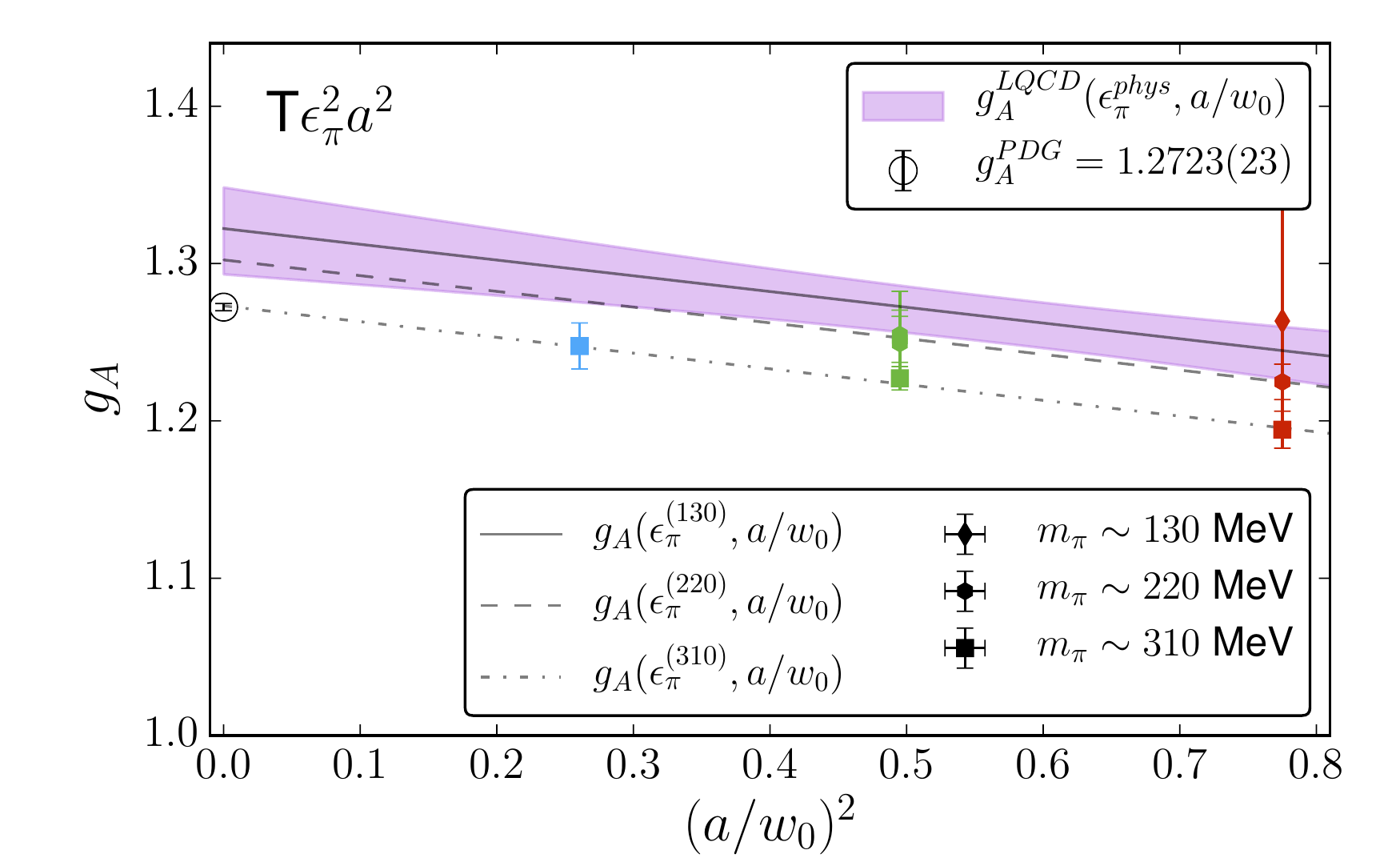}
\includegraphics[width=\columnwidth]{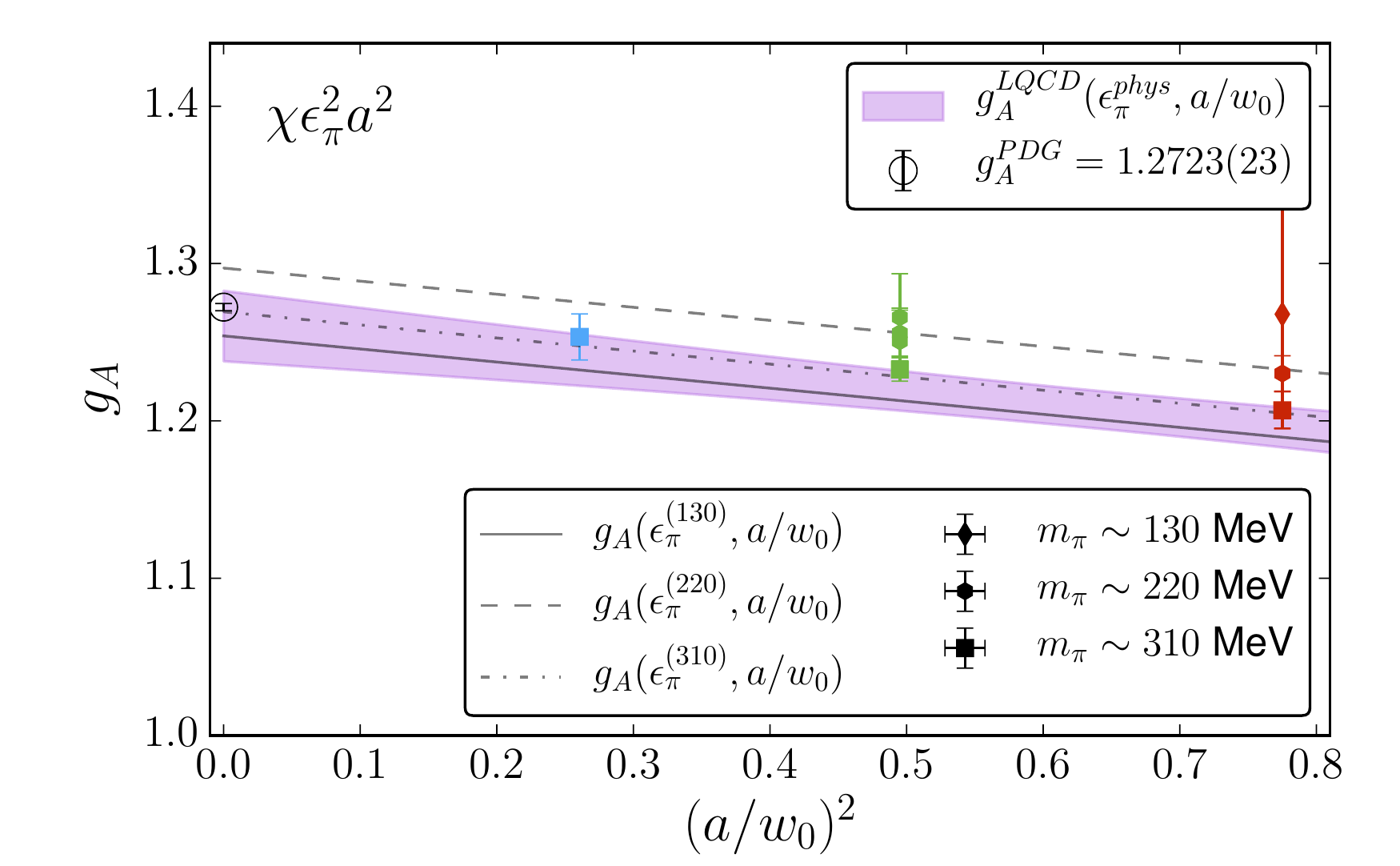}
\caption{\label{fig:gA_continuum} Continuum extrapolation of $g_A$ using the two fit ans\"{a}tze Eqs.~\eqref{eq:t_esq1_a2} (top) and \eqref{eq:x_nlo_a2} (bottom).
The values of $g_A$ in the plots have been adjusted for finite volume corrections.
}
\end{figure}

The largest uncertainty in our determination of $g_A$ results from the model extrapolation uncertainty.  The spread in the resulting distribution is driven by the highest weighted fits, Eqs.~\eqref{eq:t_esq1_a2}, \eqref{eq:t_esq1_aSa2}, \eqref{eq:x_nlo_a2} and \eqref{eq:x_nlo_aSa2} which show tension at the 1-sigma level.
The dominant source of this discrepancy arises from the pion mass extrapolation, rather than the discretization corrections.
To understand this, first observe that the discrepancy between the $\mathrm{O}(a^2)$ and $\mathrm{O}(\alpha_S a^2)$ extrapolations (the subsequent pairs in Table~\ref{tab:fit_results}), differ by less than one standard deviation.
Contrast this to the discrepancy between the final result from Eq.~\eqref{eq:t_esq1_a2} and \eqref{eq:x_nlo_a2}, for example, which differ by more than one standard deviation of each result.
At the same time, the coefficient of the discretization LEC in these two fits are consistent with each other $c_{2a}[\text{\eqref{eq:t_esq1_a2}}] = -0.101(40)$ and $c_{2a}[\text{\eqref{eq:x_nlo_a2}}] = -0.084(30)$, as displayed in Fig.~\ref{fig:gA_continuum}.
Further, at the coarsest lattice spacing, our value of $g_A$ is only 6\% different from the continuum extrapolated value and at the finest, it is 2.1\%, demonstrating a very mild continuum extrapolation.

\begin{figure}
\includegraphics[width=\columnwidth]{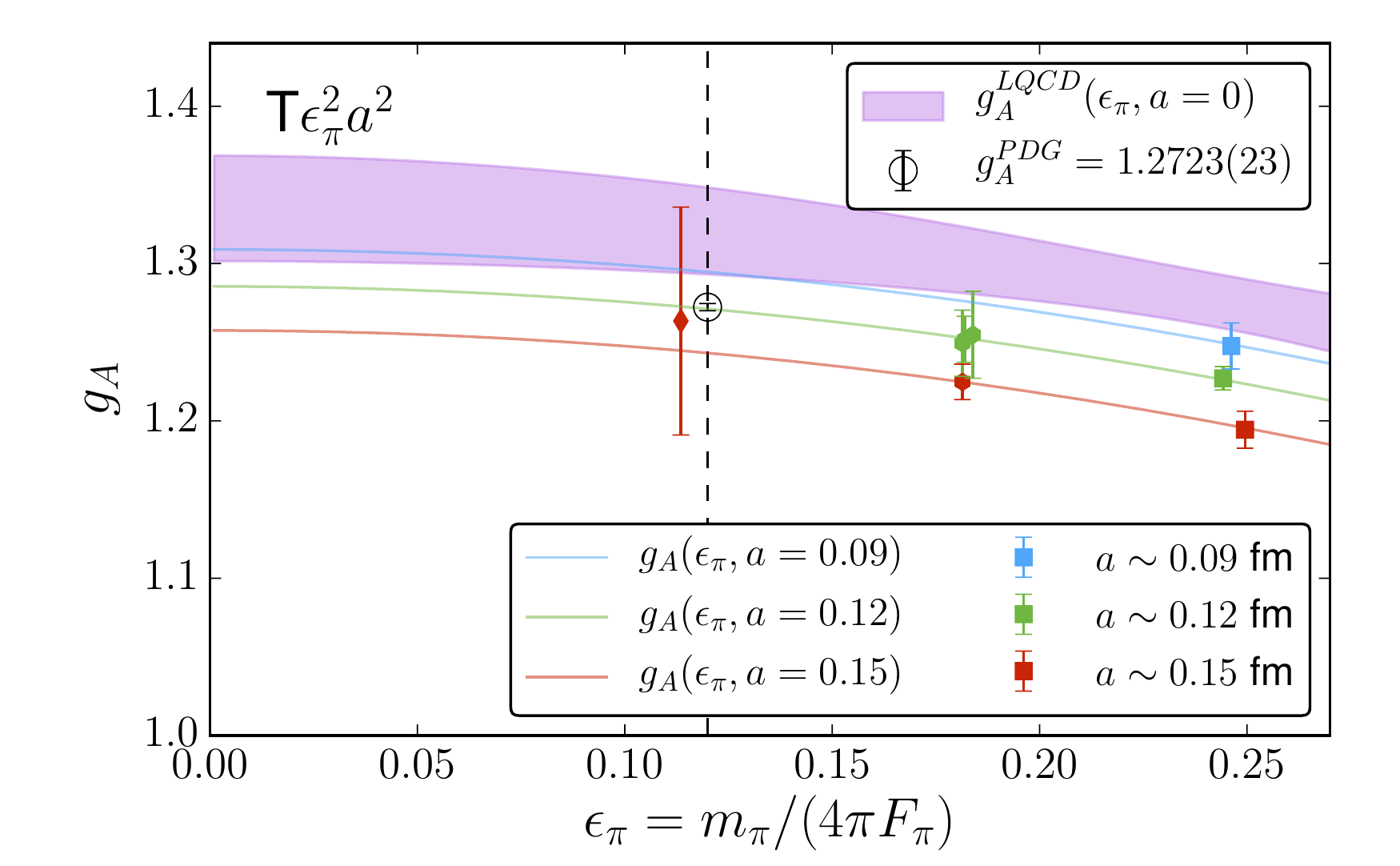}
\includegraphics[width=\columnwidth]{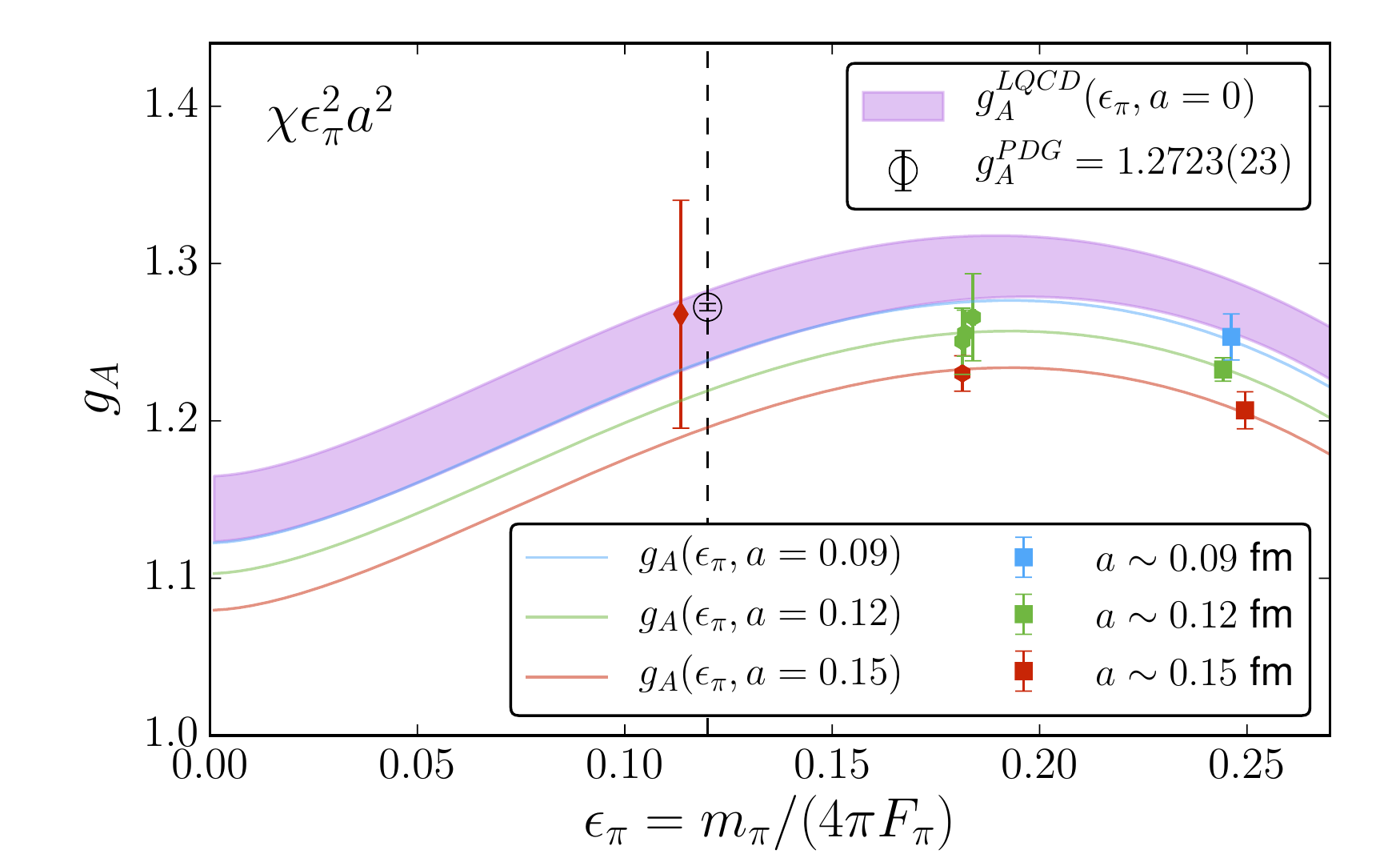}
\caption{\label{fig:gA_chiral}
Chiral extrapolation of $g_A$ resulting from the fit ans\"{a}tze Eqs.~\eqref{eq:t_esq1_a2} (top) and~\eqref{eq:x_nlo_a2} (bottom).
The values of $g_A$ in the plots have been adjusted for finite volume corrections.
}
\end{figure}

The resulting pion mass dependence ($\e_\pi$) is displayed in Fig.~\ref{fig:gA_chiral}, for the two most weighted fits.
From bottom to top, the solid red, green and blue curves are the resulting extrapolation as a function of $\e_\pi$ at fixed lattice spacing.  The filled magenta band is the 68\% confidence interval for the continuum, infinite volume extrapolated value of $g_A(\e_\pi)$.
The Taylor expansion fit results in a very mild pion mass dependence, whereas the $\chi$PT fit shows some noticeable pion mass dependence.
This curvature is generated by the $\chi$PT extrapolation formula as the coefficient of the NLO logarithm is constrained by the leading order term, generating a large logarithm.  The local counter-term contribution, proportional to $c_2$, competes with the logarithmic correction, leading to the observed strong pion mass dependence.

In Ref.~\cite{Brantley:2016our}, the first conclusive evidence for such chiral-logarithms in the baryon sector was observed and presented for the iso-vector nucleon mass.
In that case, the presence of the chiral logarithm was more prominent as the low-order Taylor expansion was incapable of parameterizing the LQCD results.
This is not the case in the present work, so such strong conclusions about the evidence of the non-analytic quark mass dependence predicted from $\chi$PT can not be currently drawn from $g_A$.
In order to reduce the pion mass extrapolation uncertainty (by ruling out or finding consistent the $\chi$PT and Taylor expansion analysis), one would need to have enough of a lever arm to constrain higher order corrections, for example by obtaining results at more pion masses as well as more precise results at the physical pion mass.

The NLO $\chi$PT analysis results in the LECs
\begin{align}
& g_0 = 1.144(21)\, ,&
&c_2 = -9.48(72)\, .&
\end{align}
The parameter covariance matrix can be extracted from the Python analysis scripts and/or resulting \texttt{sqlite} file accompanying this work, as well as the resulting parameters and parameter covariance matrices from all fits, Eqs.~\eqref{eq:t_esq0_a2}-\eqref{eq:x_nlo_aSa2}.

\subsubsection{Volume dependence}
\begin{figure}
\includegraphics[width=\columnwidth]{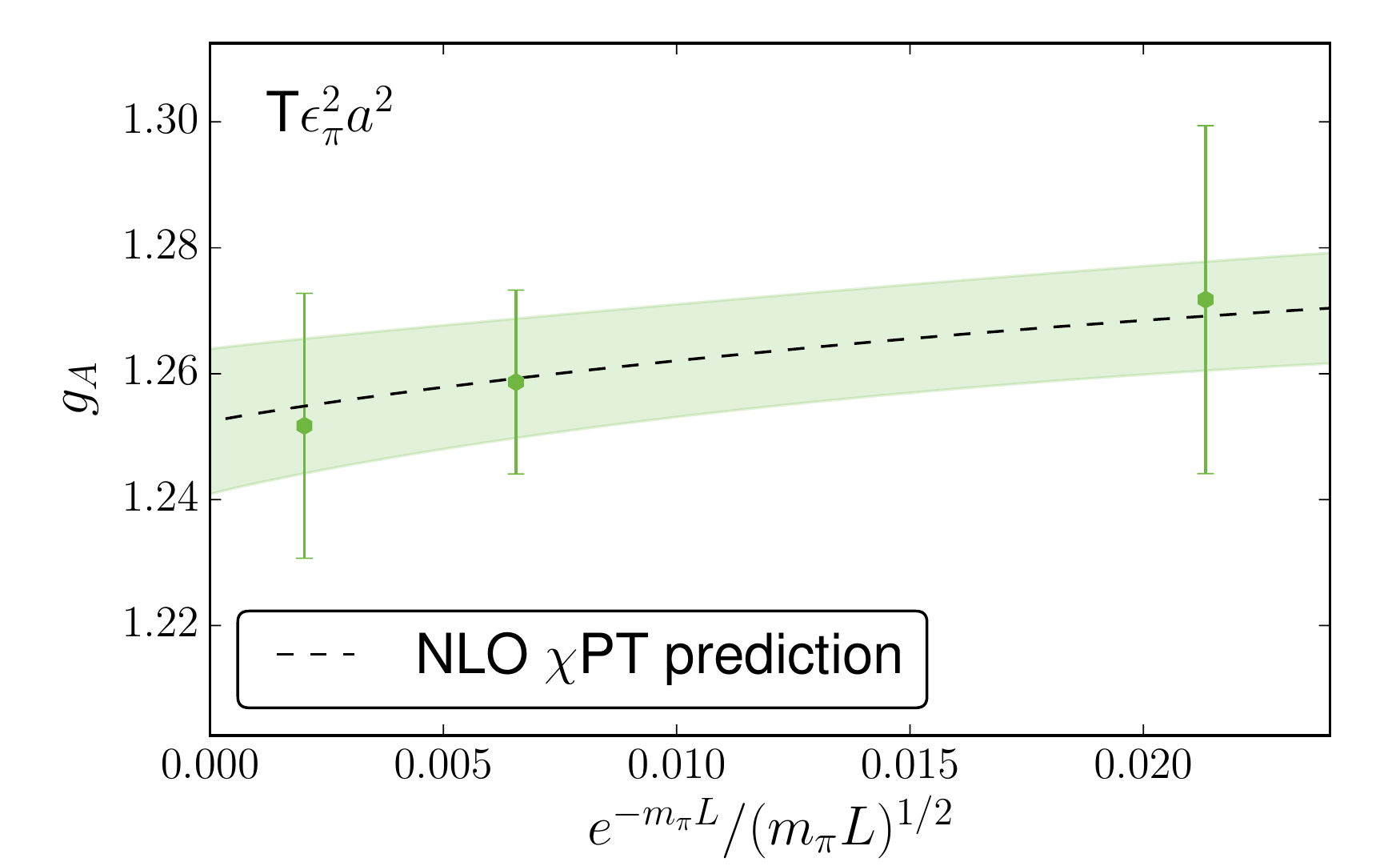}
\includegraphics[width=\columnwidth]{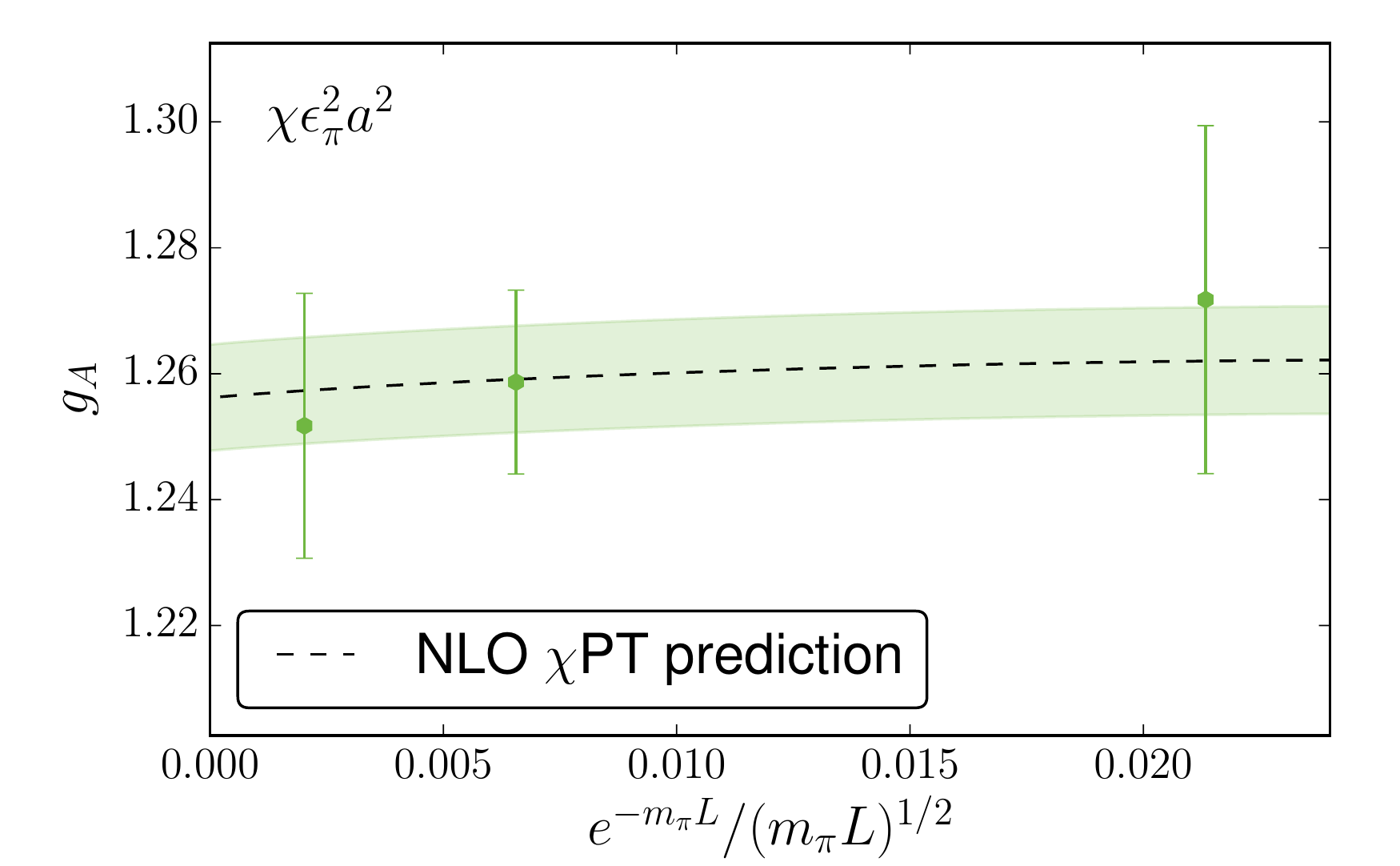}
\caption{\label{fig:gA_volume} Volume corrections compared to the predicted leading volume dependence from $SU(2)$ $\chi$PT~\cite{Beane:2004rf}.
In the Taylor expansion fit (top), the coefficient of these corrections are a free parameter, while in the $\chi$PT fit (bottom), the coefficient also serves as the leading contribution to $g_A$.
In both cases, the volume corrections agree very well with the predicted $\chi$PT formula, Eq.~\eqref{eq:gA_FV}.
}
\end{figure}

There has been some discussion in the literature that $g_A$ may be particularly susceptible to finite-volume corrections such that the leading $\chi$PT prediction for the volume dependence is grossly insufficient to explain the observed volume dependence~\cite{Jaffe:2001eb,Cohen:2001bg,Yamazaki:2008py,Yamazaki:2009zq}.
In Fig.~\ref{fig:gA_volume}, we display the result of our dedicated volume study.
In the Taylor expansion fit, Eq.~\eqref{eq:t_esq1_a2}, the coefficient of the volume corrections is determined to be $g_0^L = 1.42(53)$.
In the $\chi$PT extrapolation enhanced by discretization corrections, Eq.~\eqref{eq:x_nlo_a2}, the coefficient of the volume correction is the same as that which appears in the infinite volume extrapolation, which is determined to be $g_0 = 1.144(21)$.
We conclude that the leading volume corrections are in extremely good agreement with the numerical LQCD results and that the coefficient of the volume corrections as determined by these various fits are also consistent.

\subsubsection{Truncation error}
The NNLO $\chi$PT corrections to $g_A$ scale as $\e_\pi^3$.
Using EFT power-counting arguments, the estimate of such corrections is of order $(\e_\pi^{phys})^3 \sim 0.002$ which is less than 10\% of our statistical uncertainty. However, tension between the NLO Taylor expansion and $\chi$PT suggests that the truncation leads to error comparable to our statistical uncertainty. We fully account for this error in our final AIC averaged result. 

At the physical pion mass, the total extrapolation from our coarsest lattice spacing to the continuum limit, using the $a^2$ ans\"{a}tze, is less than 5\% of the central value of $g_A$. 
Assuming the dimensionless coefficient of the $a^4$ contribution is similar in magnitude to that of the $a^2$, these higher order contributions are na\"{i}vely 5\% of 5\% ($\sim 0.25$\%) at the coarsest spacing.
This is comparable to the $~\sim0.3\%$ difference between the $a^2$ and $\alpha_S a^2$ extrapolations in both the Taylor expansion and $\chi$PT fits at LO and NLO.
While we are unable to constrain the $a^4$ coefficients with only three lattice spacings, these observations, and the mild continuum extrapolation suggests this study is not necessary.
We include as an uncertainty the difference between the $a^2$ and $\alpha_S a^2$ in our final AIC averaged result, which we find to be a reasonable estimate of higher order discretization corrections.

\subsection{$|V_{ud}|$ from neutron decay}
Knowledge of $g_A$, together with experimentally measured value of the neutron lifetime $\t_n$, can be used to determine the value of the CKM (Cabbibo-Kobayashi-Maskawa) matrix element $|V_{ud}|$,
\begin{equation}
|V_{ud}|^2 = \frac{4908.7(1.9)\text{sec}}{\t_n(1+3g_A^2)},
\end{equation}
where the uncertainty in the numerator comes from electroweak RC (radiative corrections)~\cite{Czarnecki:2004cw,Marciano:2005ec}. Using the world average value~\cite{Olive:2016xmw}
\begin{equation}
\t_n = 880.3(1.1)~\text{sec}
\end{equation}
and the AIC averaged value of $g_A$ determined from our lattice calculation, we obtain
\begin{equation}
|V_{ud}| = 0.9697(6)_{\t_n}(233)_{g_A}(2)_{\text{RC}}.
\end{equation}
This result is consistent at the percent level of the PDG value, $0.9758(6)_{\t_n}(15)_{g_A}(2)_{\text{RC}}$, but still an order-of-magnitude less precise.

An even more precise determination of $|V_{ud}|$ may be obtained experimentally from decay rates of super-allowed nuclear beta-decays, giving a value of $0.97417(5)_{\text{exp.}}(9)_{\text{nucl.dep.}}(18)_{\text{RC}}$. This result however, introduces model-dependent uncertainty. $|V_{ud}|$ may be determined to similar precision by calculating, using LQCD, the ratio of pseudoscalar decay constants $F_K/F_\pi$ and $|V_{us}|$ from $K\rightarrow \pi$ semileptonic decay.  When these values are combined with the experimentally determined ratio of $|V_{us}/V_{ud}|F_K/F_\pi$, one arrives at the FLAG (Flavor Lattice Averaging Group) value of $|V_{ud}|=0.97440(18)$\footnote{$N_f=2+1+1$ average from January 2017 update.}~\cite{Aoki2017}.  Here we present an alternative way to constrain $|V_{ud}|$. However, effort beyond the scope of this work is required for the hadronic uncertainty to be comparable to the uncertainty in the neutron lifetime measurement.

\section{Conclusion \label{Sec:conclusion}}
We have performed a calculation of the nucleon axial charge, $g_A$, using lattice QCD methods. We utilize a new method motivated by the Feynman-Hellmann theorem~\cite{Bouchard:2016heu} for computing matrix elements, which gives us good control over finite temporal effects at a modest computational cost. 
Of important note, we achieve precise results with only $\mathrm{O}(4-5\text{K})$ thousand stochastic samples on all but 2 of the ensembles used in this work, substantially less than the tens to many tens-of-thousands of samples used with more traditional computational strategies.

We have performed extrapolations in lattice spacing, volume, and pion mass, and report a final value of \begin{equation*}
g_A = \ga\, .
\end{equation*}
This calculation is only the second lattice QCD determination of $g_A$ with all systematics accounted for, and is the first to do so and agree with the experimental determination within error bars. Because the previous calculation by the PNDME collaboration was performed using the same lattice ensembles, it will be important future work to determine the source of discrepancy between the results.

The first error we quote includes all statistical and fitting systematic uncertainties and the second encodes systematic uncertainties due to the extrapolations to the physical point. We use a variety of different models to perform our extrapolations, and use a weighted average utilizing the AIC to determine the systematic uncertainty from all extrapolations. The majority of the uncertainty results from the extrapolation in pion mass, and further reducing this uncertainty will require additional pion mass points in order to pin down higher order contributions to both the chiral perturbation theory and Taylor expansion fits. We find very mild dependence on the lattice spacing, and moderate finite volume effects which match the expected corrections as determined from effective theory. Finally, we use our result, in combination with the experimentally determined neutron lifetime, to determine the CKM matrix element $|V_{ud}|$, which we also find to be consistent with the PDG value at the percent level.

\bigskip
Interested readers can download the Python analysis scripts used for this work and an \texttt{hdf5} file with the relevant numerical results at the \texttt{github} address:

{\centering\url{https://github.com/callat-qcd/project_gA}\par}

Correlation functions and bootstrapped correlator fit results are stored in a \texttt{PostgreSQL} database hosted at NERSC, serving as the centralized up-to-date resource for our main analysis. \texttt{PostgreSQL} serves as an ideal database for large structured datasets common in Lattice QCD calculations (\textit{e.g.} gauge configurations, propagators, correlation functions), while added flexibility through native implementation of JavaScript Object Notation (\texttt{JSON}) accomodates unstructured data in the form of analysis results (\textit{e.g.} varying fit ansatz). With the publication of this paper, we release tables from the \texttt{PostgreSQL} database relevant to this project for the general public in the above mentioned \texttt{hdf5} file.

\acknowledgements
We gratefully acknowledge the MILC Collaboration for use of the dynamical HISQ ensembles~\cite{Bazavov:2012xda}.
We thank Carleton DeTar and Doug Toussaint for help compiling and using the MILC code at LLNL and understanding how to write source fields from Chroma that can be read by MILC for the construction of the mixed-meson correlation functions.
We also thank Claude Bernard for useful correspondence regarding scale setting and taste violations with the HISQ action.
We thank Peter Lepage for discussions about the \texttt{lsqfit}, \texttt{gvar} Python packages that are used extensively for this project, as well as his insights into statistical inference.
We also thank Wayne Tai Lee for discussions about model selection, uncertainty estimation, and his willingness to share his expertise in the field of statistics.
Part of this work was performed at the Kavli Institute for Theoretical Physics supported by NSF Grant No. PHY-1125915.

The software used for this work was built on top of the Chroma software suite~\cite{Edwards:2004sx} and the highly optimized QCD GPU library QUDA~\cite{Clark:2009wm,Babich:2011np}.
We also utilized the highly efficient HDF5 I/O Library~\cite{hdf5} with an interface to HDF5 in the USQCD QDP++ package that was added with SciDAC 3 support (CalLat)~\cite{Kurth:2015mqa}, as well as the MILC software for solving for HISQ propagators.
Finally, the HPC jobs were efficiently managed with a \texttt{bash} job manager, \texttt{METAQ}~\cite{Berkowitz:2017vcp}, capable of intelligently backfilling idle nodes in sets of nodes bundled into larger jobs submitted to HPC systems.  \texttt{METAQ} was developed with SciDAC 3 support (CalLat) and is available on \texttt{github}.
The numerical calculations in this work were performed at:
Lawrence Livermore National Laboratory on the Surface and RZhasGPU GPU clusters as well as the Cab CPU and Vulcan BG/Q clusters;
and the Oak Ridge Leadership Computing Facility at the Oak Ridge National Laboratory, which is supported by the Office of Science of the U.S. Department of Energy under Contract No. DE-AC05-00OR22725, on the OLCF Titan machine through a DOE INCITE award (CalLat).
We thank the Lawrence Livermore National Laboratory (LLNL) Institutional Computing Grand Challenge program for the computing allocation.

This work was performed with support from LDRD funding from LLNL 13-ERD-023 (EB, ER, PV) and LDRD funding from LBNL (AWL);
and by the RIKEN Special Postdoctoral Researcher program (ER).
The work of NG was supported in part by Leverhulme Trust (grant RPG-2014-118).
This work was also performed under the auspices of the U.S. Department of Energy by Lawrence Livermore National Laboratory under Contract DE-AC52-07NA27344 (EB, ER, PV);
under contract DE-AC05-06OR23177, under which Jefferson Science Associates, LLC, manages and operates the Jefferson Lab (BJ, KO);
under contract DE-AC02-05CH11231, which the Regents of the University of California manage and operate Lawrence Berkeley National Laboratory and the National Energy Research Scientific Computing Center (DB, CCC, TK, HMC, AWL);
This work was further performed under the auspices of the U.S. Department of Energy, Office of Science, Office of Nuclear Physics under contracts:
DE-FG02-04ER41302 (CMB, KNO);
DE-SC00046548 (AN);
DE-SC0015376, Double-Beta Decay Topical Collaboration (DB, HMC, AWL);
by the Office of Advanced Scientific Computing Research, Scientific Discovery through Advanced Computing (SciDAC) program under Award Number KB0301052 (EB, TK, AWL);
and by the DOE Early Career Research Program, Office of Nuclear Physics under FWP NQCDAWL (DB, HMC, CCC, AWL).
This work is supported in part by the DFG and the NSFC through funds provided to the Sino-German CRC 110 ``Symmetries and the Emergence of Structure in QCD'' (EB).

\bibliography{c51_bib}

\appendix

\onecolumngrid
\newpage
\section{Autocorrelation study}
\label{app:autocorrelation}
\begin{figure*}[h]
\includegraphics[width=0.43\textwidth]{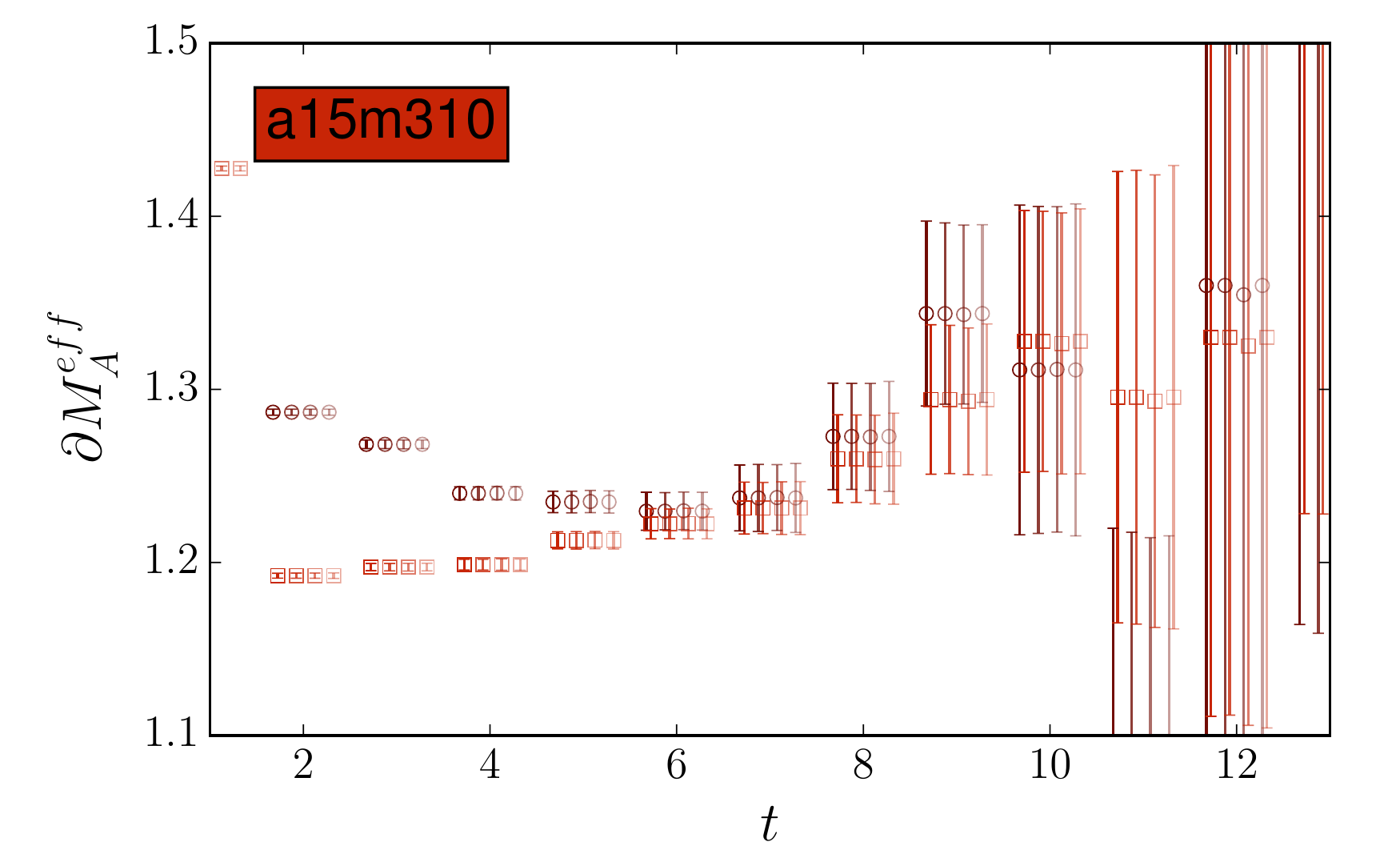}
\includegraphics[width=0.43\textwidth]{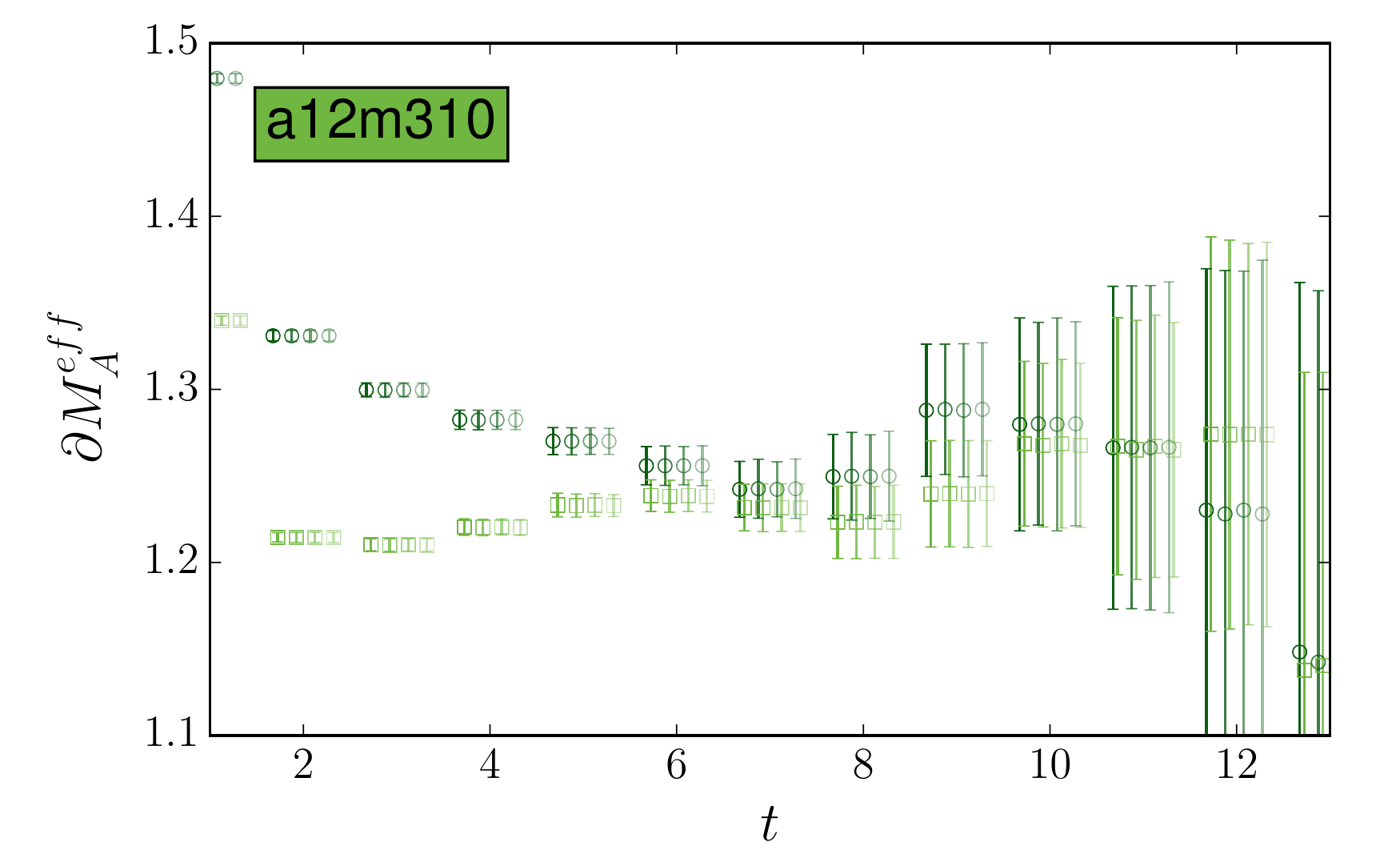}
\includegraphics[width=0.43\textwidth]{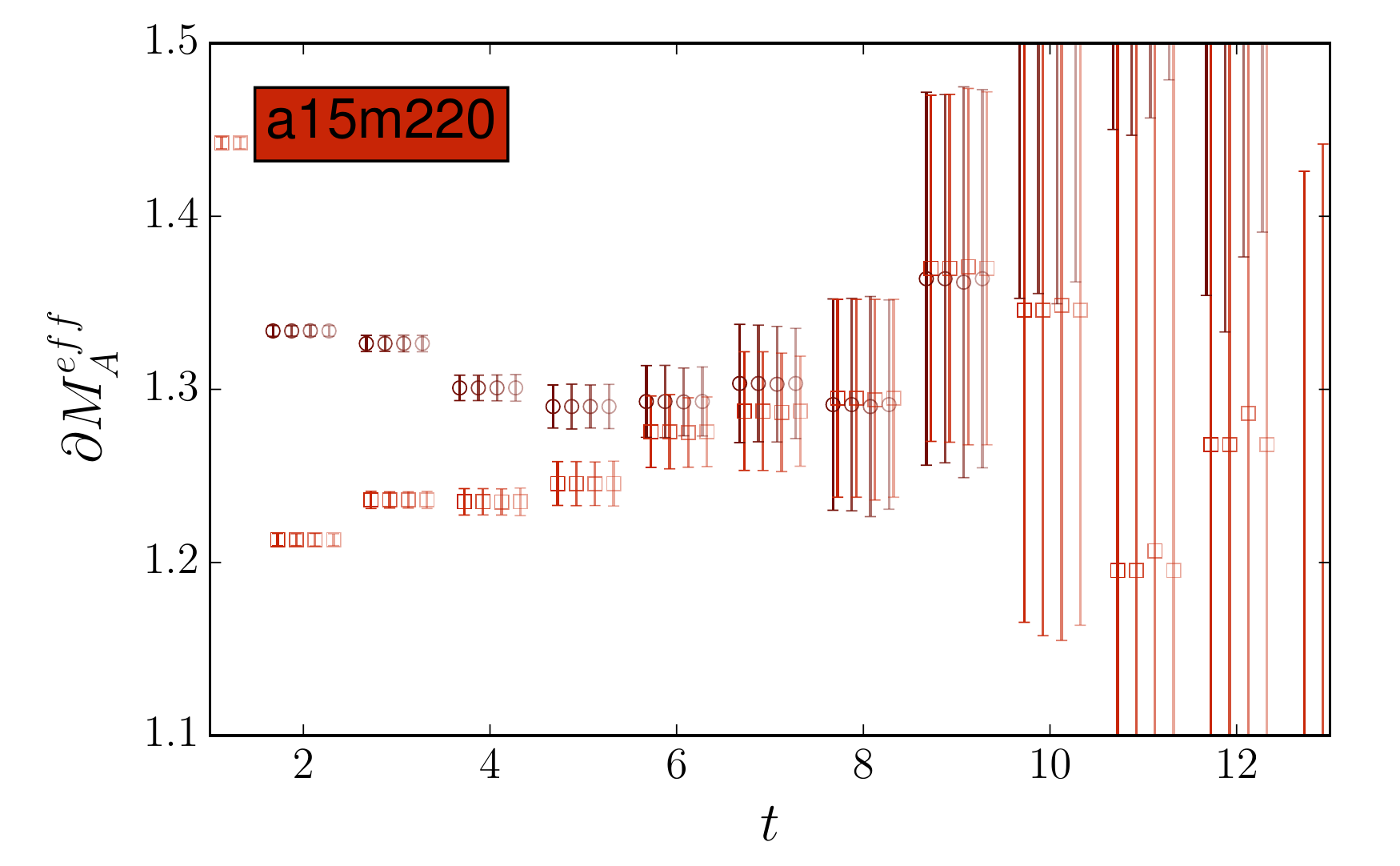}
\includegraphics[width=0.43\textwidth]{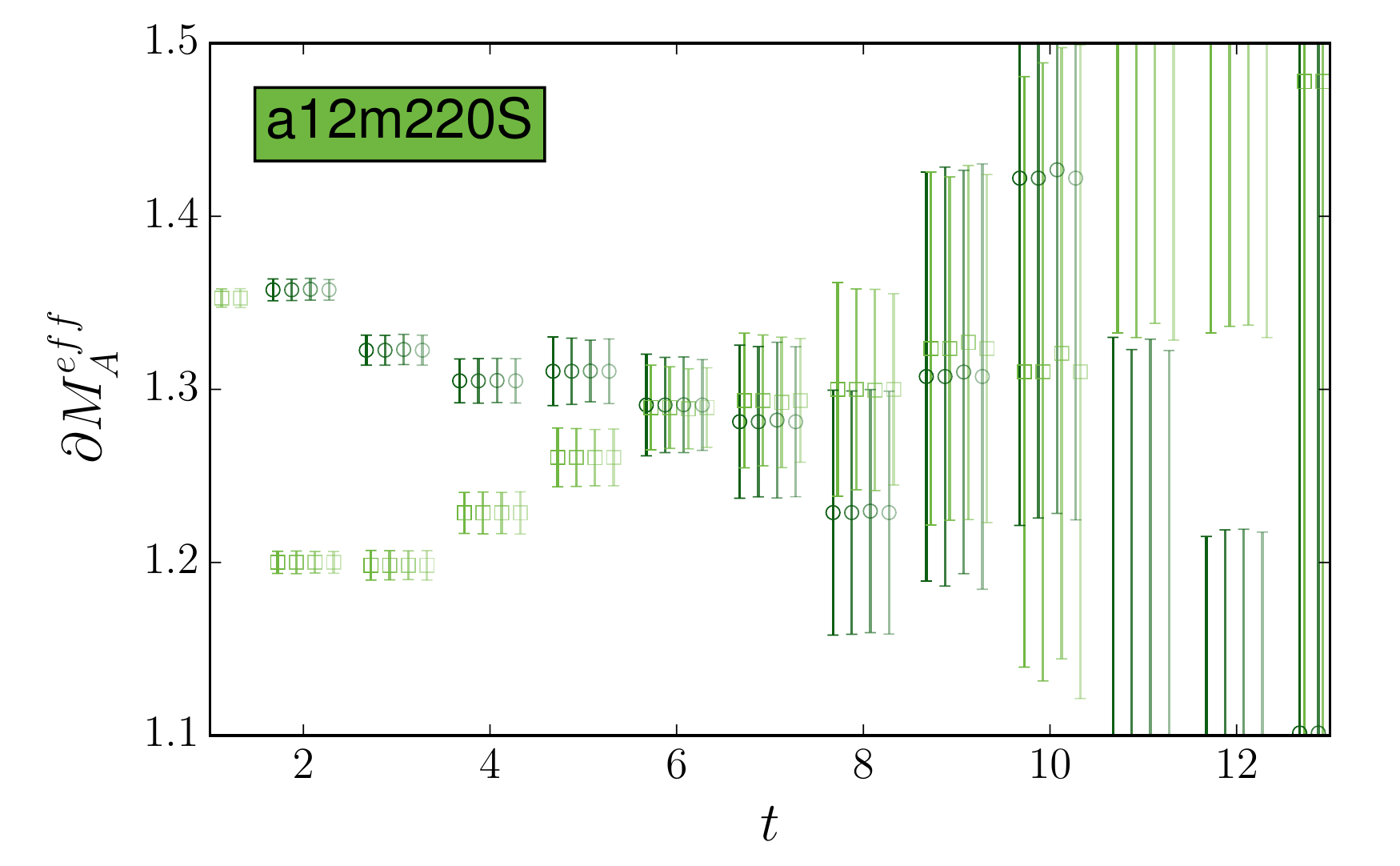}
\includegraphics[width=0.43\textwidth]{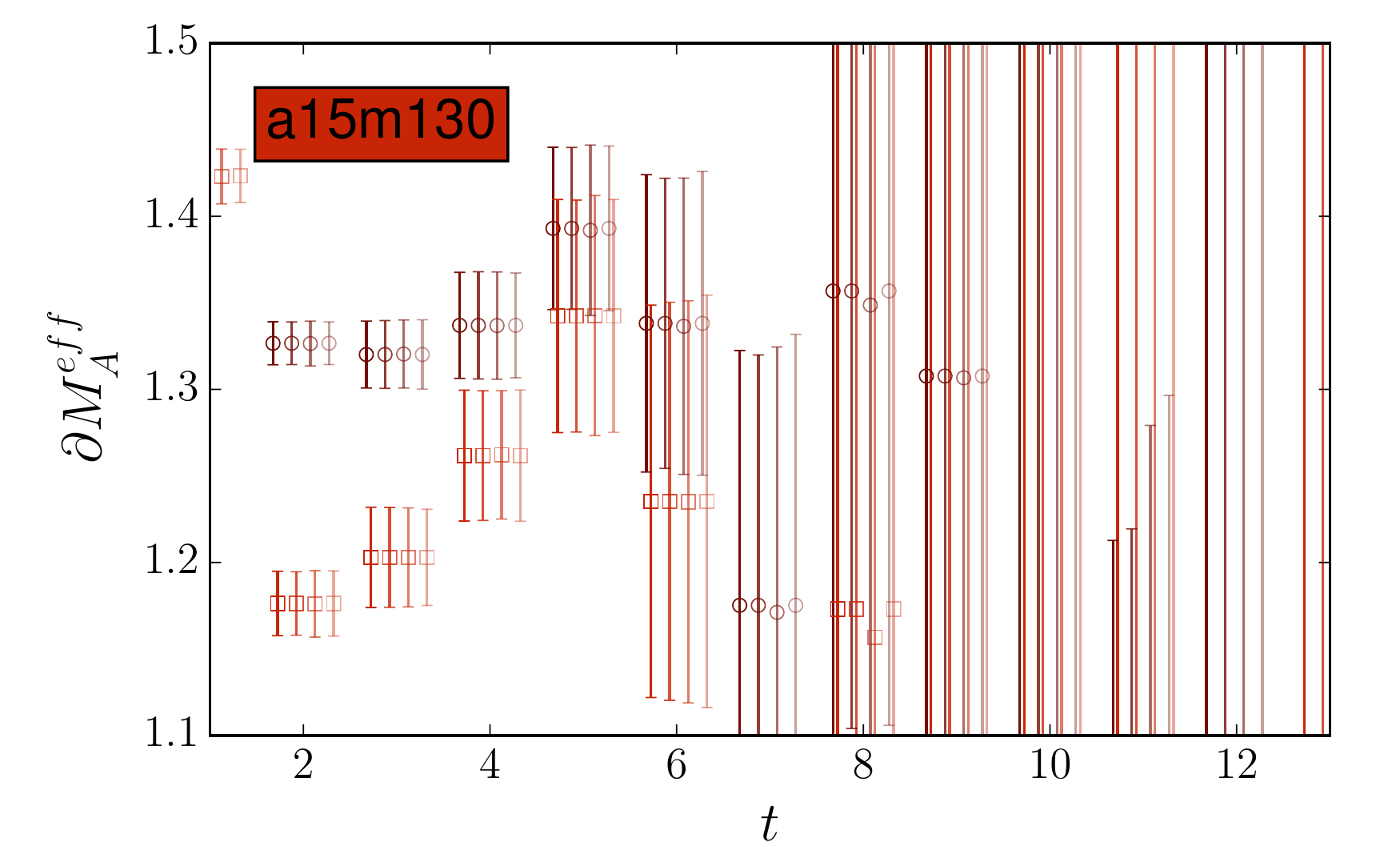}
\includegraphics[width=0.43\textwidth]{auto_a12m220.pdf}
\includegraphics[width=0.43\textwidth]{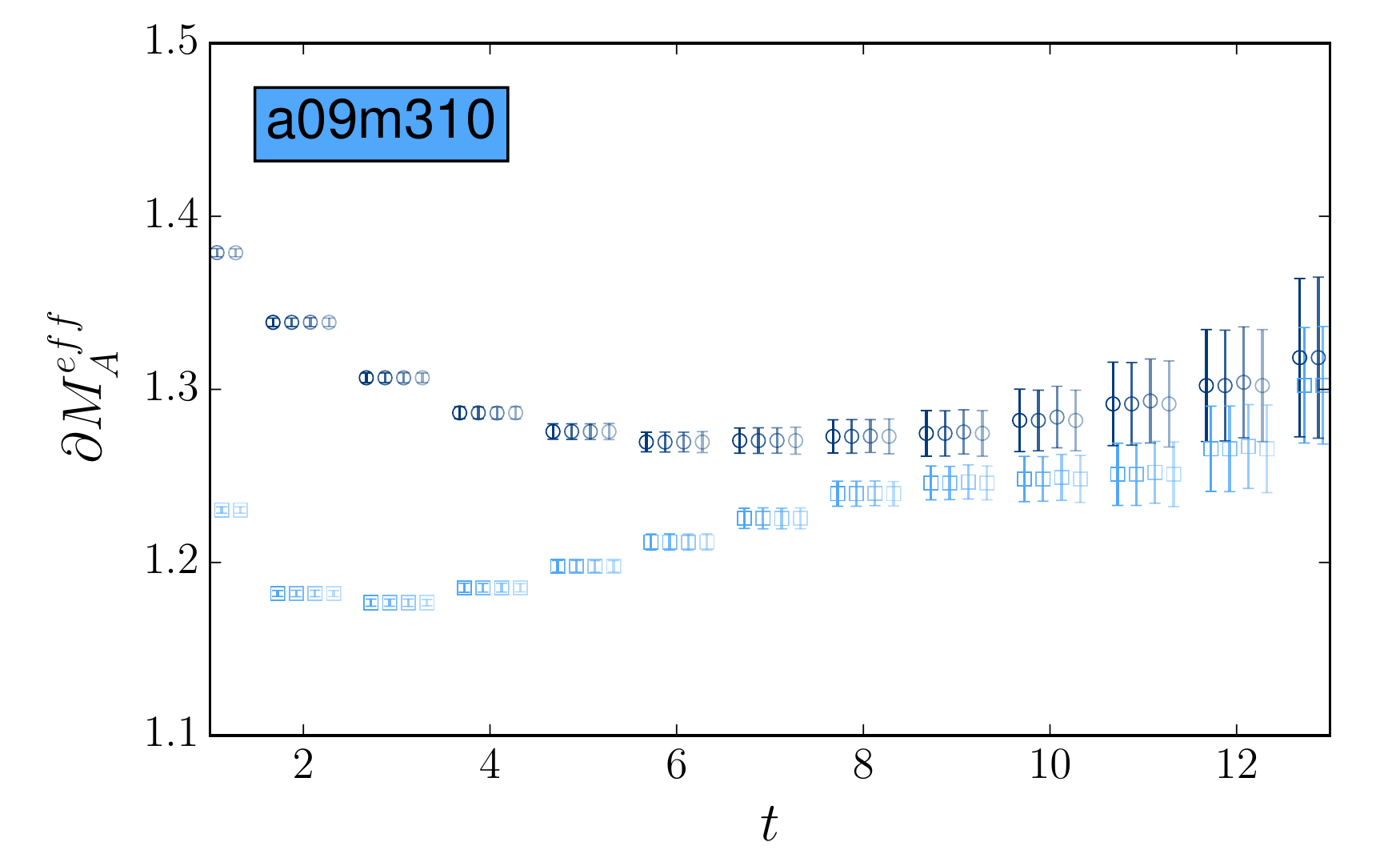}
\includegraphics[width=0.43\textwidth]{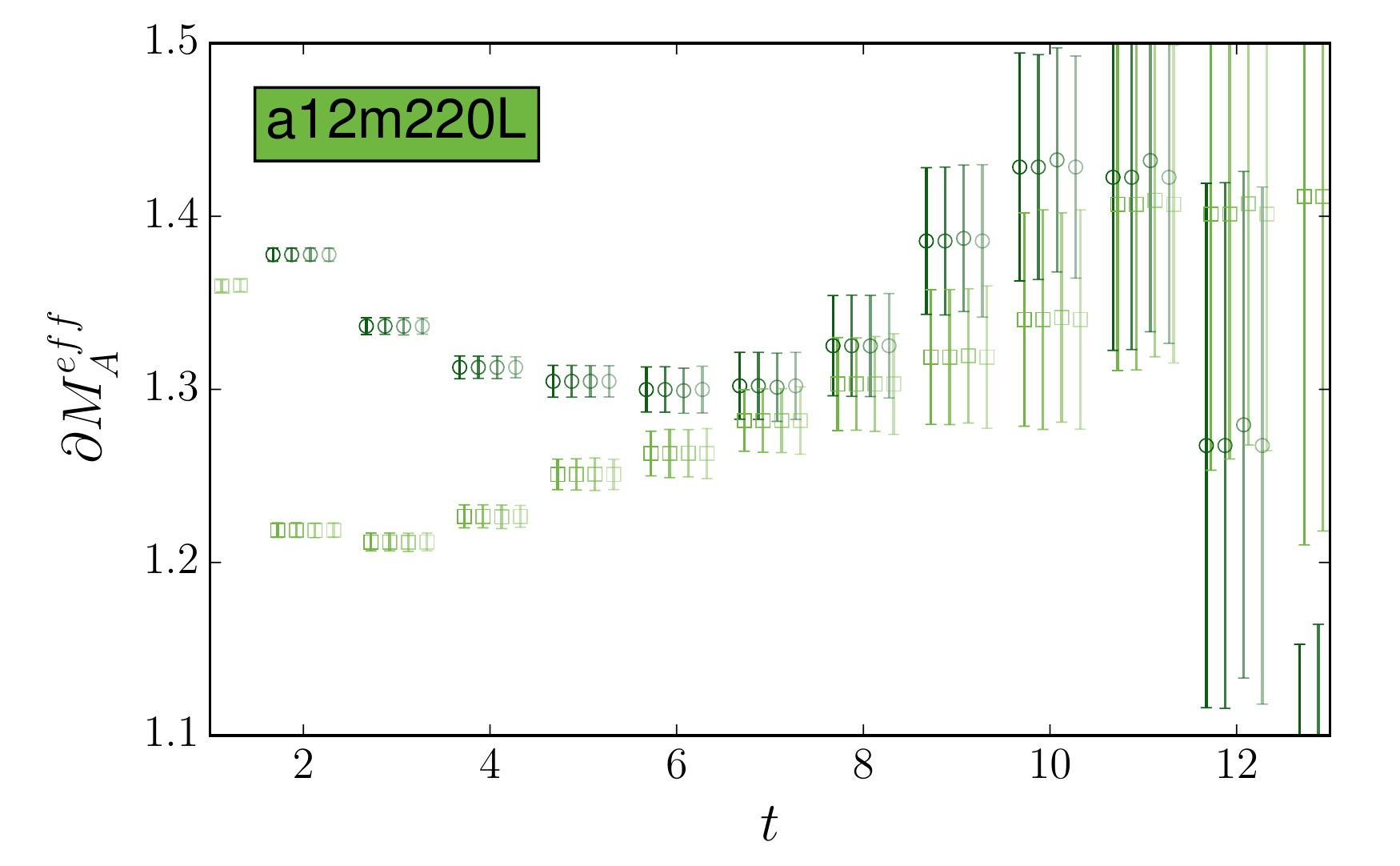}
\caption{Autocorrelation study for all ensembles used in this work. Plots have the same color scheme as Fig.~\ref{fig:autocorrelation}.}
 \label{fig:autocorrelation_complete}
\end{figure*}
\FloatBarrier

\newpage
\section{Bootstrap correlator fits}
\label{app:bs_correlator_fits}
\begin{figure*}[h]
\includegraphics[width=0.43\textwidth]{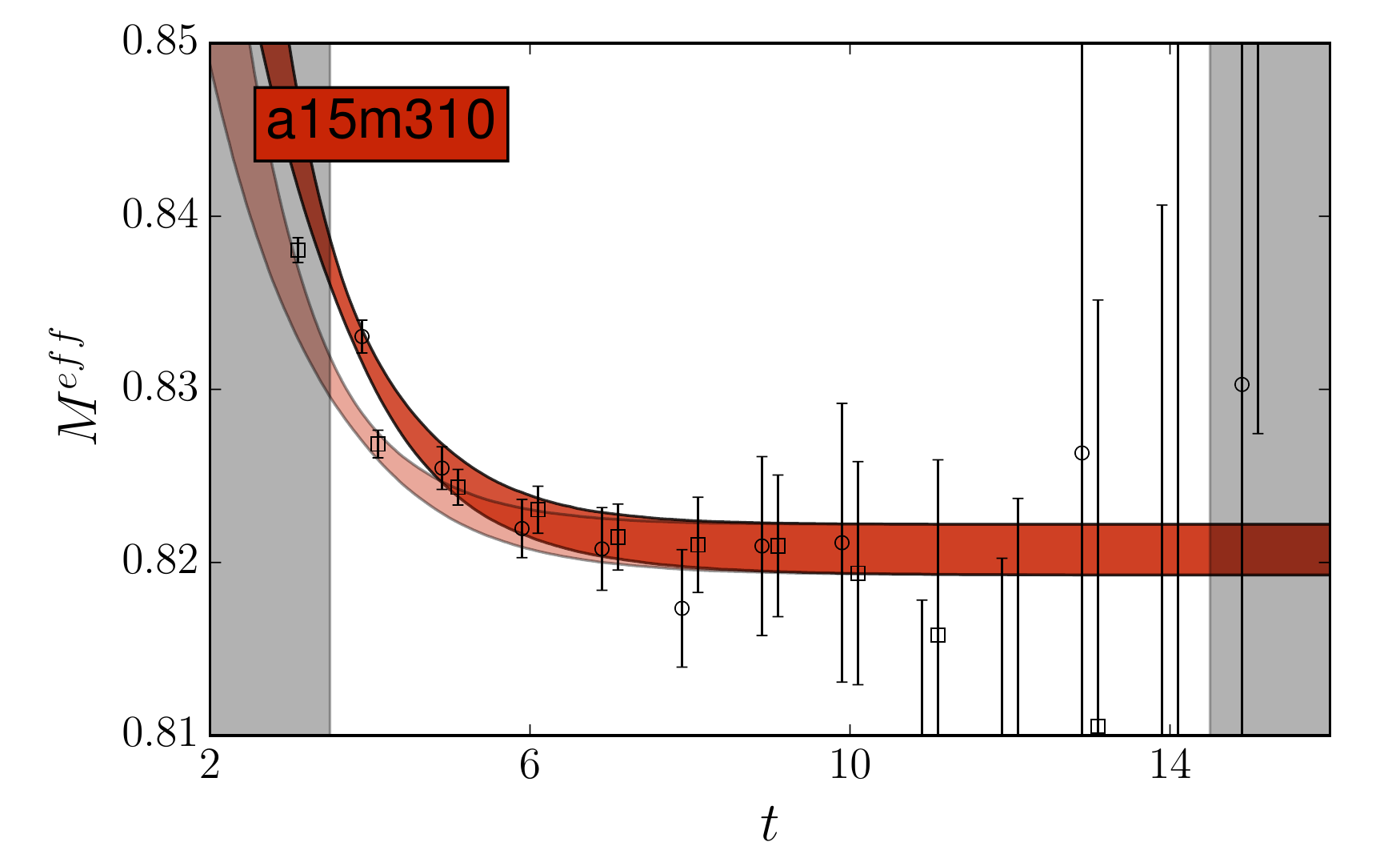}
\includegraphics[width=0.43\textwidth]{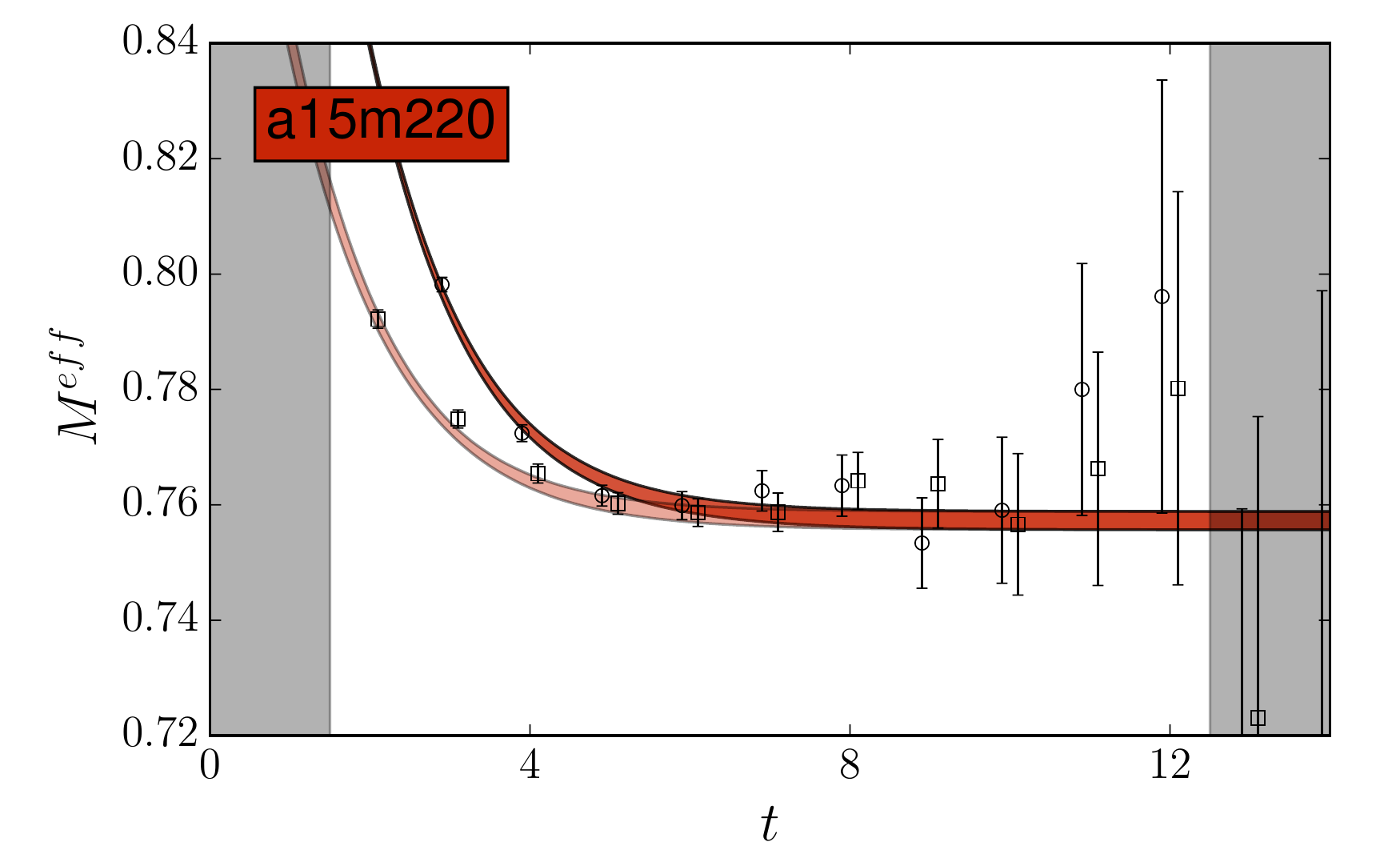}
\includegraphics[width=0.43\textwidth]{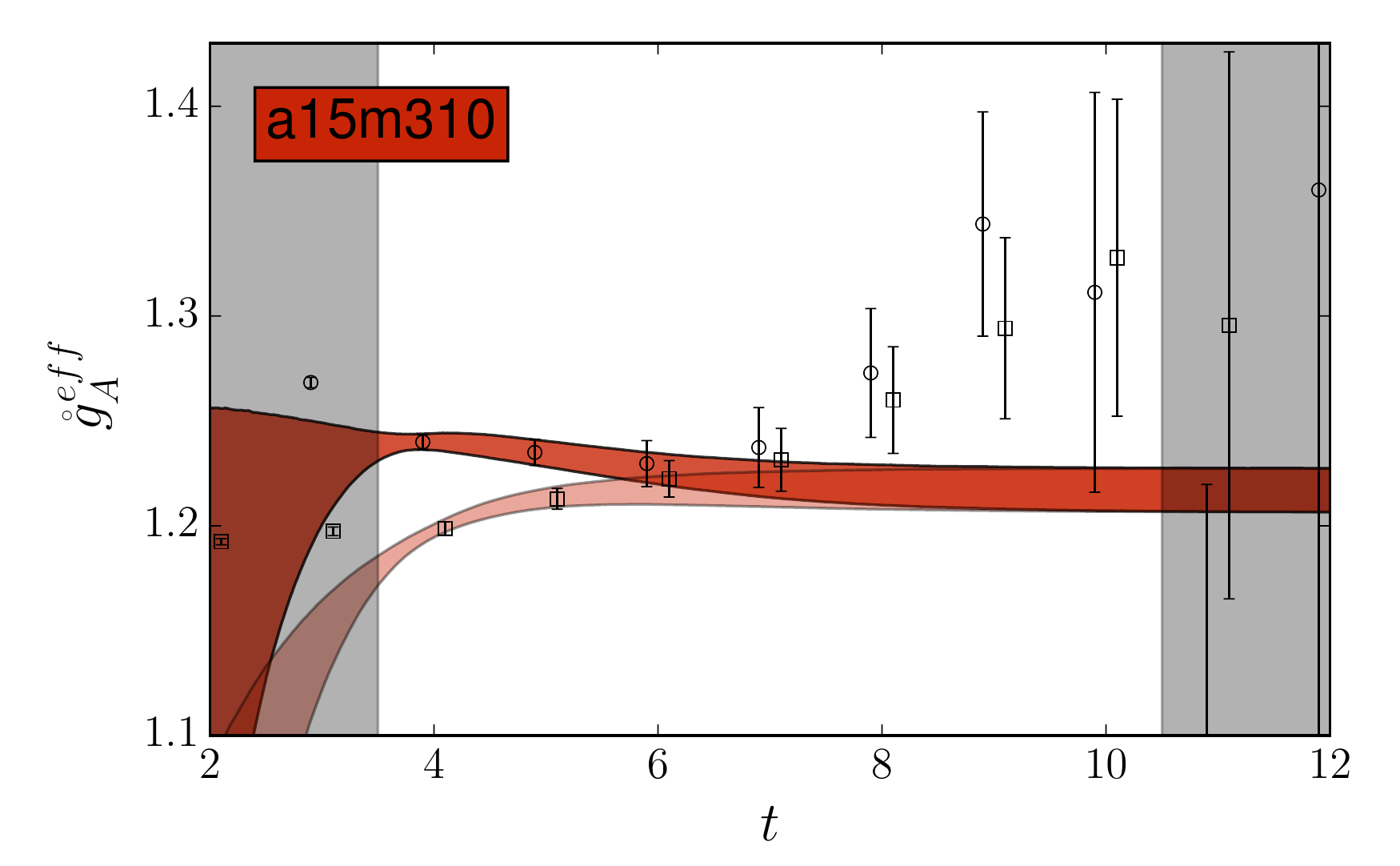}
\includegraphics[width=0.43\textwidth]{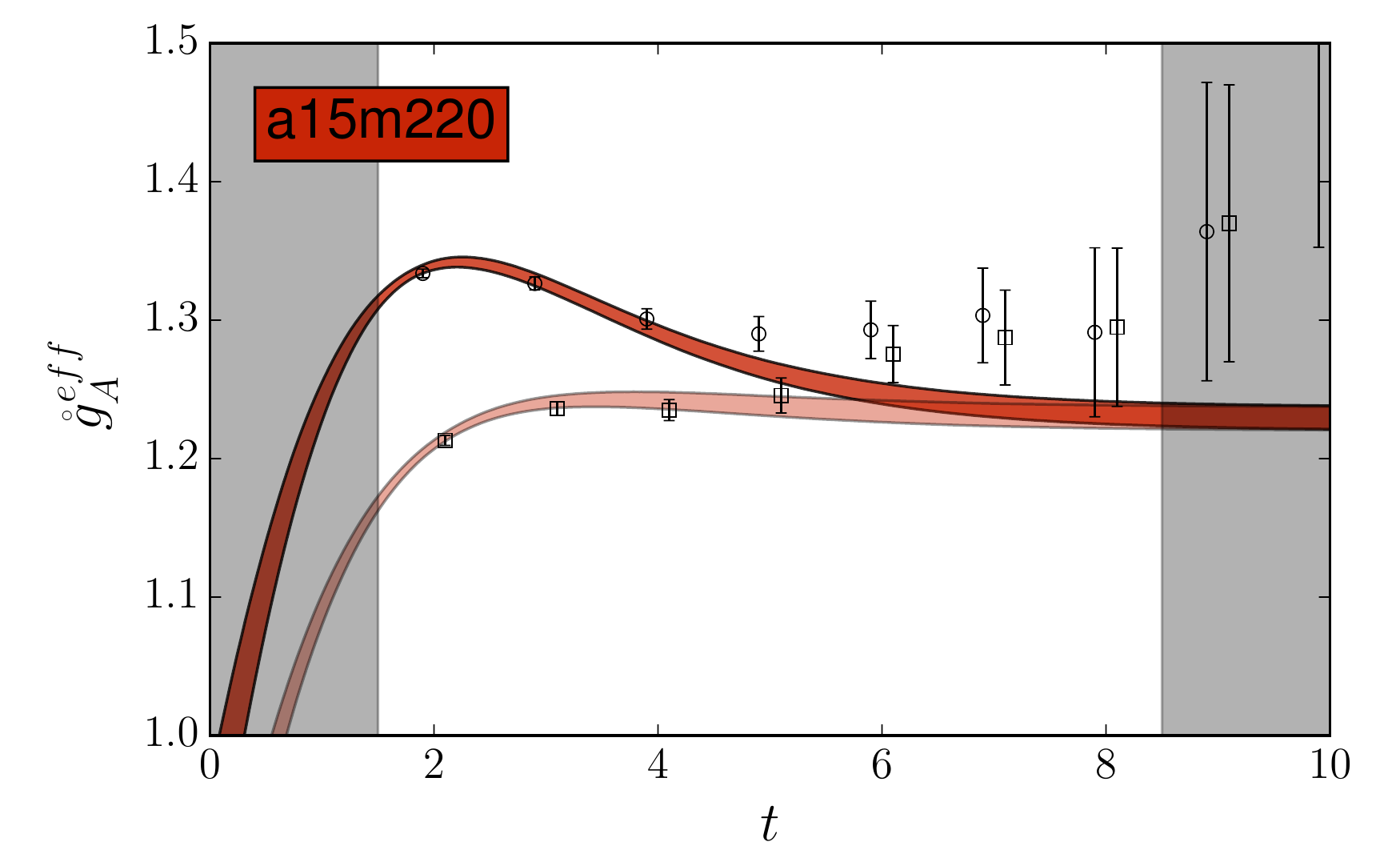}
\includegraphics[width=0.43\textwidth]{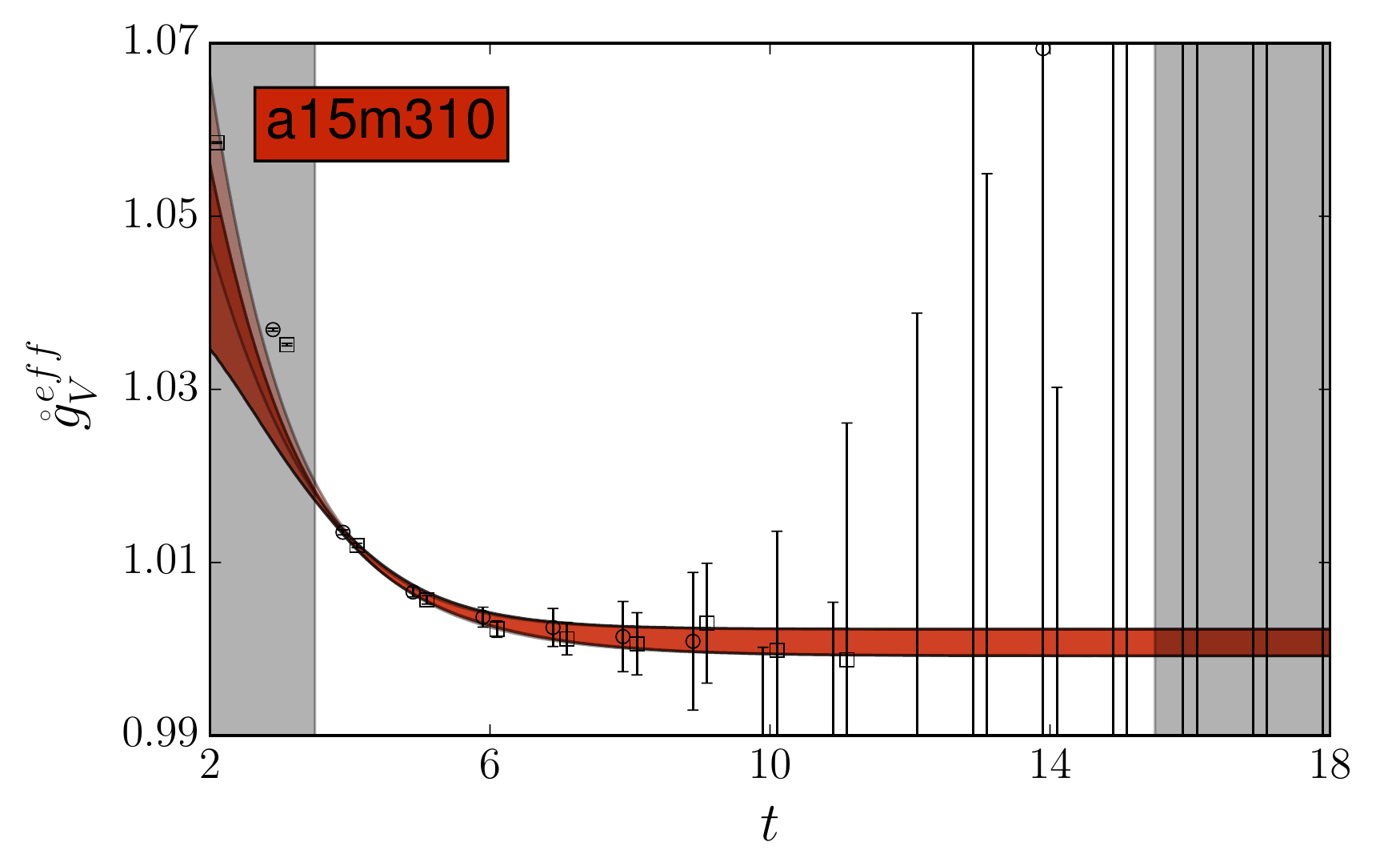}
\includegraphics[width=0.43\textwidth]{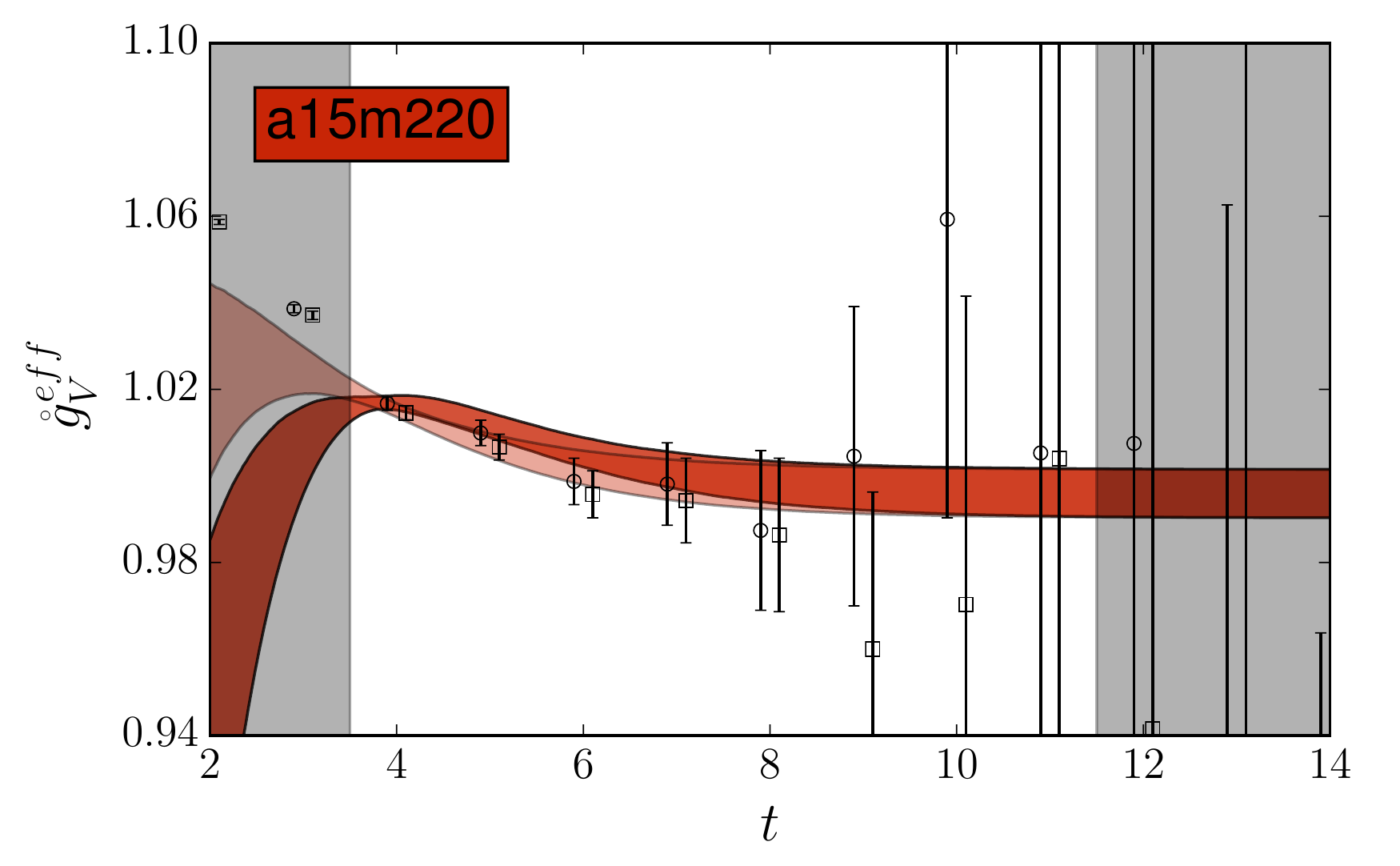}
\includegraphics[width=0.43\textwidth]{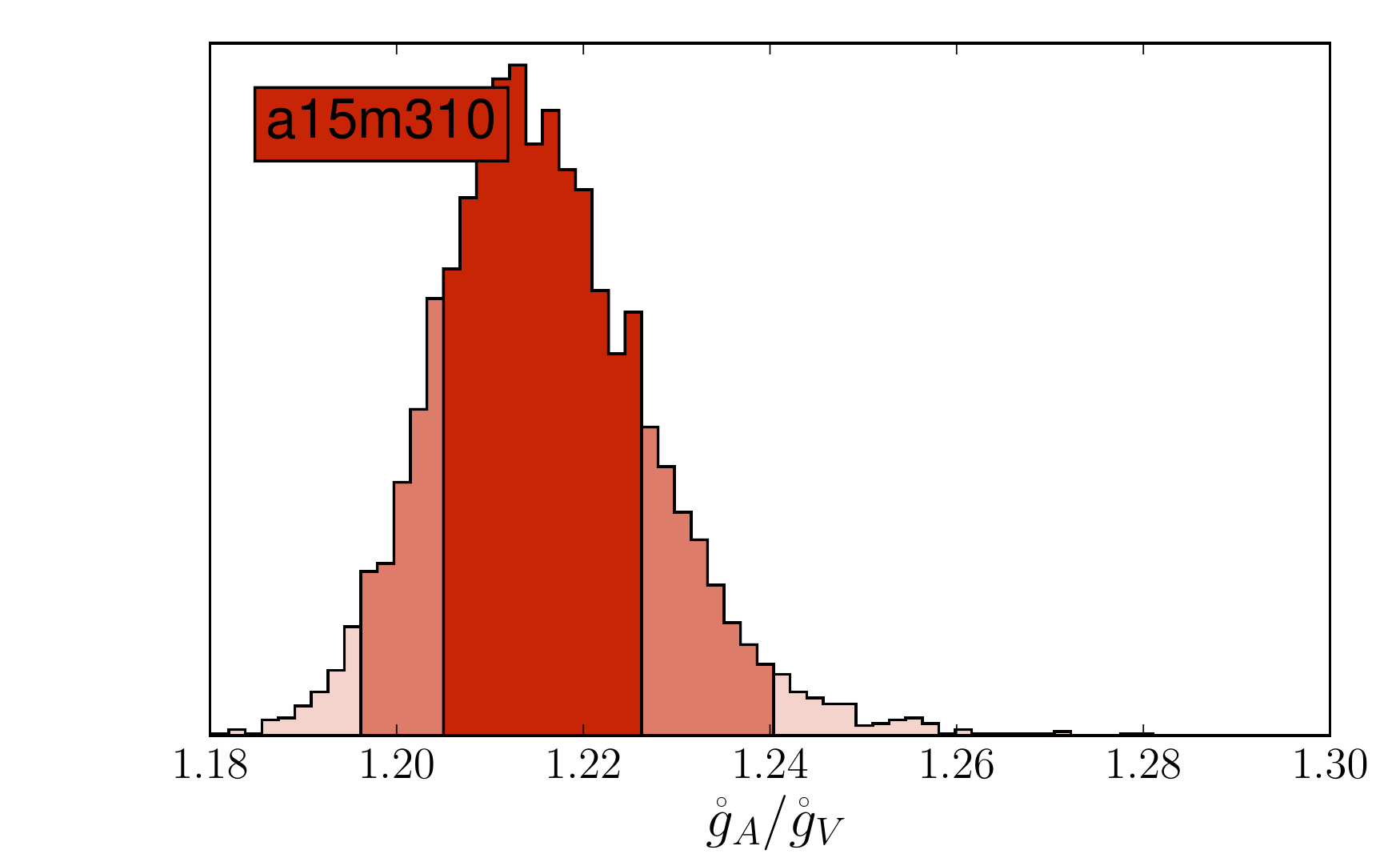}
\includegraphics[width=0.43\textwidth]{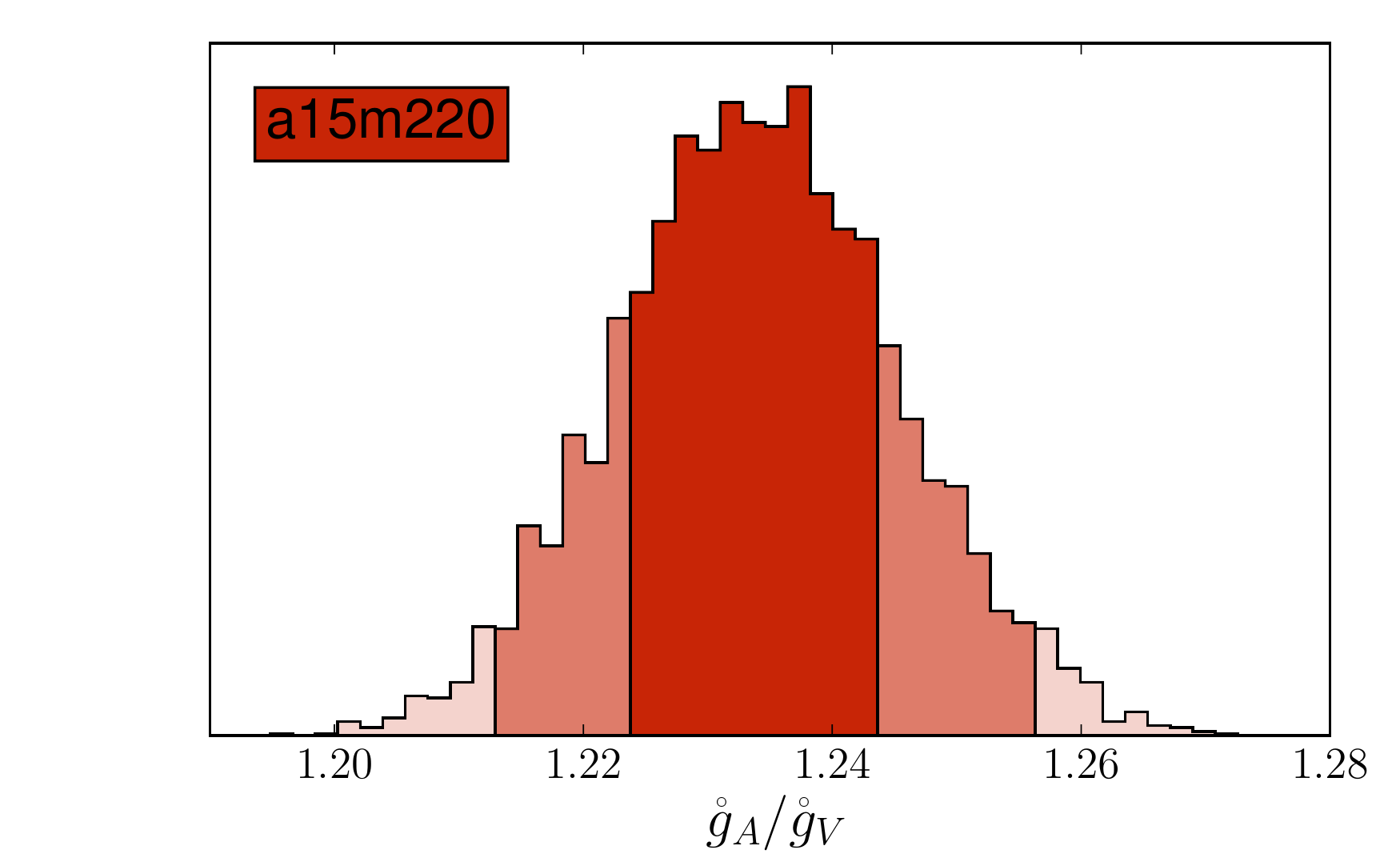}
\caption{Unbiased bootstrap fit curves with 68\% confidence intervals. Results from one simultaneous fit are represented in each column. The resulting biased bootstrap histograms for $\mathring{g}_{A}/\mathring{g}_{V}$ follow at the bottom. In the histograms, regions mark the 68\% and 95\% confidence interval.}
 \label{fig:a15m310a15m220_curve}
\end{figure*}

\begin{figure*}[h]
\includegraphics[width=0.43\textwidth]{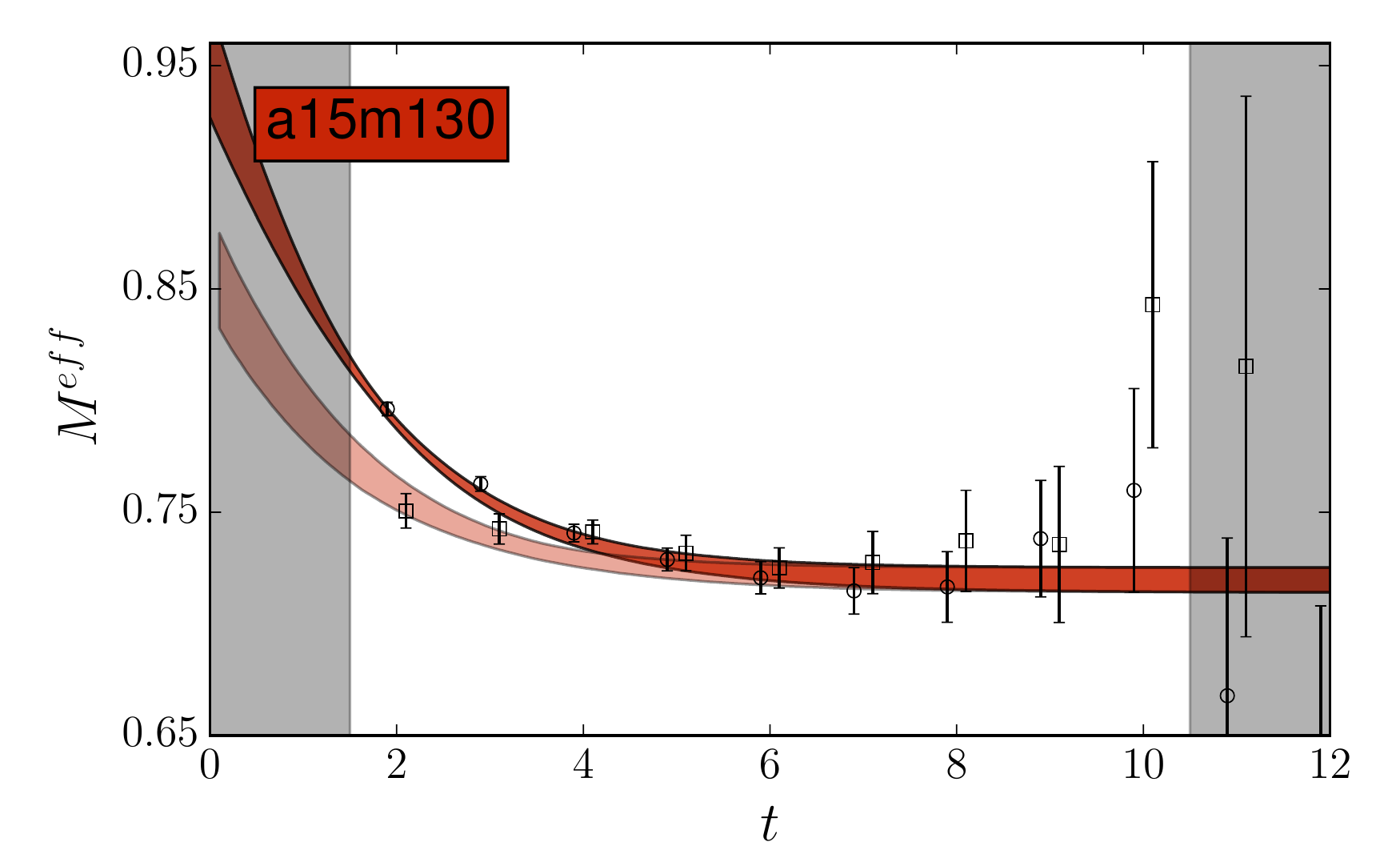}
\includegraphics[width=0.43\textwidth]{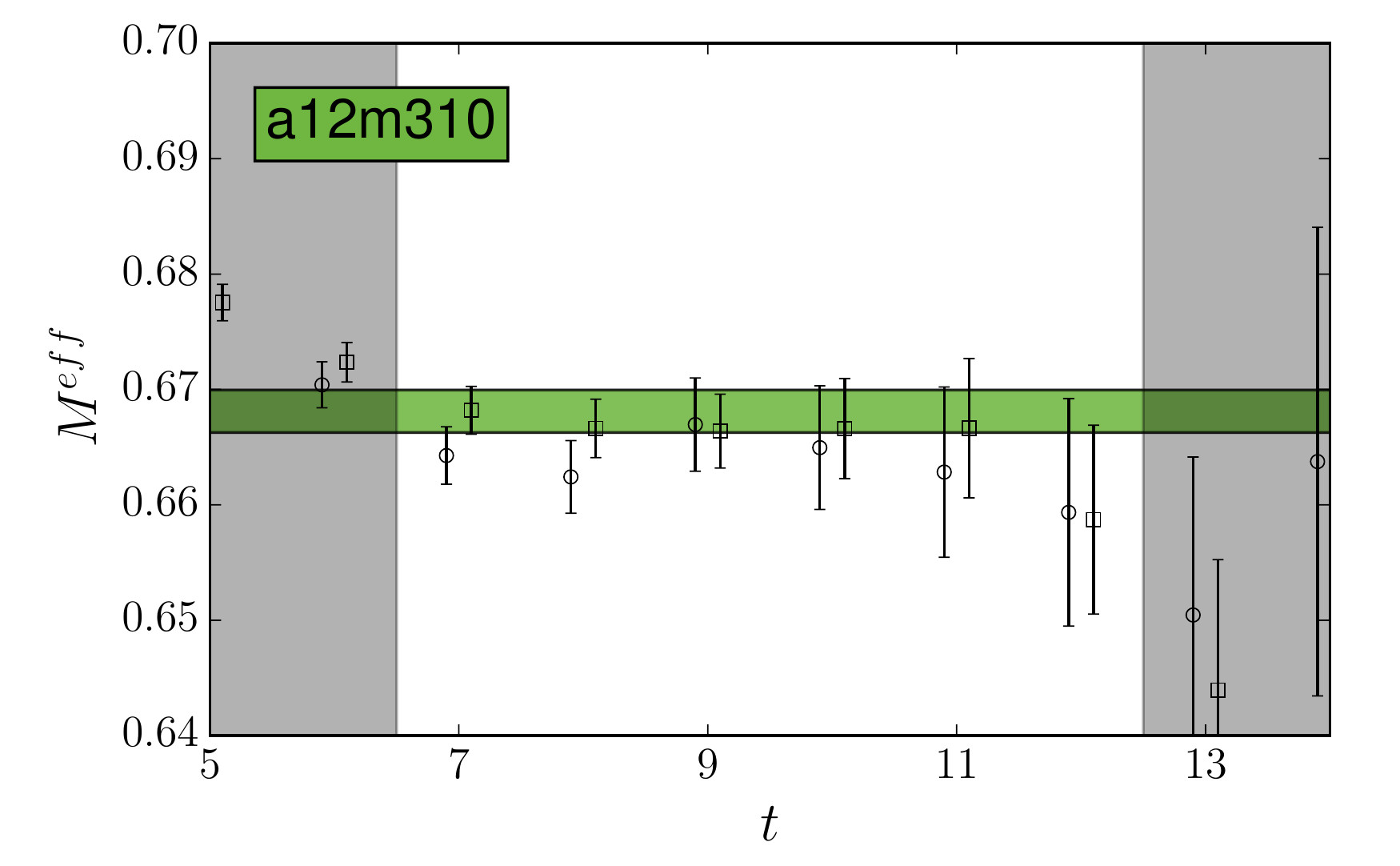}
\includegraphics[width=0.43\textwidth]{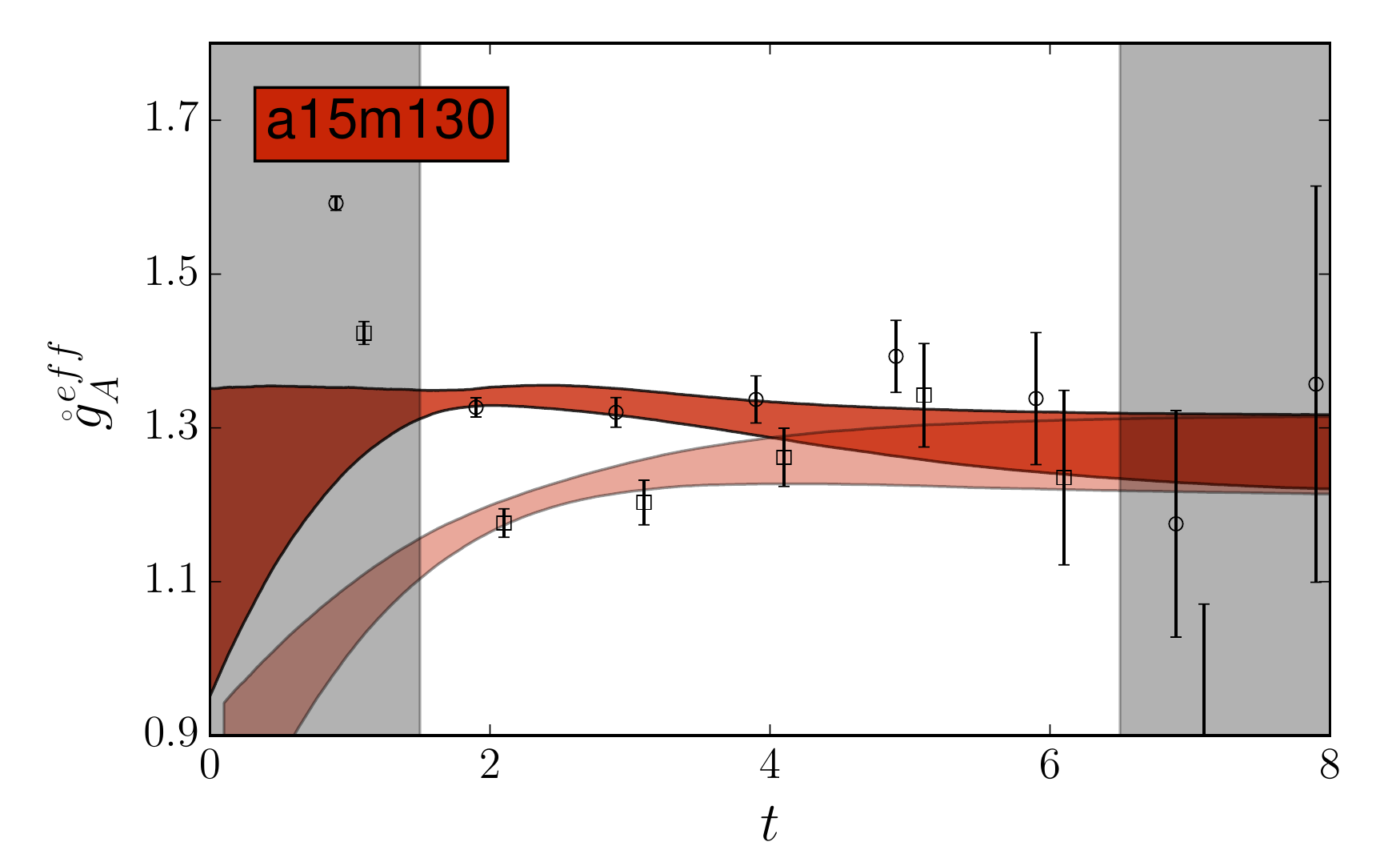}
\includegraphics[width=0.43\textwidth]{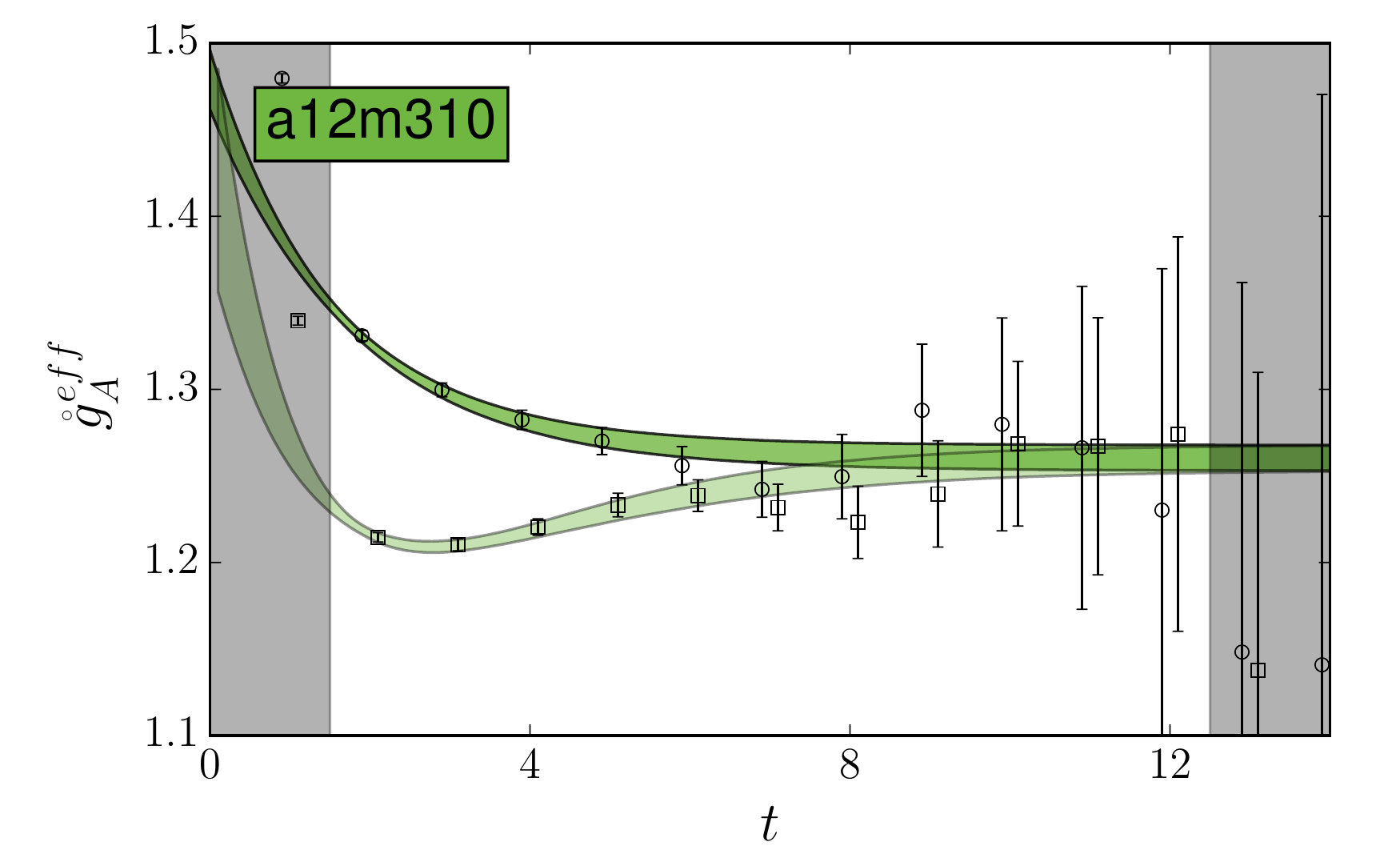}
\includegraphics[width=0.43\textwidth]{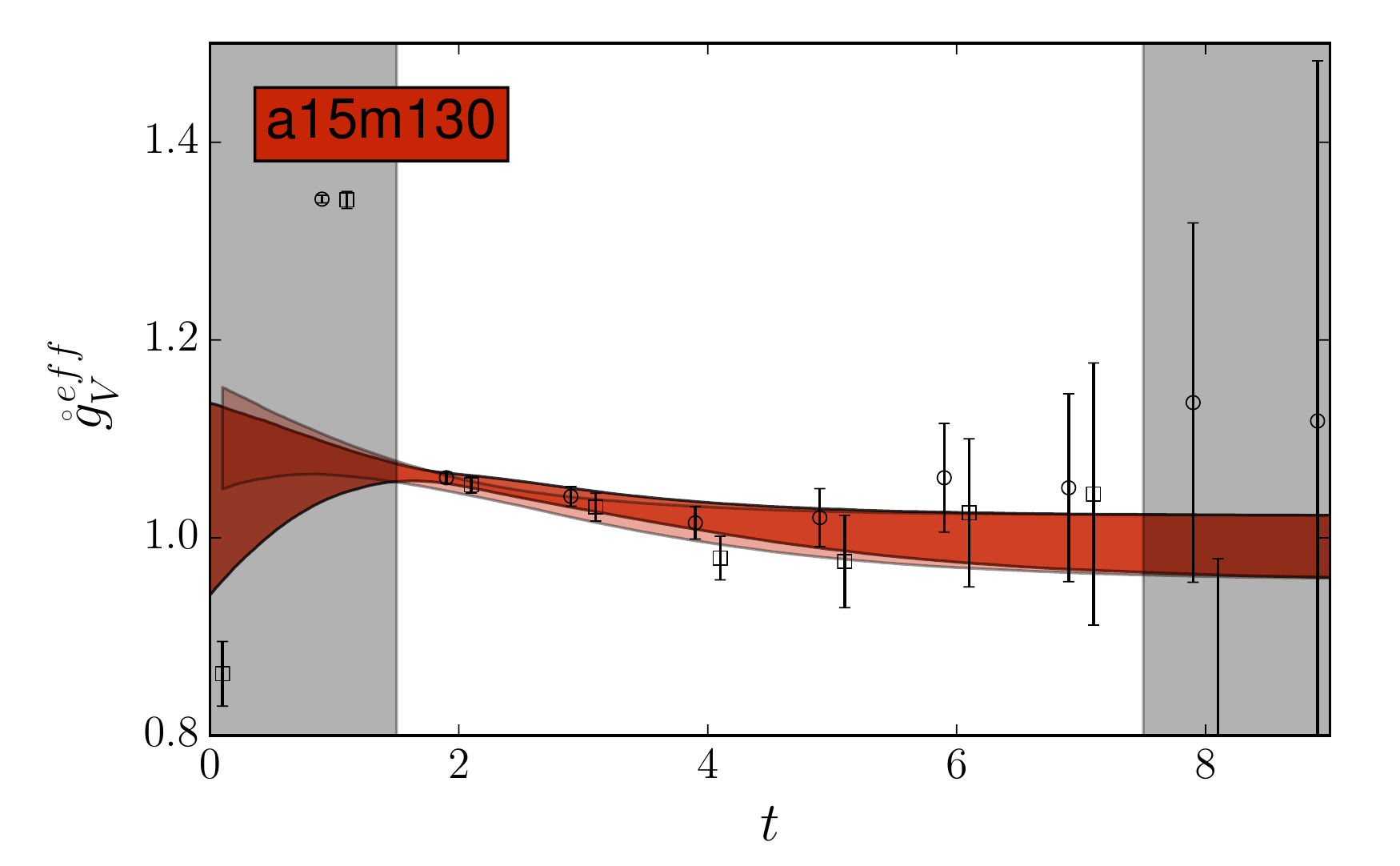}
\includegraphics[width=0.43\textwidth]{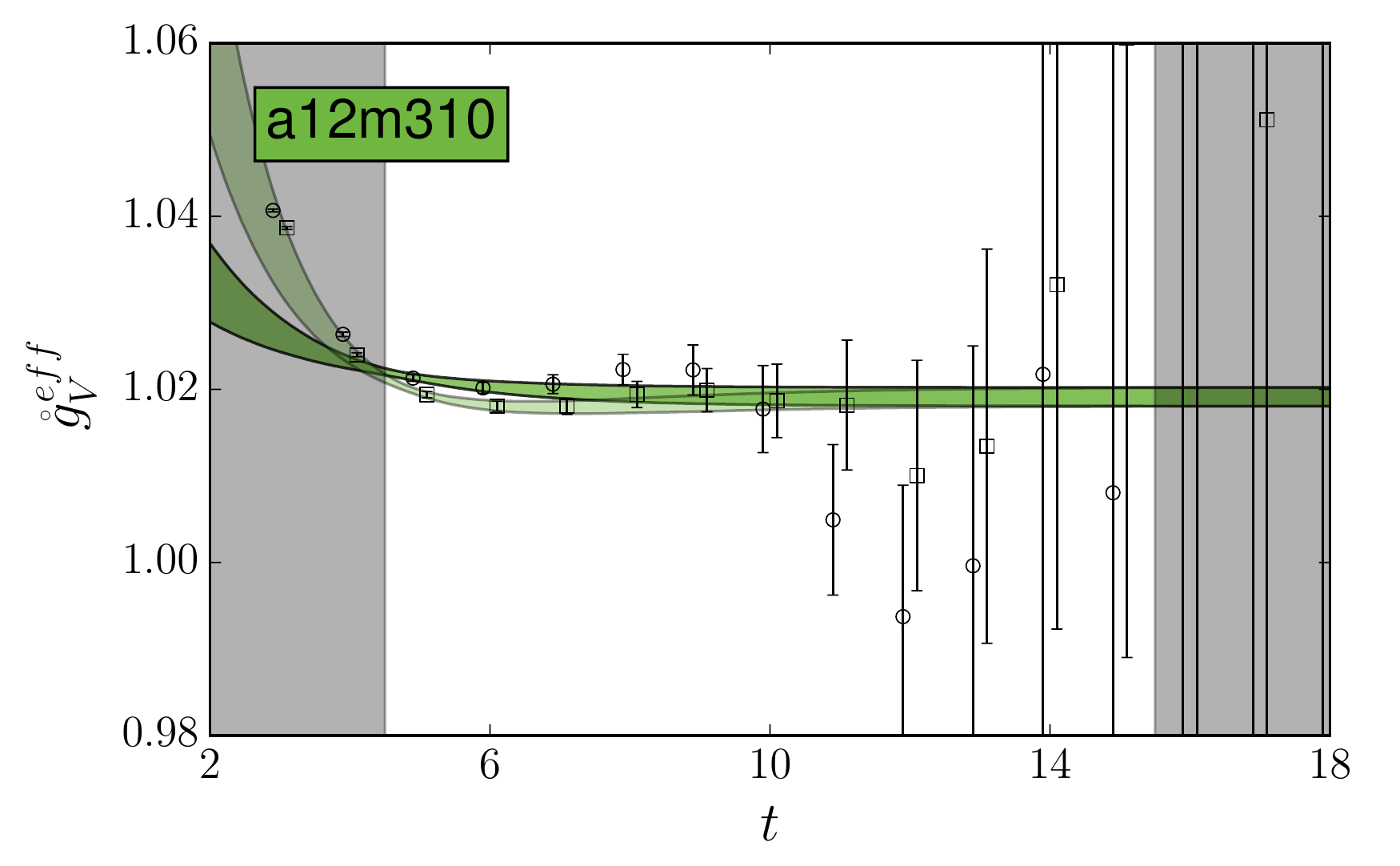}
\includegraphics[width=0.43\textwidth]{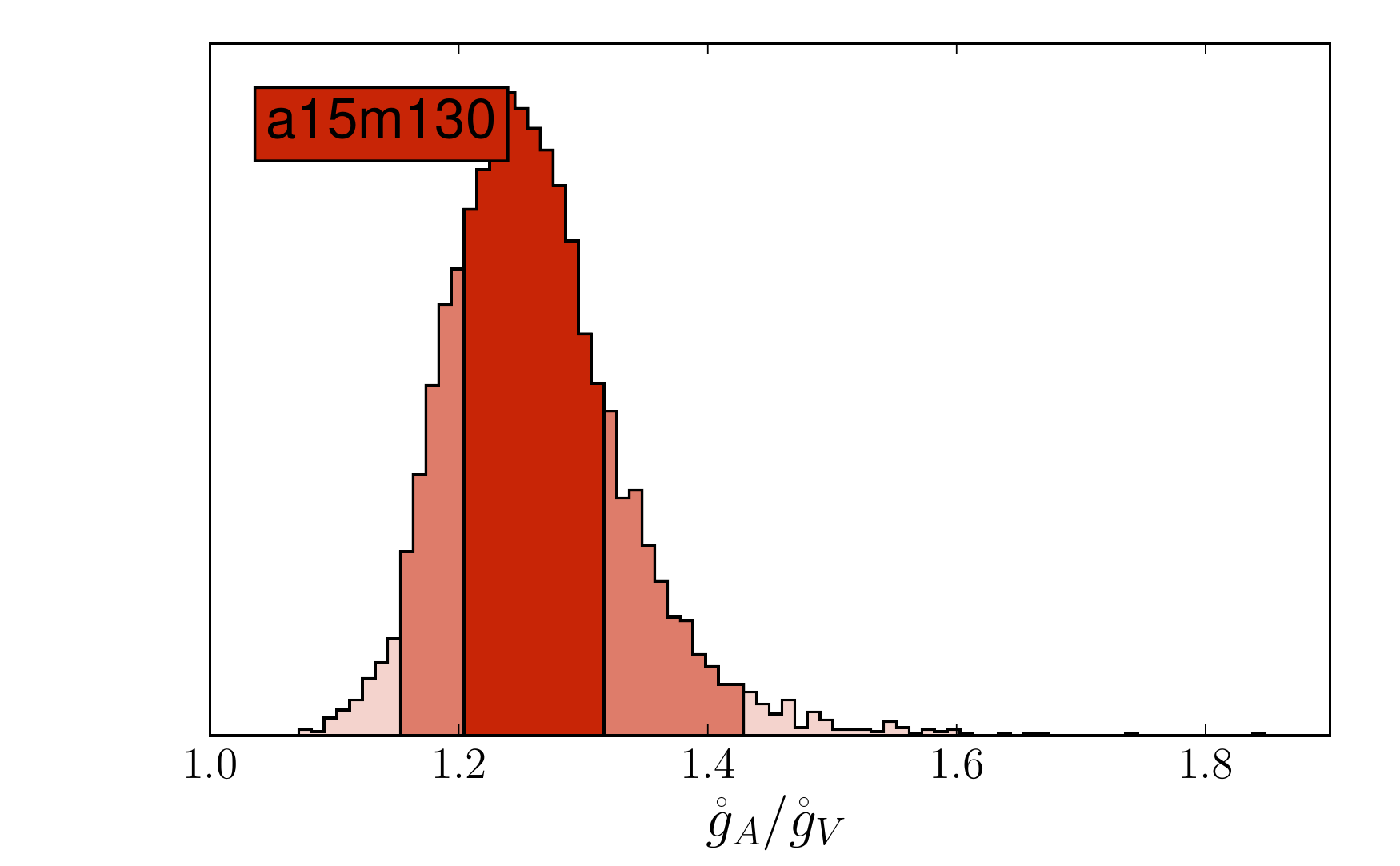}
\includegraphics[width=0.43\textwidth]{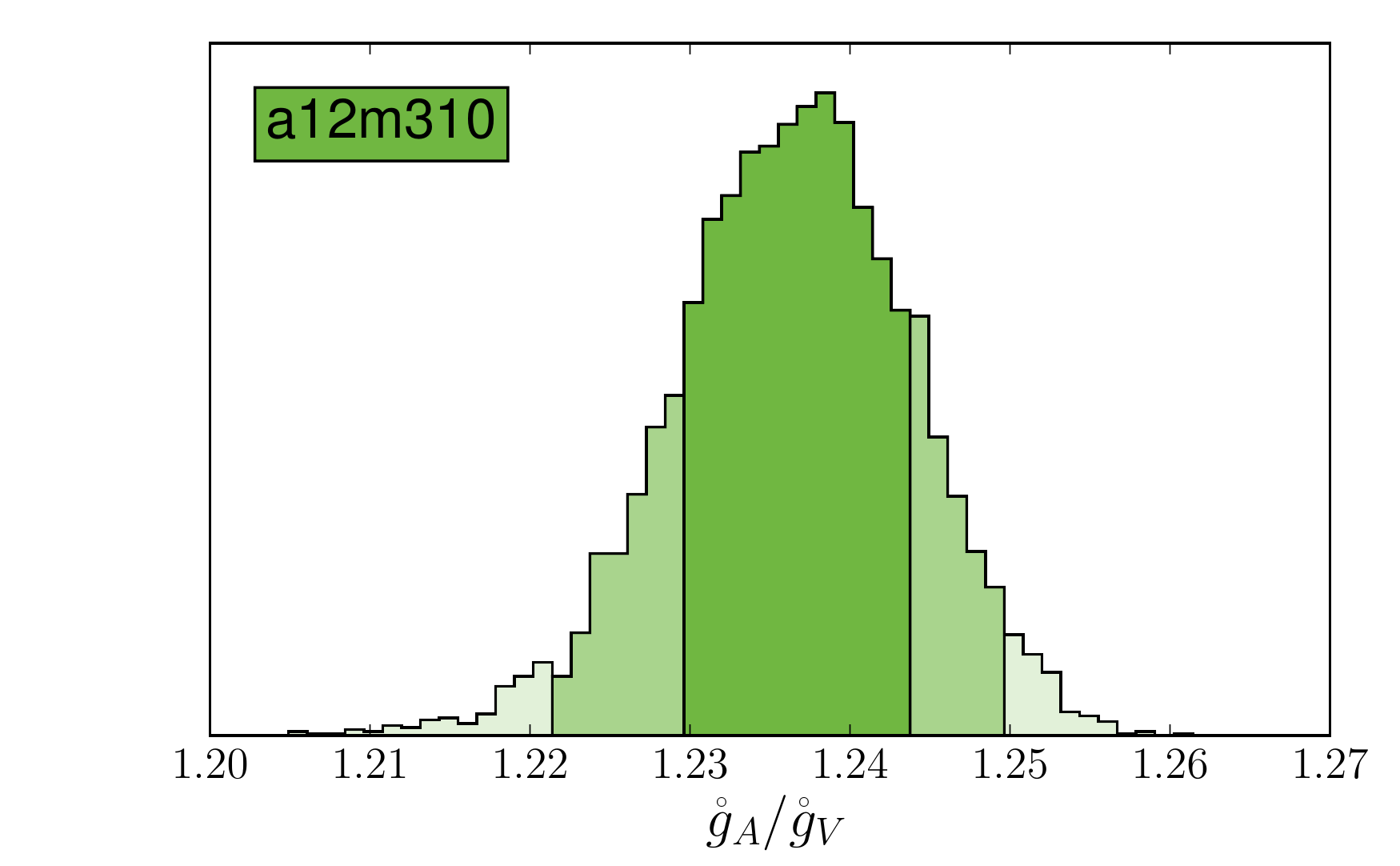}
\caption{Same as Fig.~\ref{fig:a15m310a15m220_curve} for the a15m130 and a12m310 ensembles.}
 \label{fig:a15m130a12m310_curve}
\end{figure*}

\begin{figure*}[h]
\includegraphics[width=0.43\textwidth]{meff_a12m220.pdf}
\includegraphics[width=0.43\textwidth]{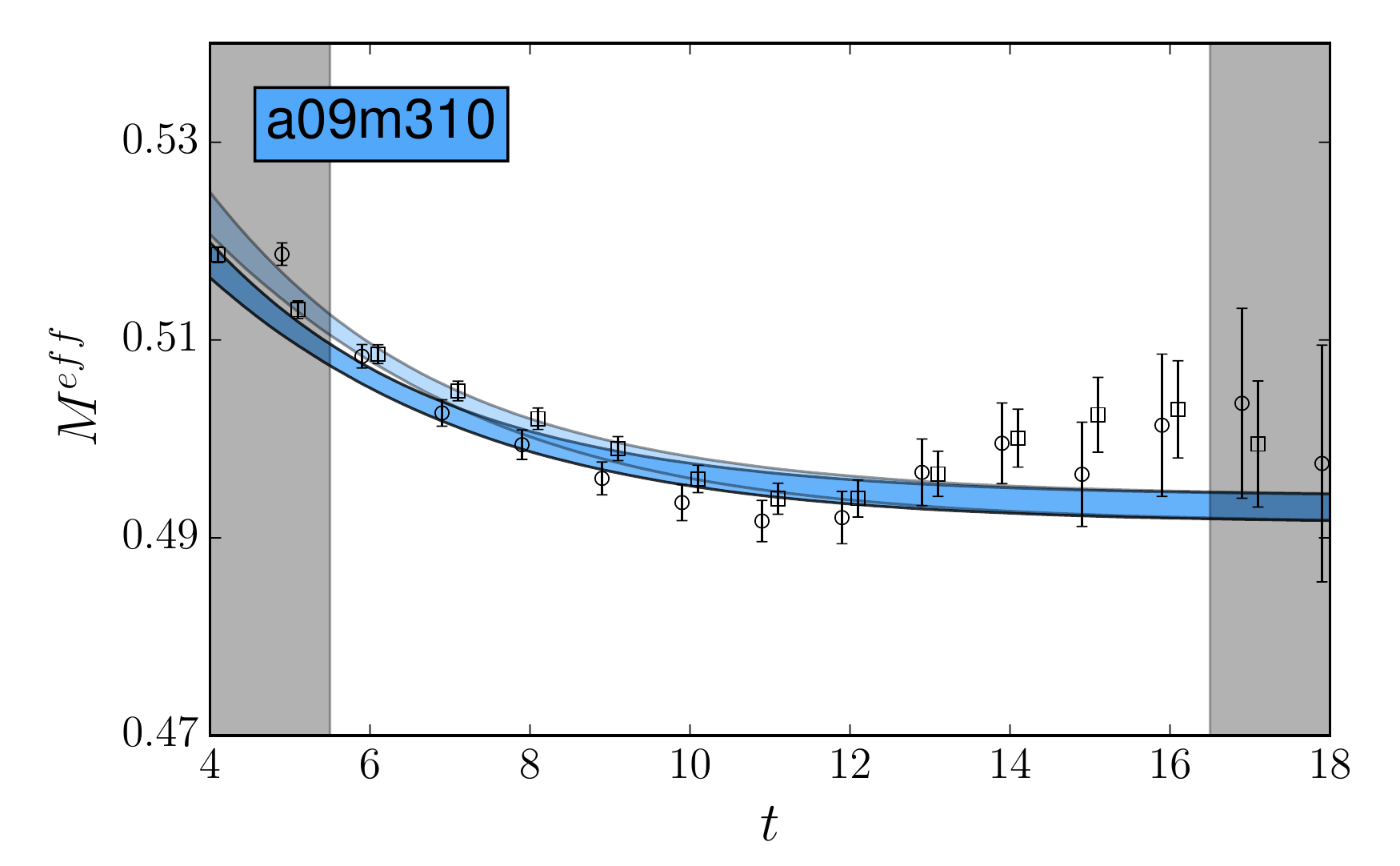}
\includegraphics[width=0.43\textwidth]{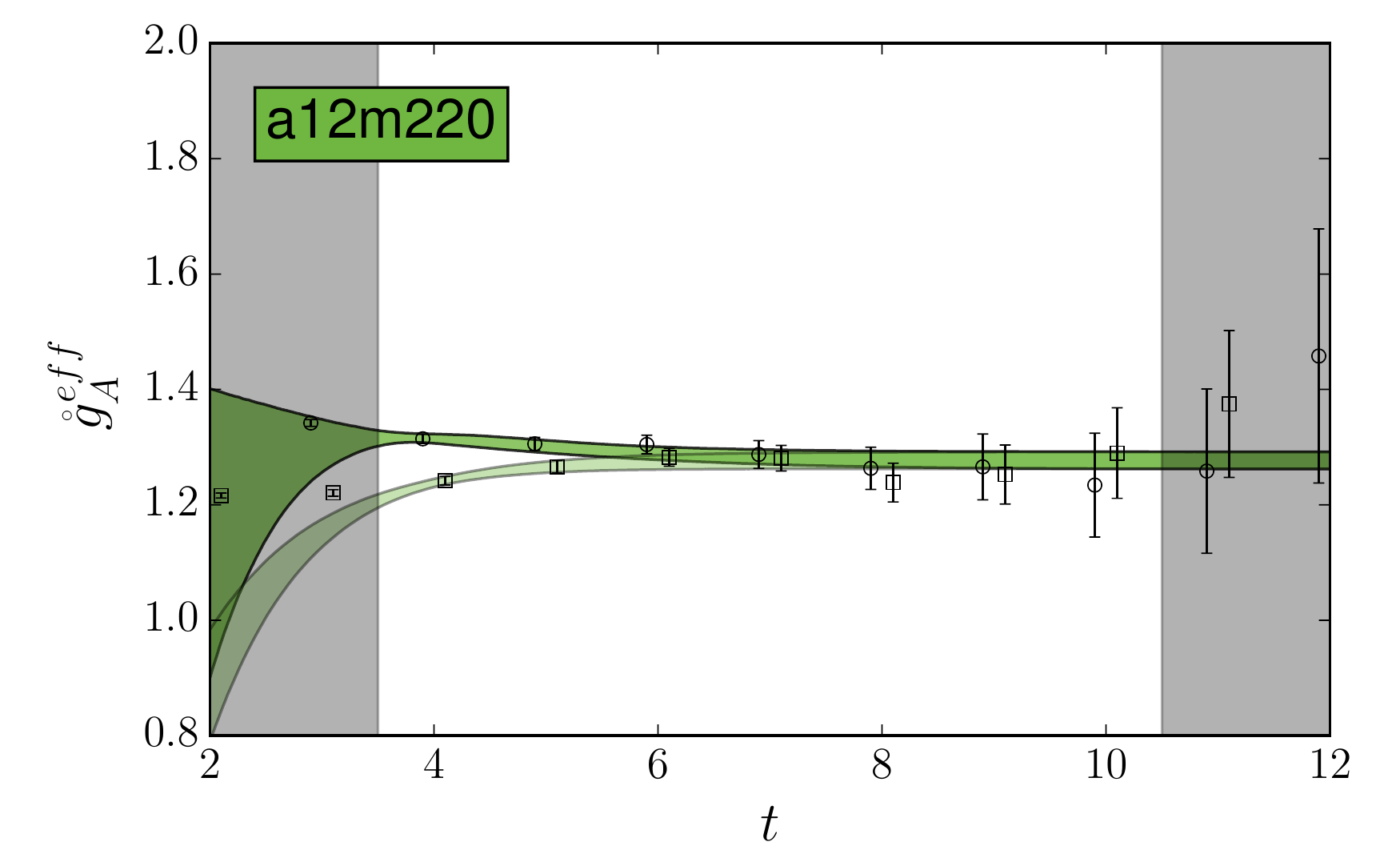}
\includegraphics[width=0.43\textwidth]{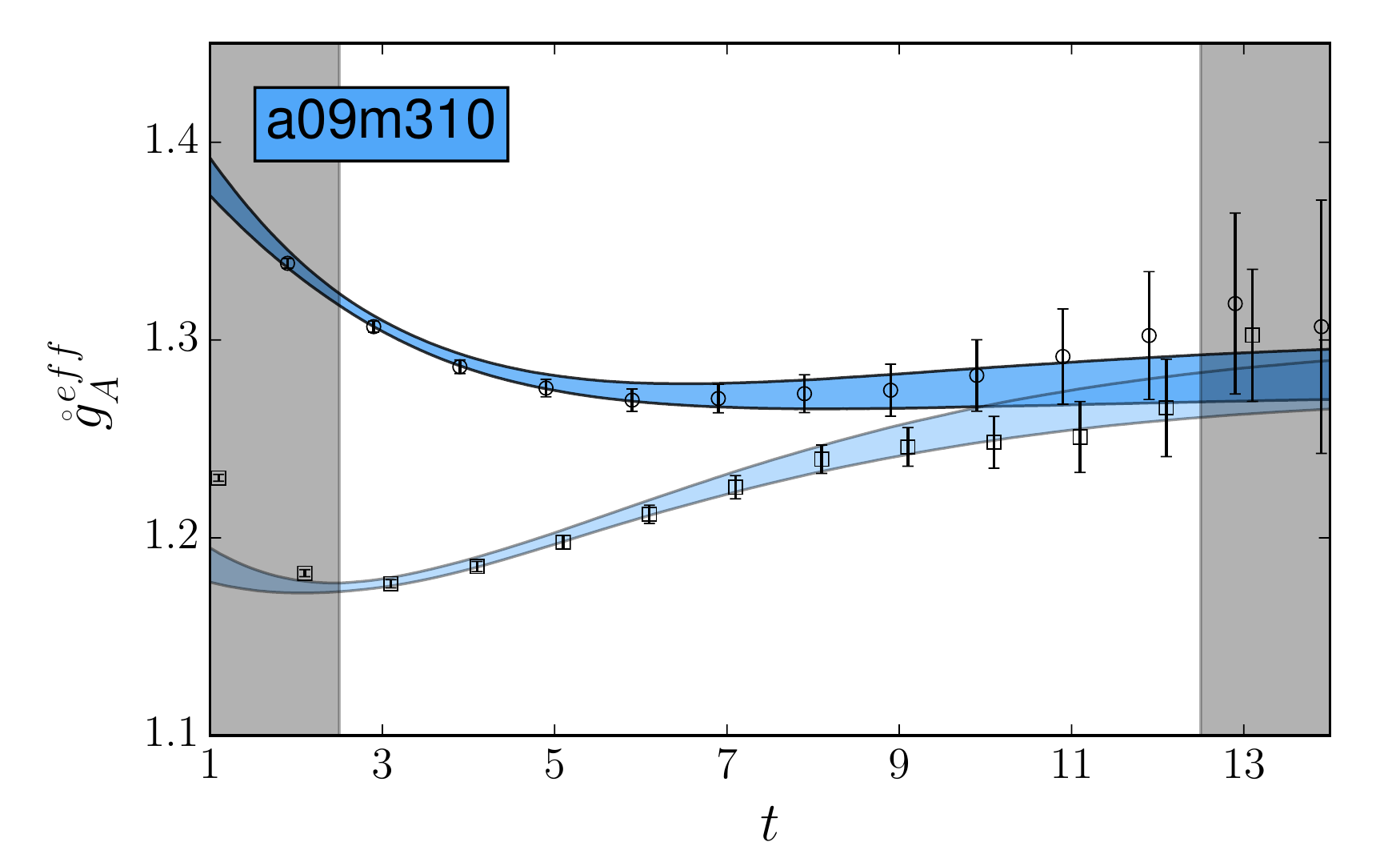}
\includegraphics[width=0.43\textwidth]{gV_a12m220.pdf}
\includegraphics[width=0.43\textwidth]{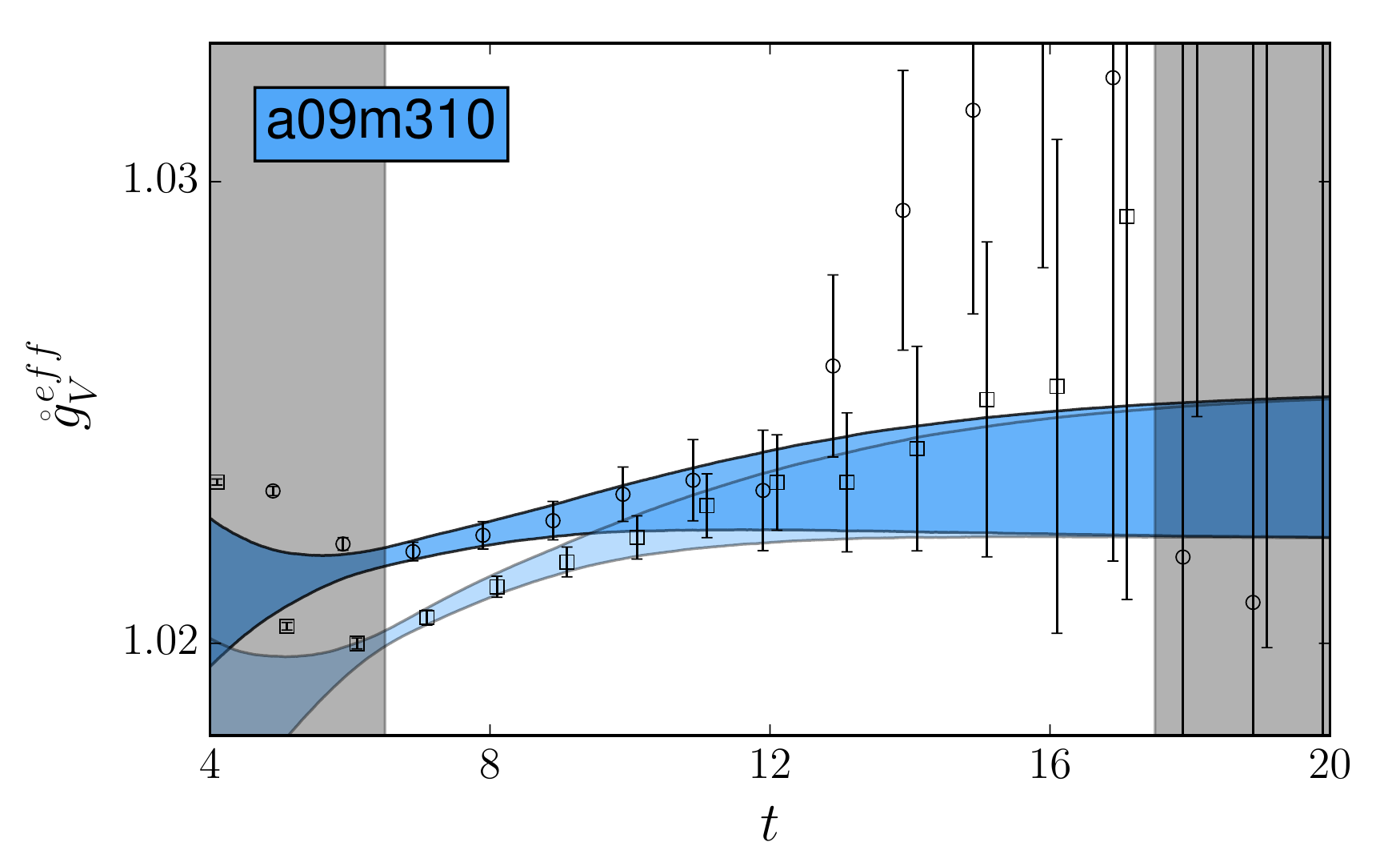}
\includegraphics[width=0.43\textwidth]{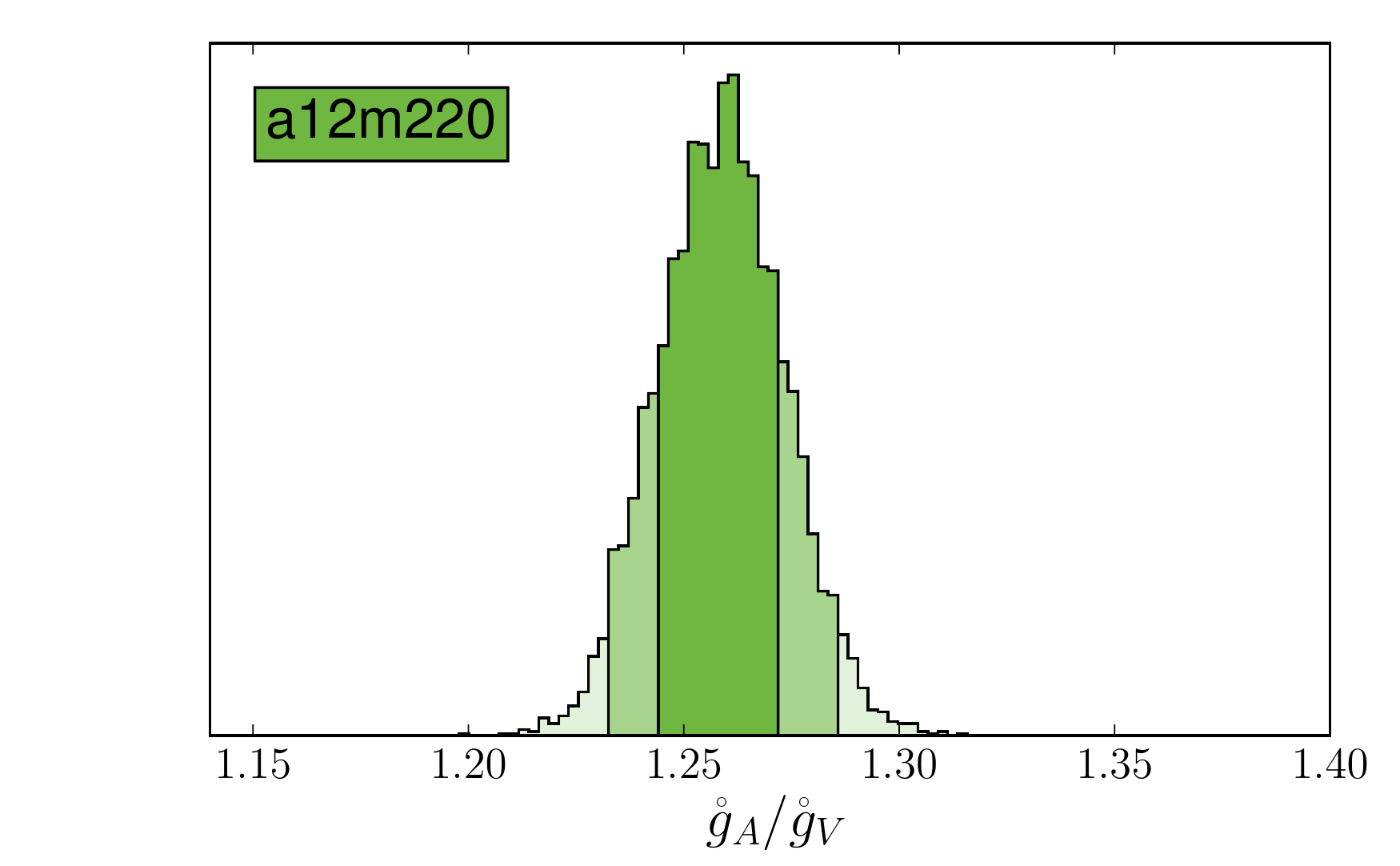}
\includegraphics[width=0.43\textwidth]{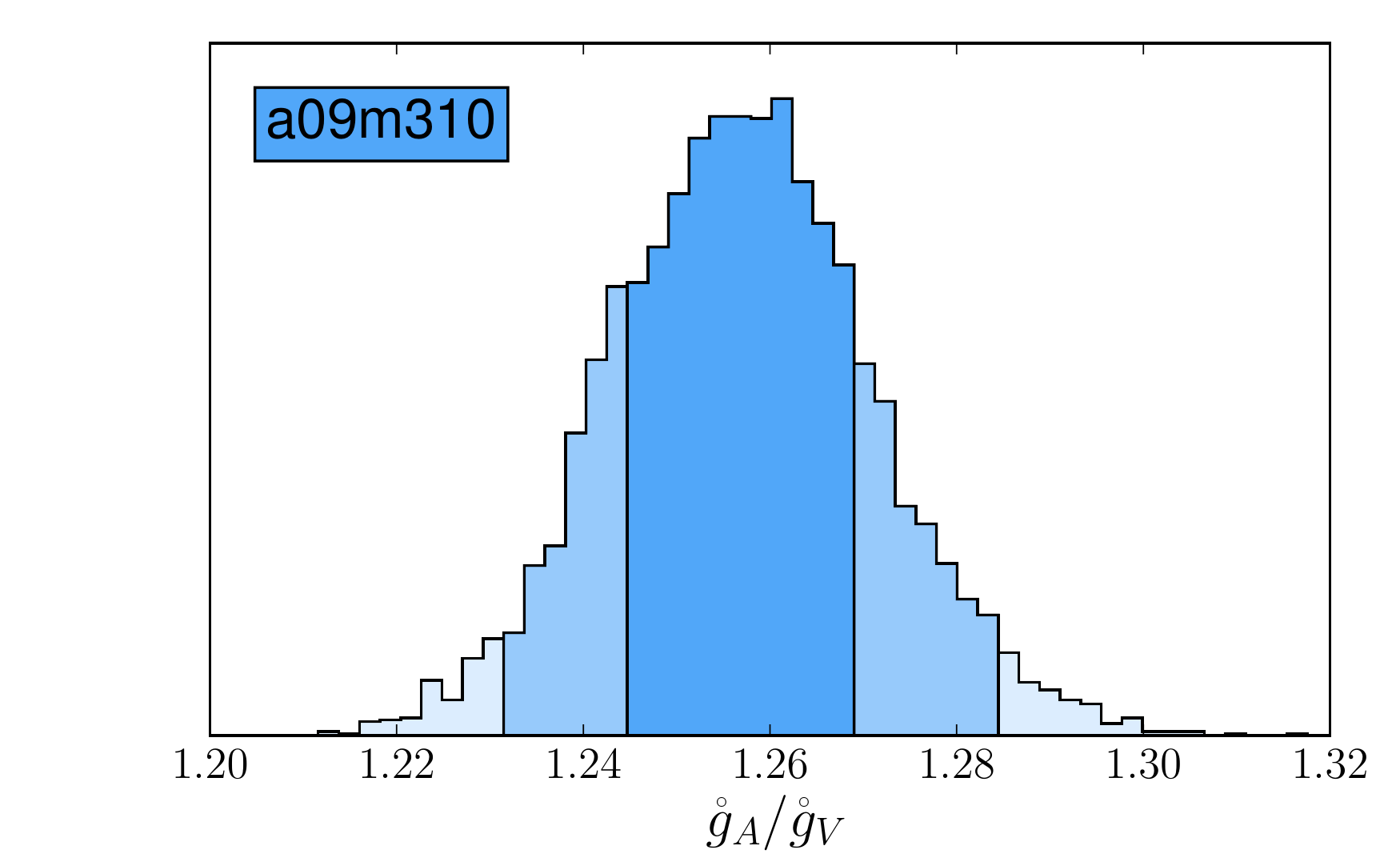}
\caption{Same as Fig.~\ref{fig:a15m310a15m220_curve}. The ordinate for $M^{eff}$, $\mathring{g}_A^{eff}$, and $\mathring{g}_V^{eff}$ and the abscissa of the histogram for the a12m220 ensemble is manually set to be the same as Fig.~\ref{fig:a12m220SL_curve}.}
 \label{fig:a12m220a09m310_curve}
\end{figure*}

\begin{figure*}[h]
\includegraphics[width=0.43\textwidth]{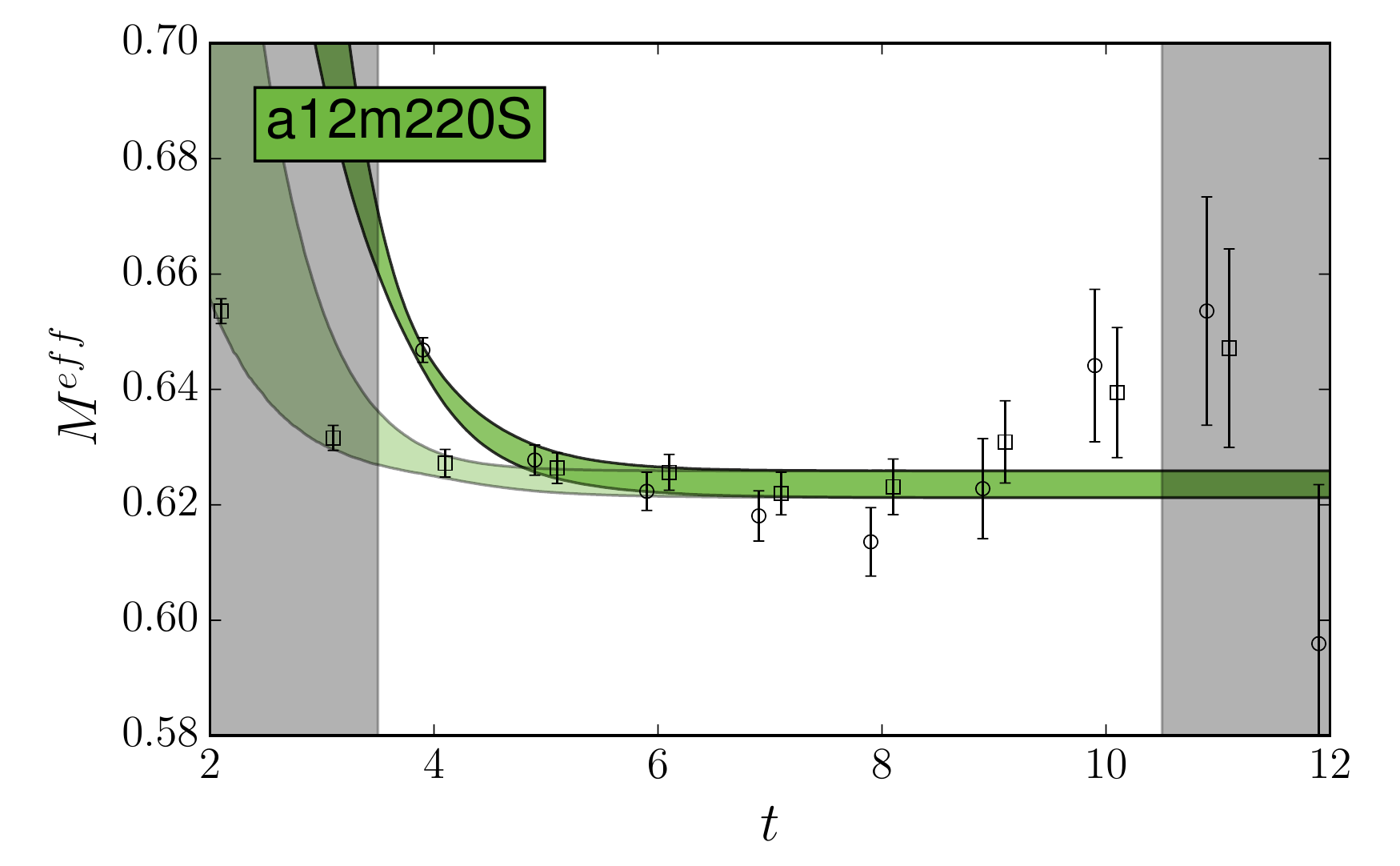}
\includegraphics[width=0.43\textwidth]{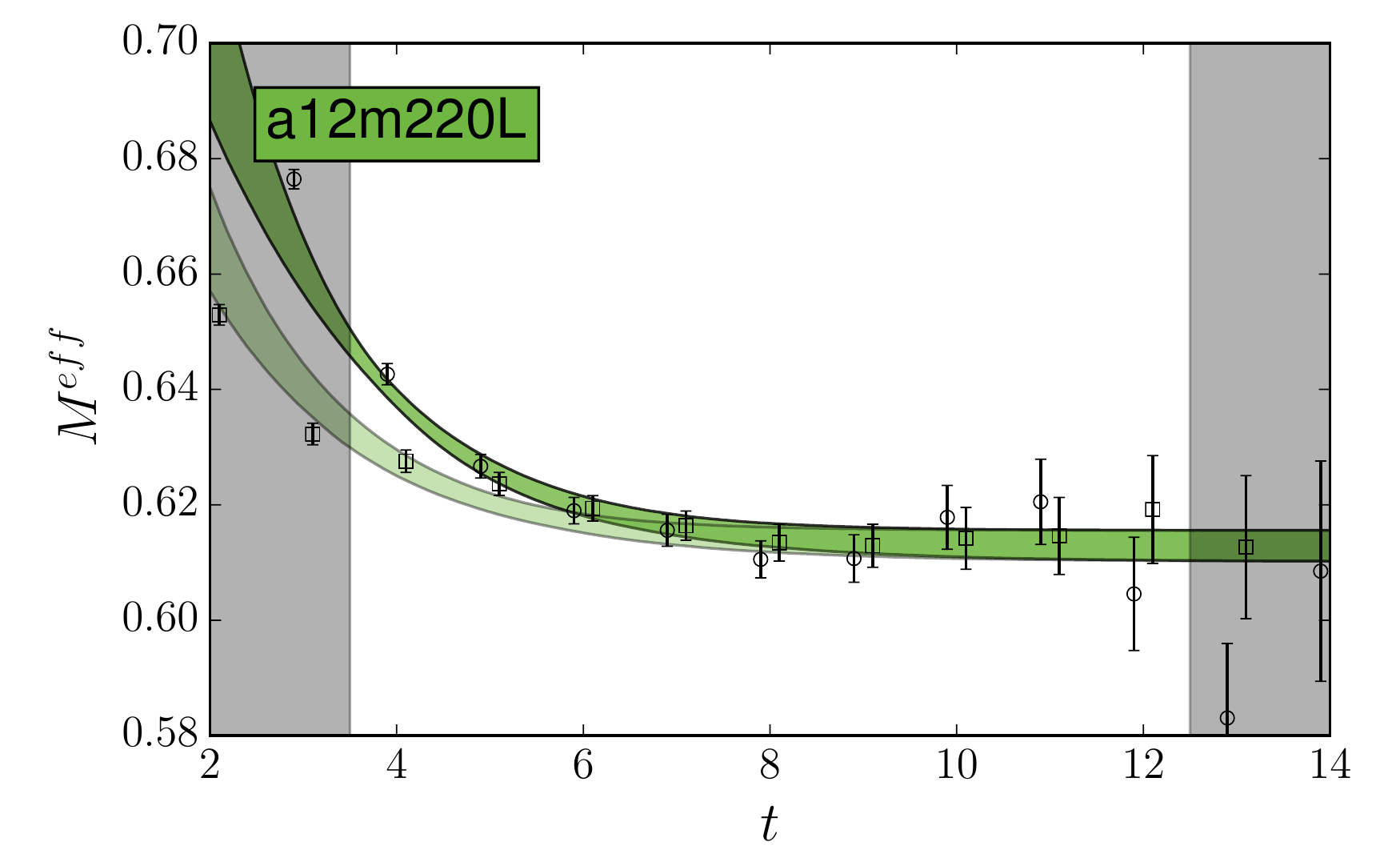}
\includegraphics[width=0.43\textwidth]{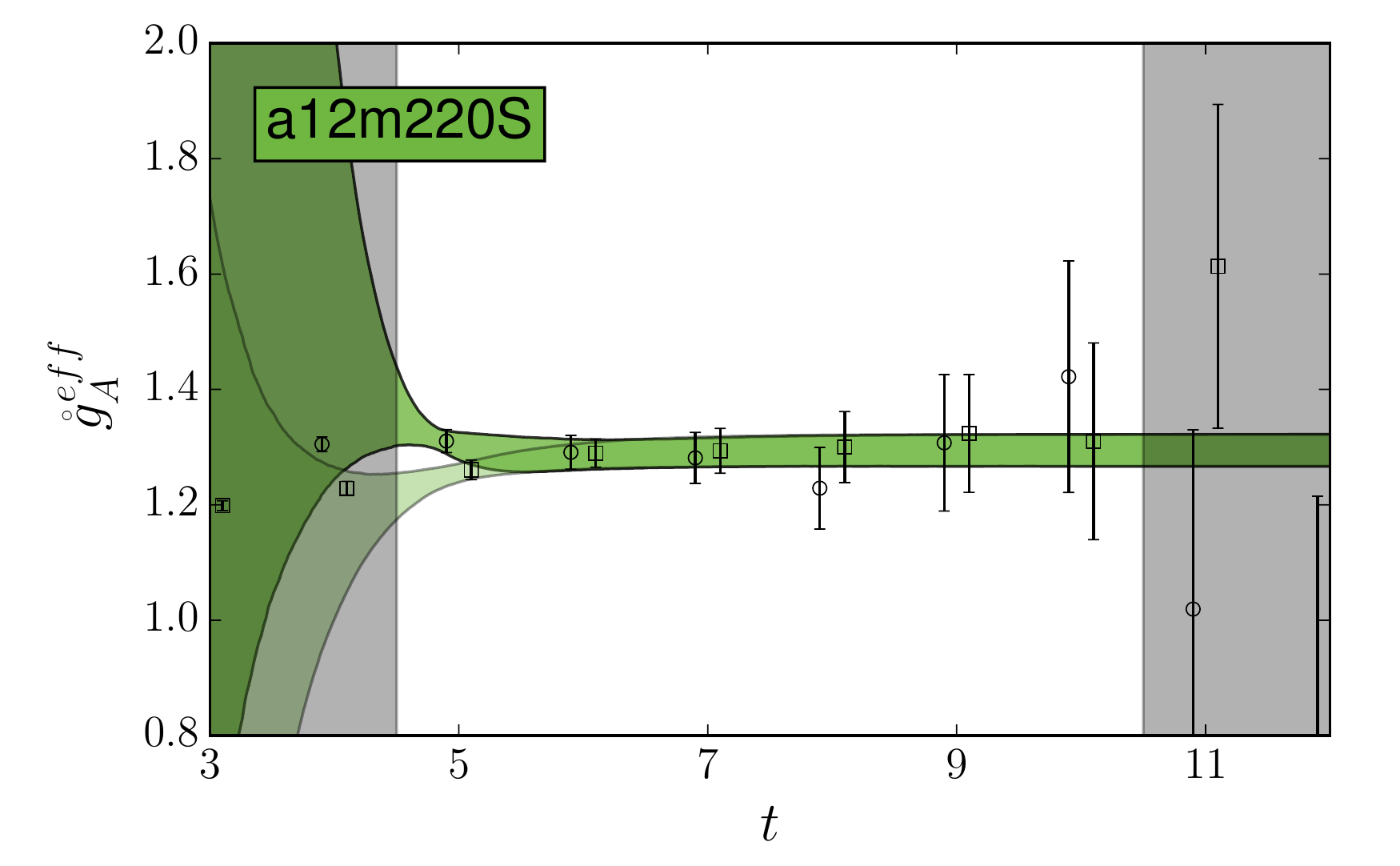}
\includegraphics[width=0.43\textwidth]{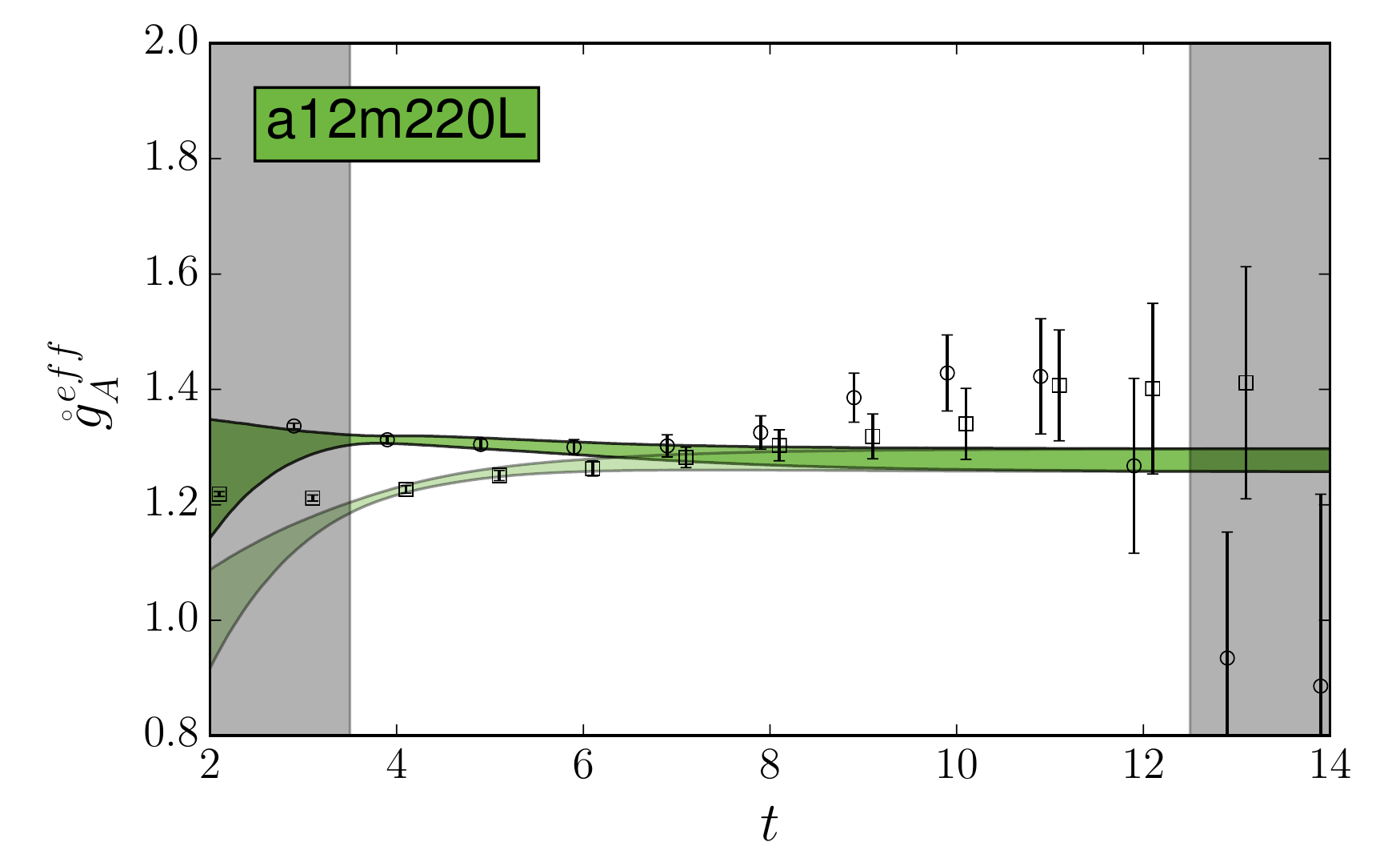}
\includegraphics[width=0.43\textwidth]{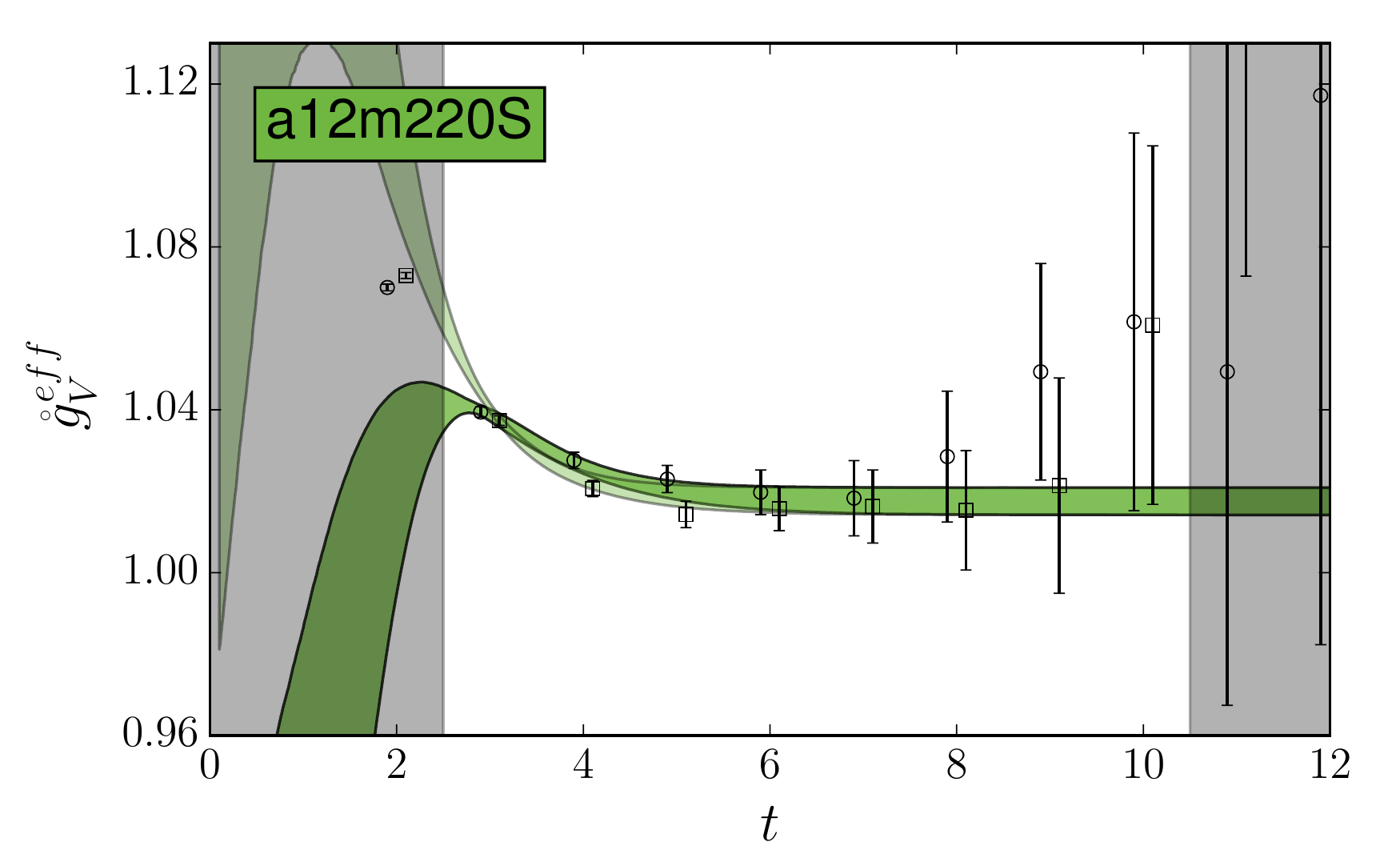}
\includegraphics[width=0.43\textwidth]{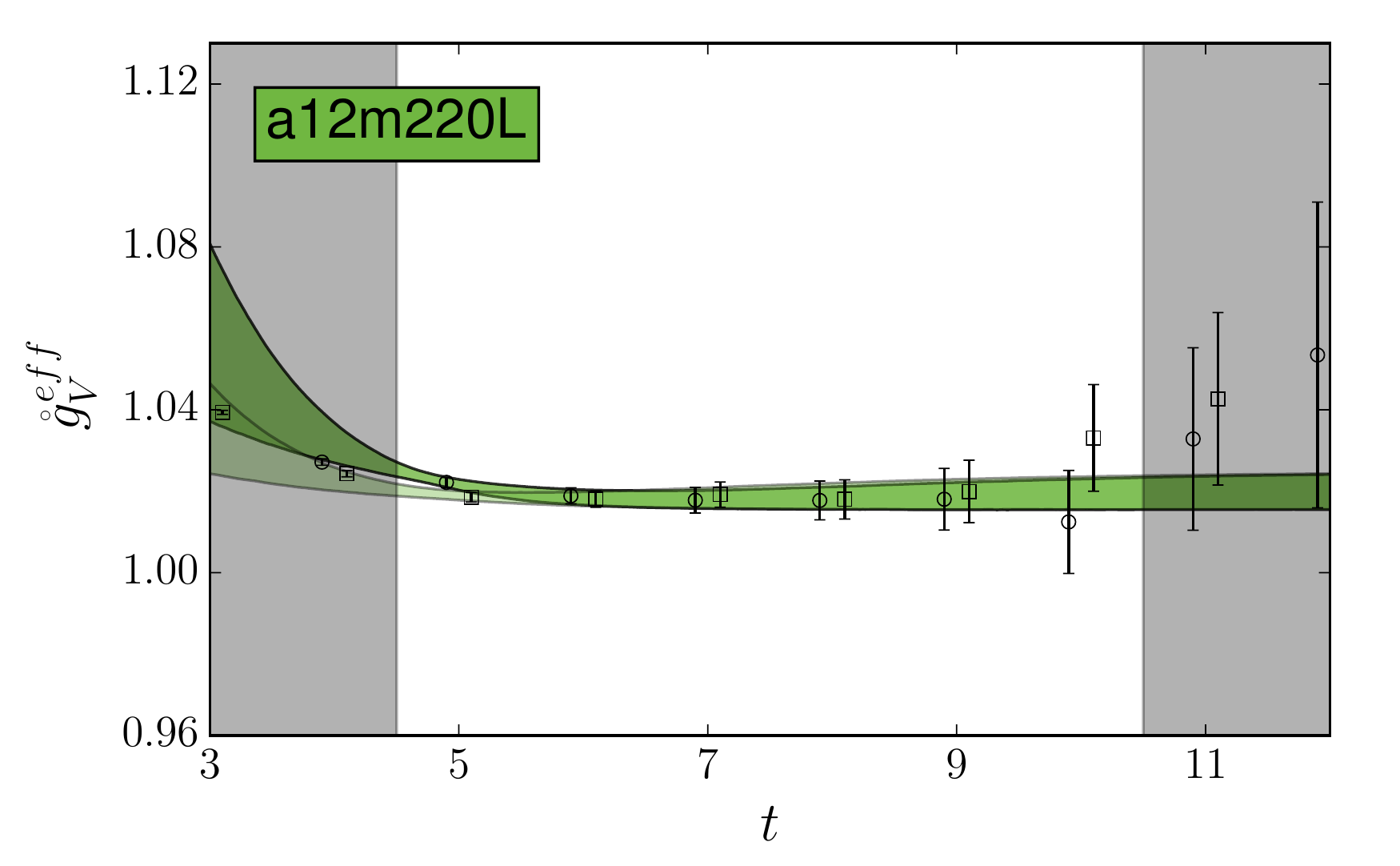}
\includegraphics[width=0.43\textwidth]{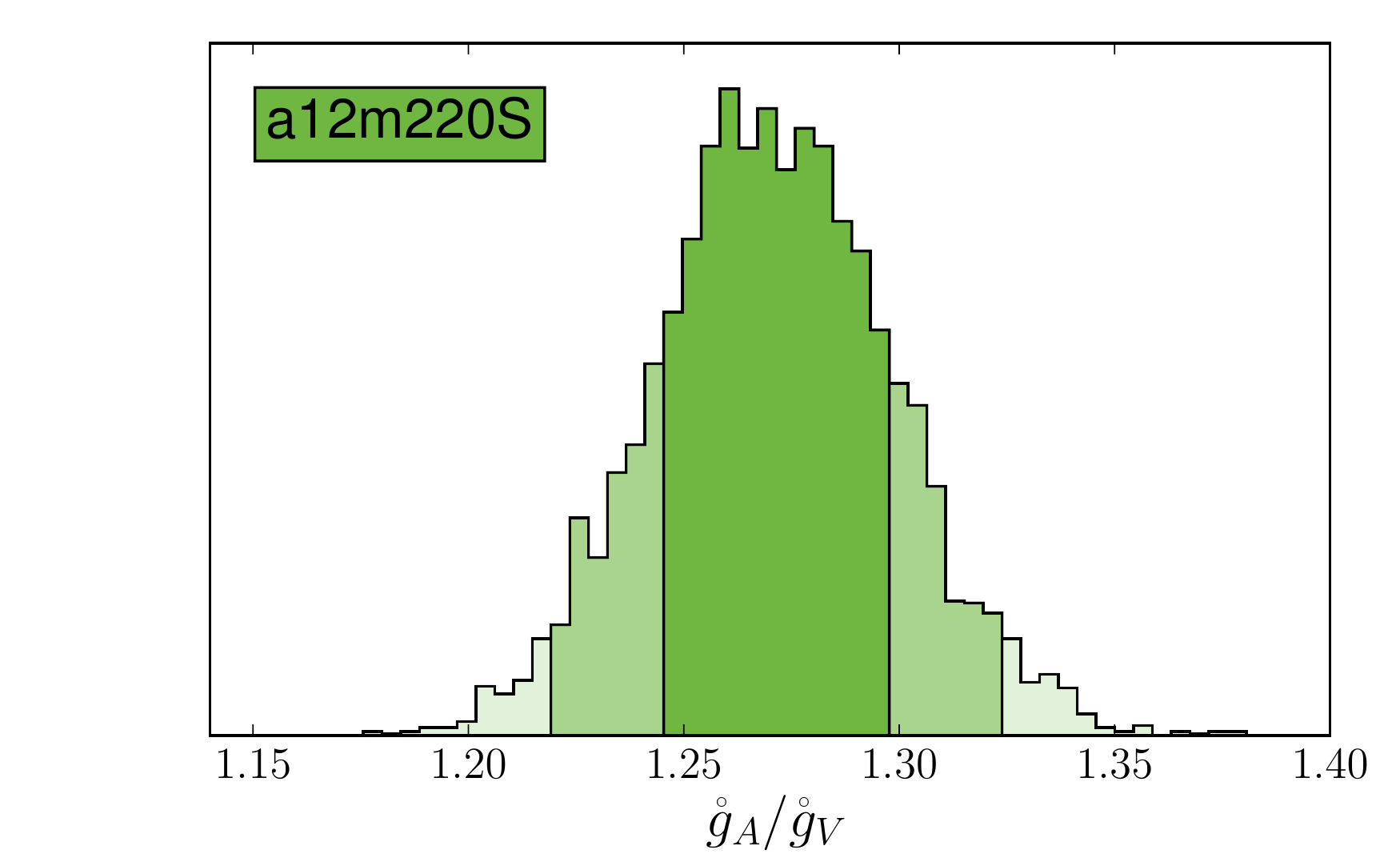}
\includegraphics[width=0.43\textwidth]{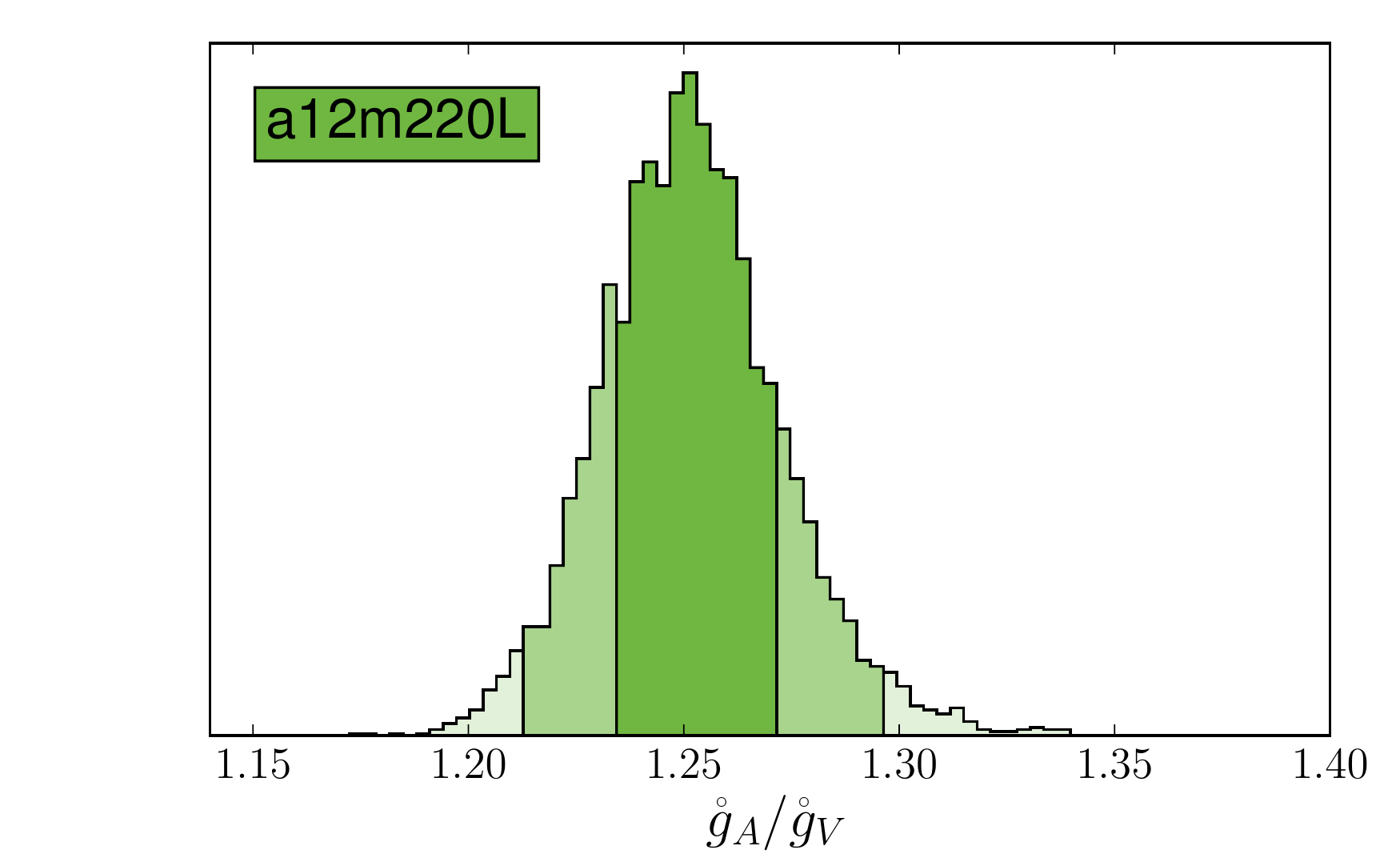}
\caption{Same as Fig.~\ref{fig:a15m310a15m220_curve}. The ordinate for $M^{eff}$, $\mathring{g}_A^{eff}$, and $\mathring{g}_V^{eff}$ and the abscissa of the histogram for the a12m220S and a12m220L ensembles are manually set to be the same as the a12m220 ensemble in Fig.~\ref{fig:a12m220a09m310_curve}.}
 \label{fig:a12m220SL_curve}
\end{figure*}
\FloatBarrier
\newpage
\section{Correlator stability plots}
\label{app:stability}
\begin{figure*}[h]
\includegraphics[width=0.43\textwidth]{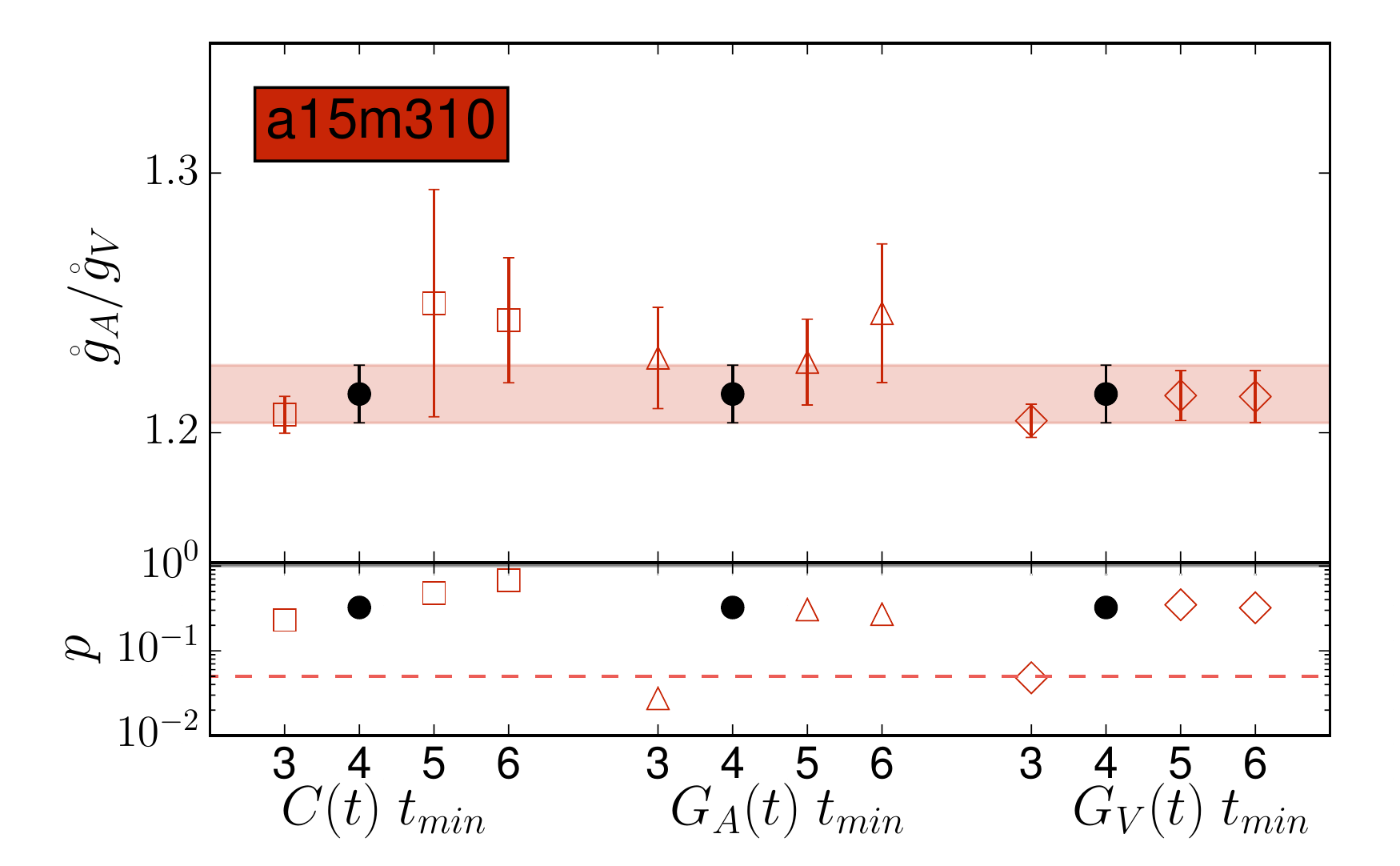}
\includegraphics[width=0.43\textwidth]{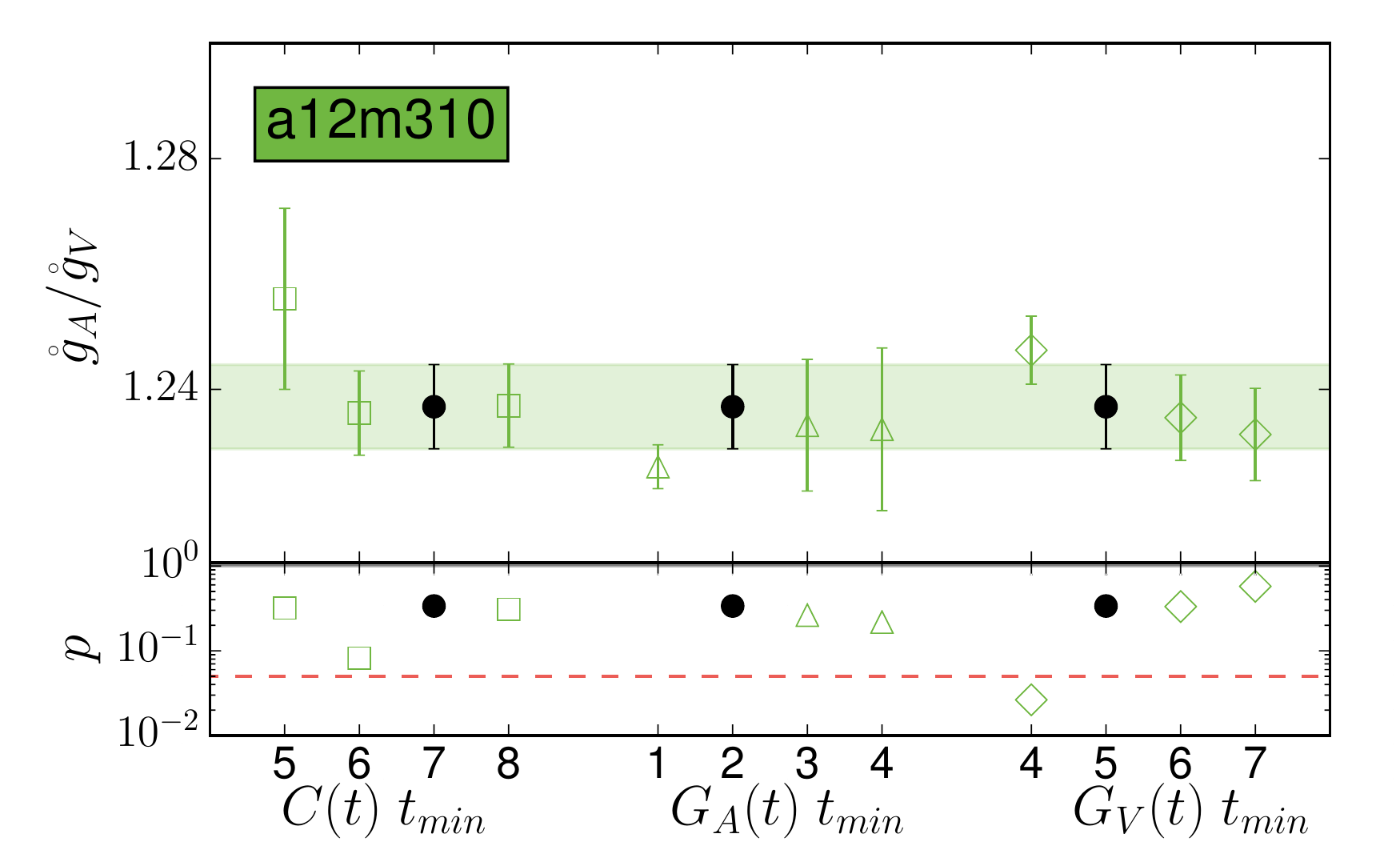}
\includegraphics[width=0.43\textwidth]{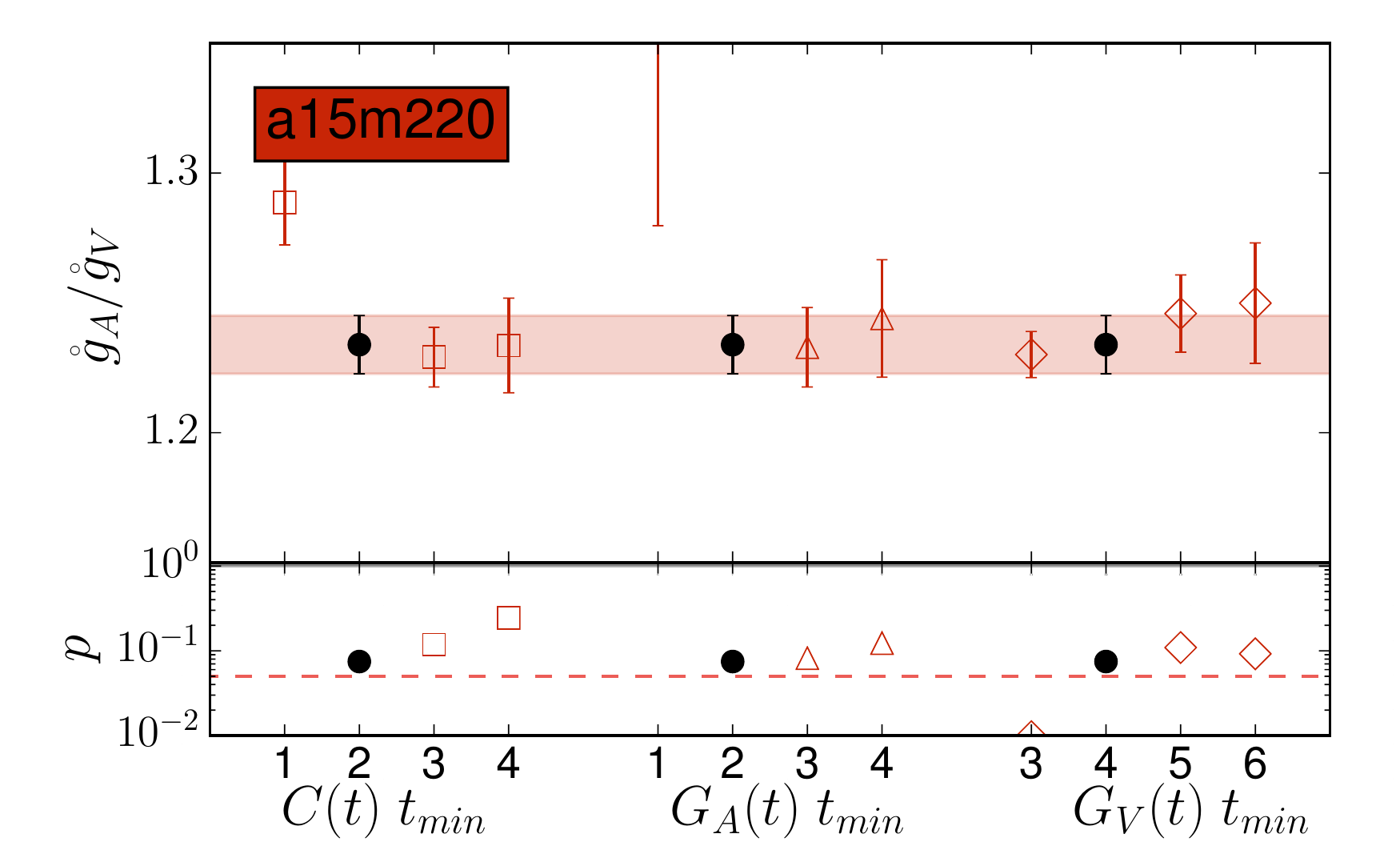}
\includegraphics[width=0.43\textwidth]{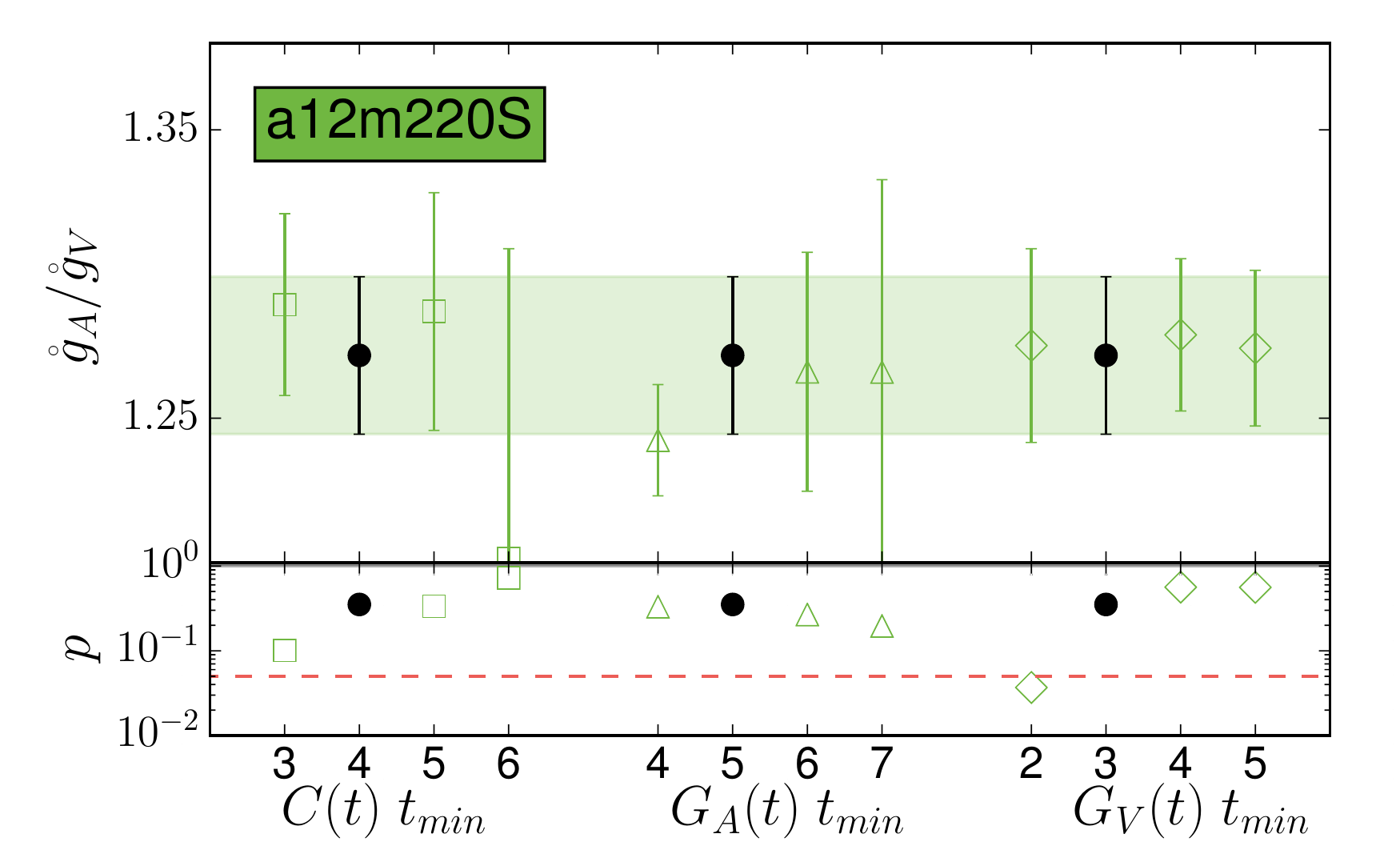}
\includegraphics[width=0.43\textwidth]{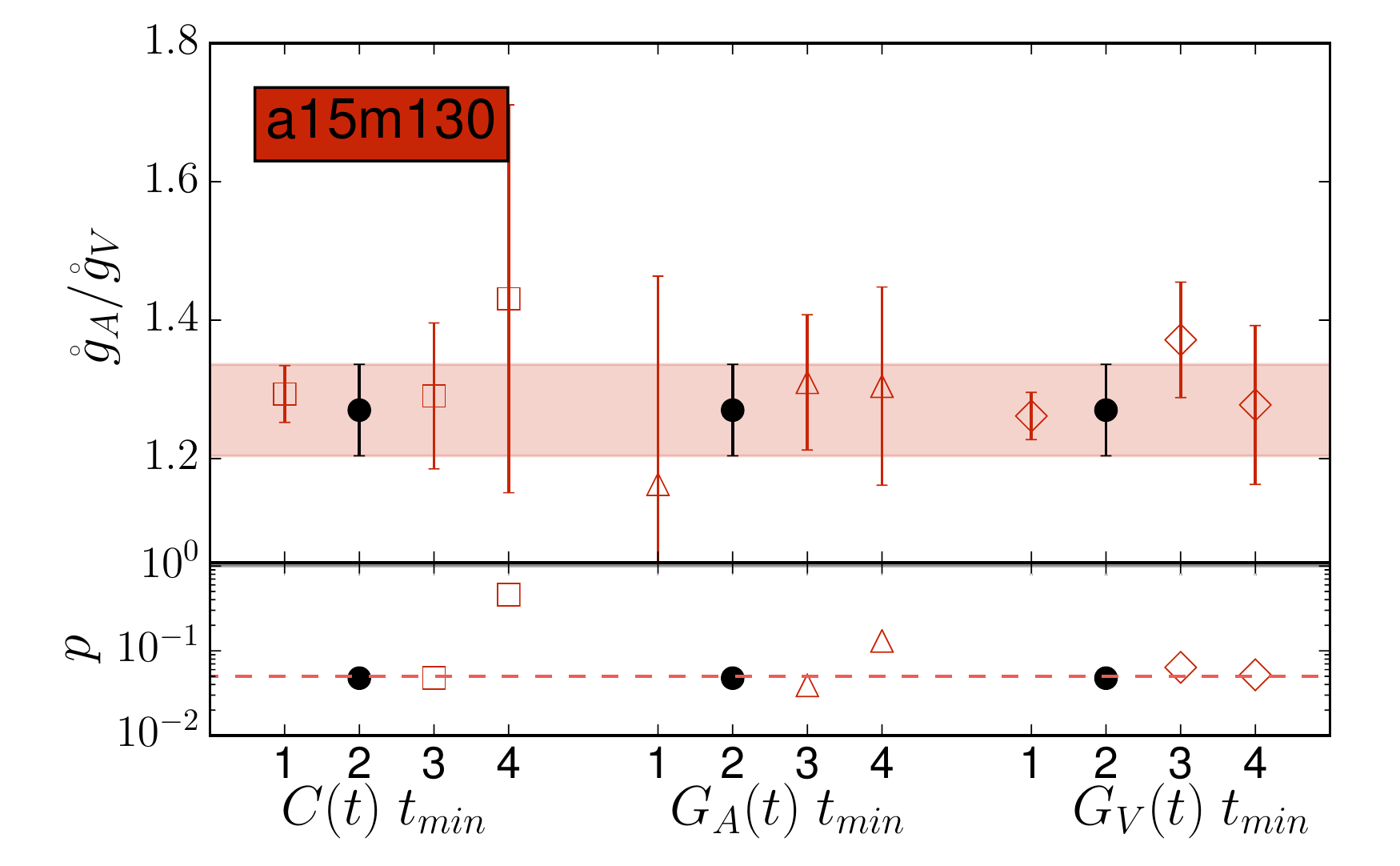}
\includegraphics[width=0.43\textwidth]{stability_a12m220.pdf}
\includegraphics[width=0.43\textwidth]{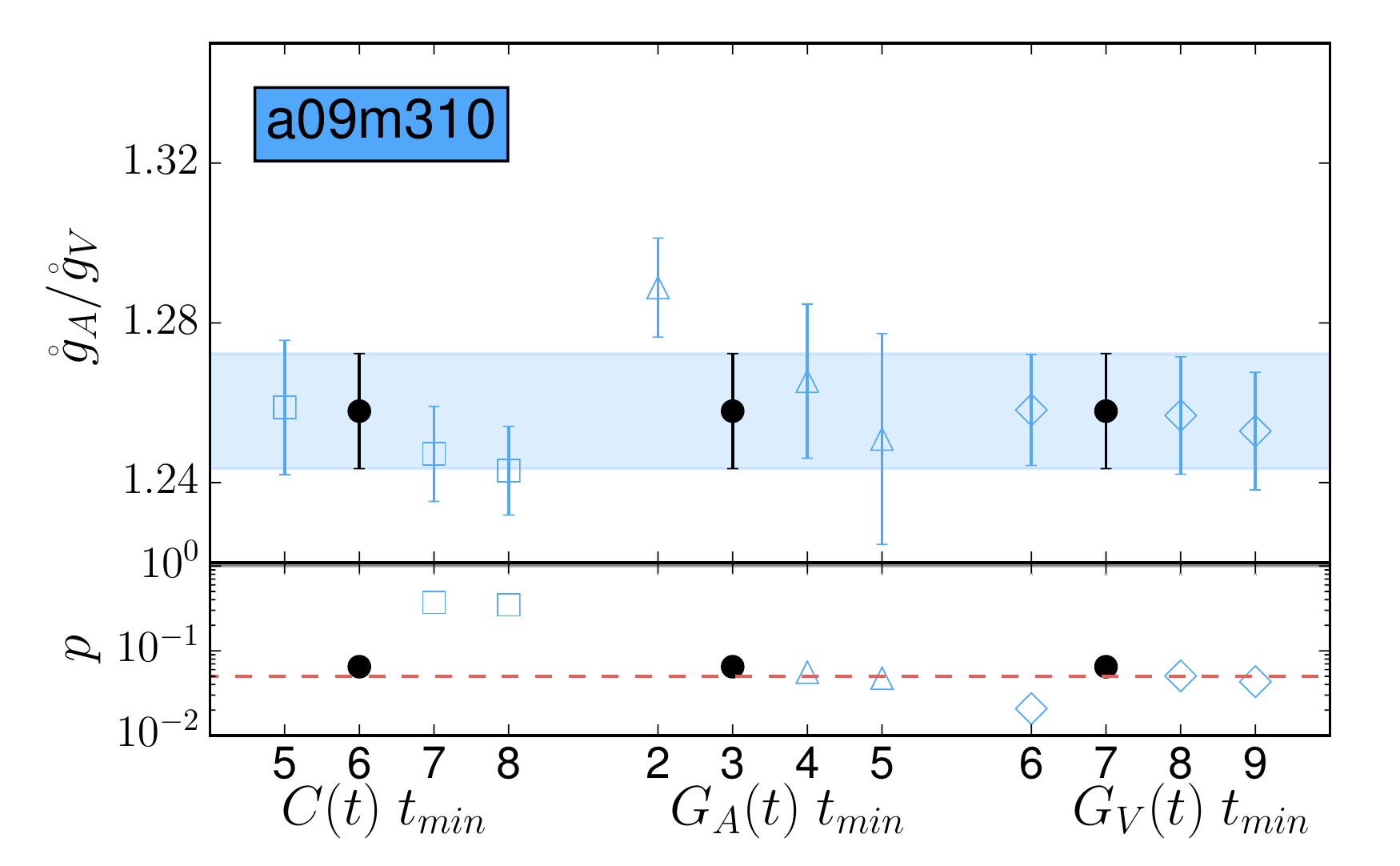}
\includegraphics[width=0.43\textwidth]{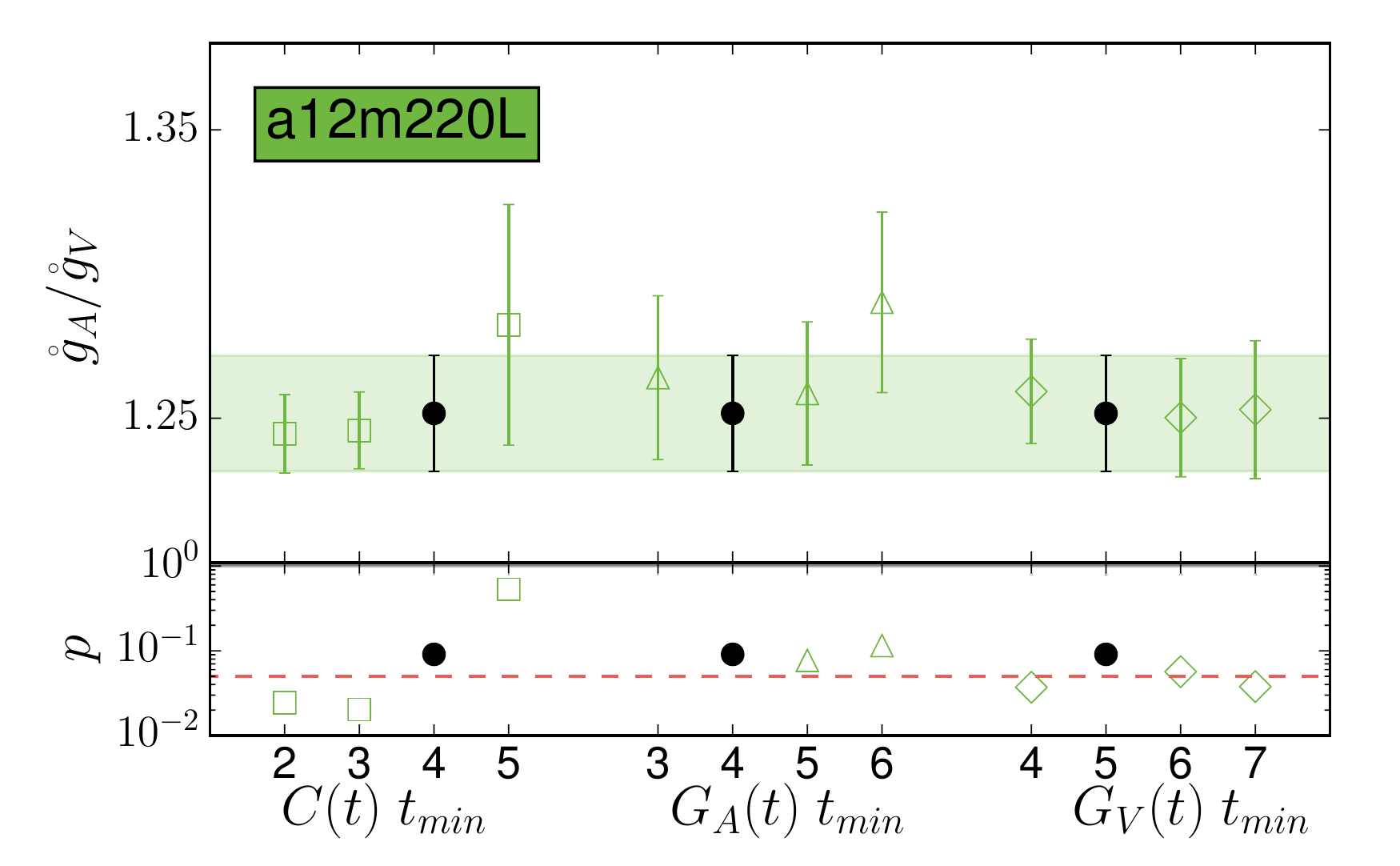}
\caption{Stability plot of $\mathring{g}_A/\mathring{g}_V$ for all ensembles. Solid symbols accompanied by shaded bands are the preferred fits. Varying fit regions for the two-point correlator ($\medsquare$), and axial ($\medtriangleup$), and vector ($\meddiamond$) effective deriatives are presented. Corresponding $p$-values are presented, with the dashed red line at $p=0.05$ discriminating statistical significance of the fit results.}
 \label{fig:stability_all}
\end{figure*}
\FloatBarrier
\newpage
\section{Bootstrap histograms of $\e_\pi$}
\label{app:Fpi_bs}
\begin{figure*}[h]
\includegraphics[width=0.43\textwidth]{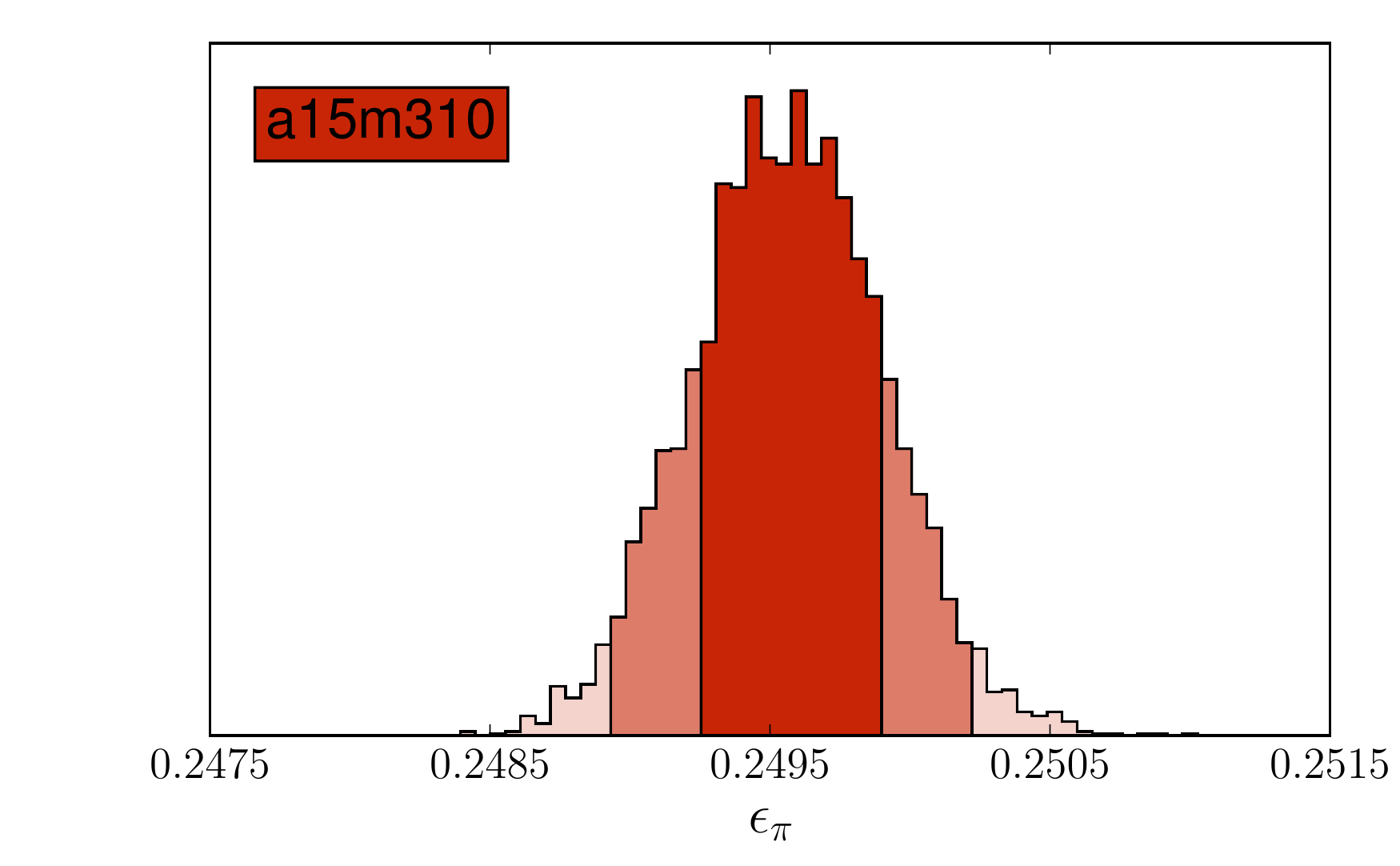}
\includegraphics[width=0.43\textwidth]{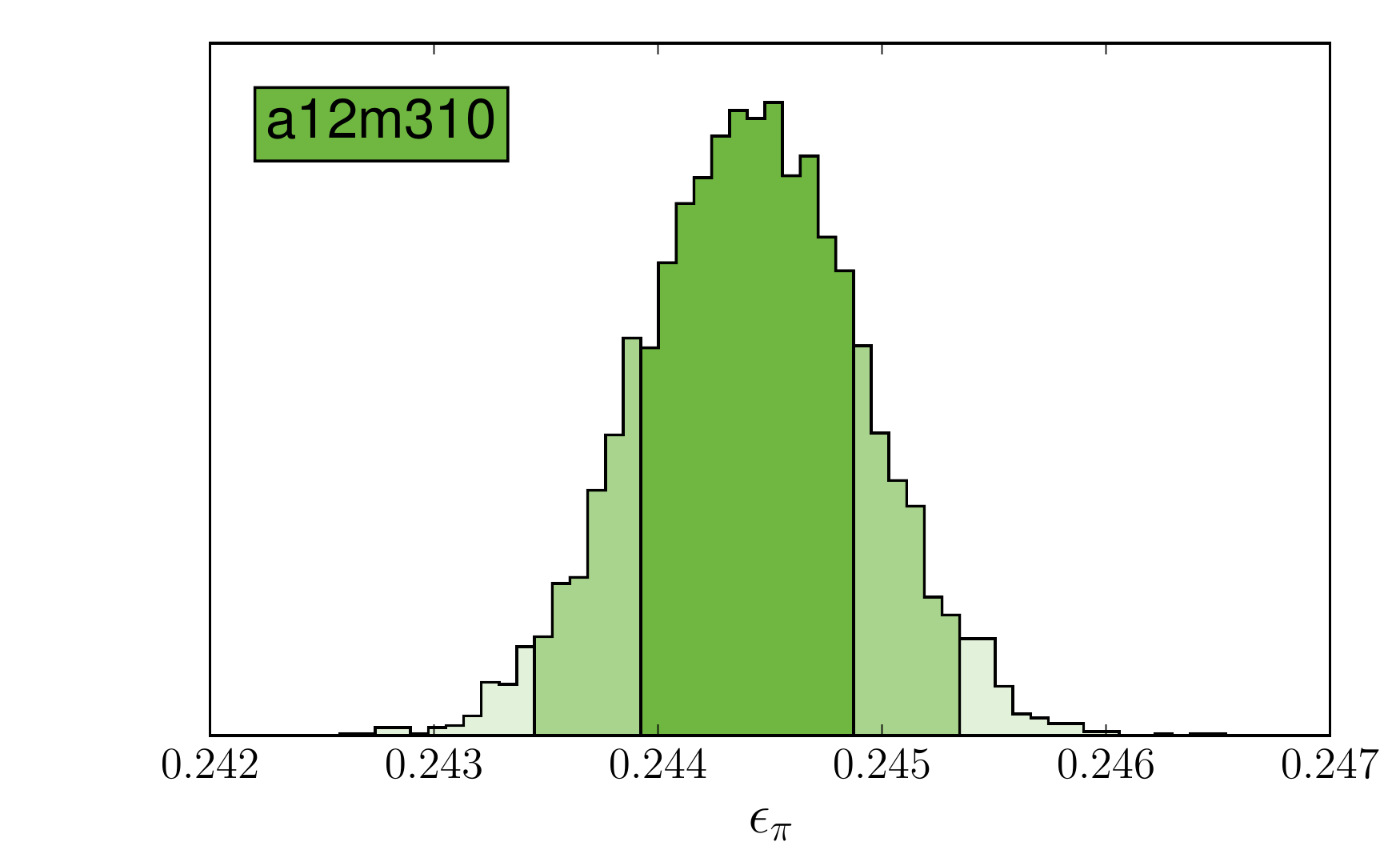}
\includegraphics[width=0.43\textwidth]{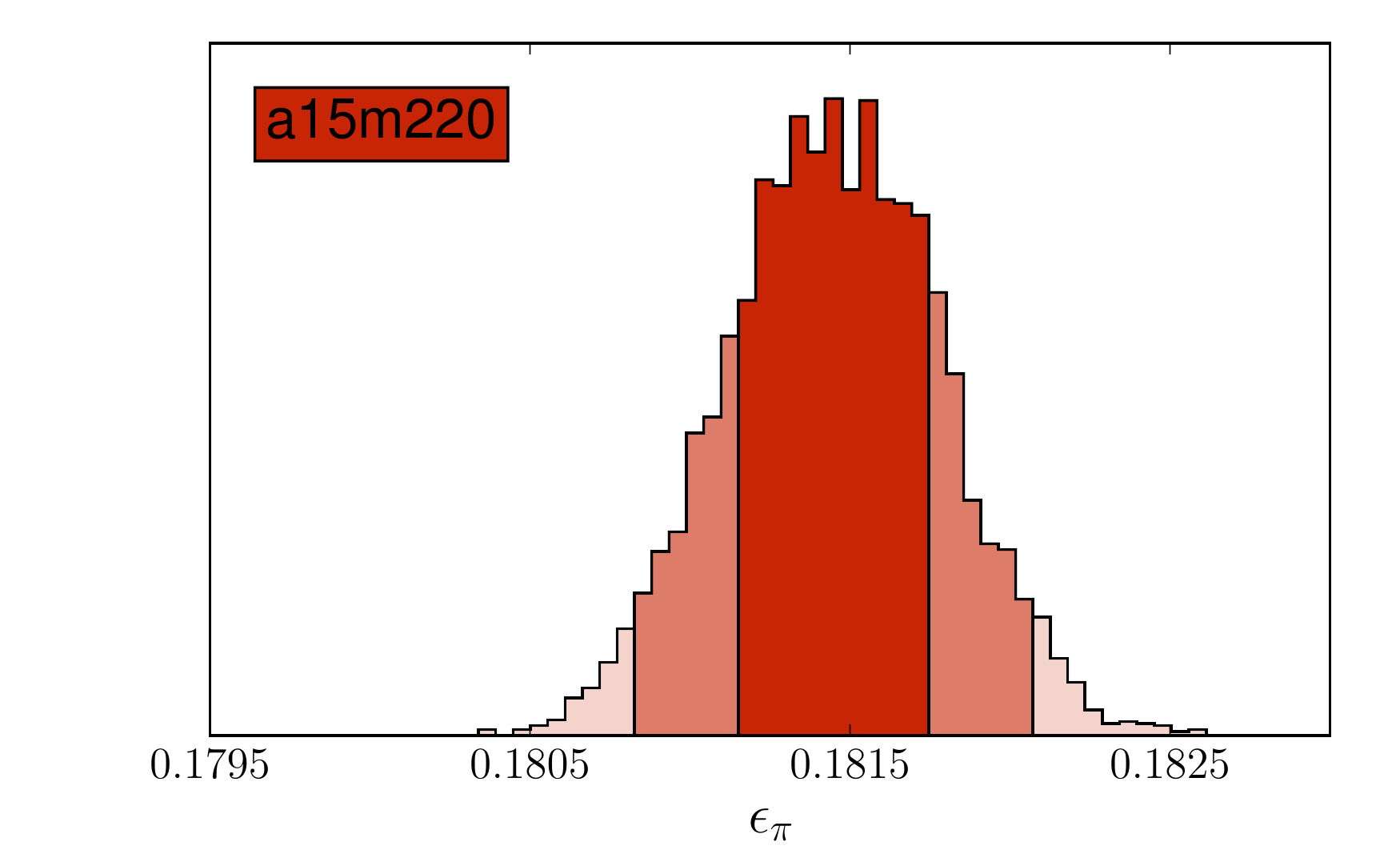}
\includegraphics[width=0.43\textwidth]{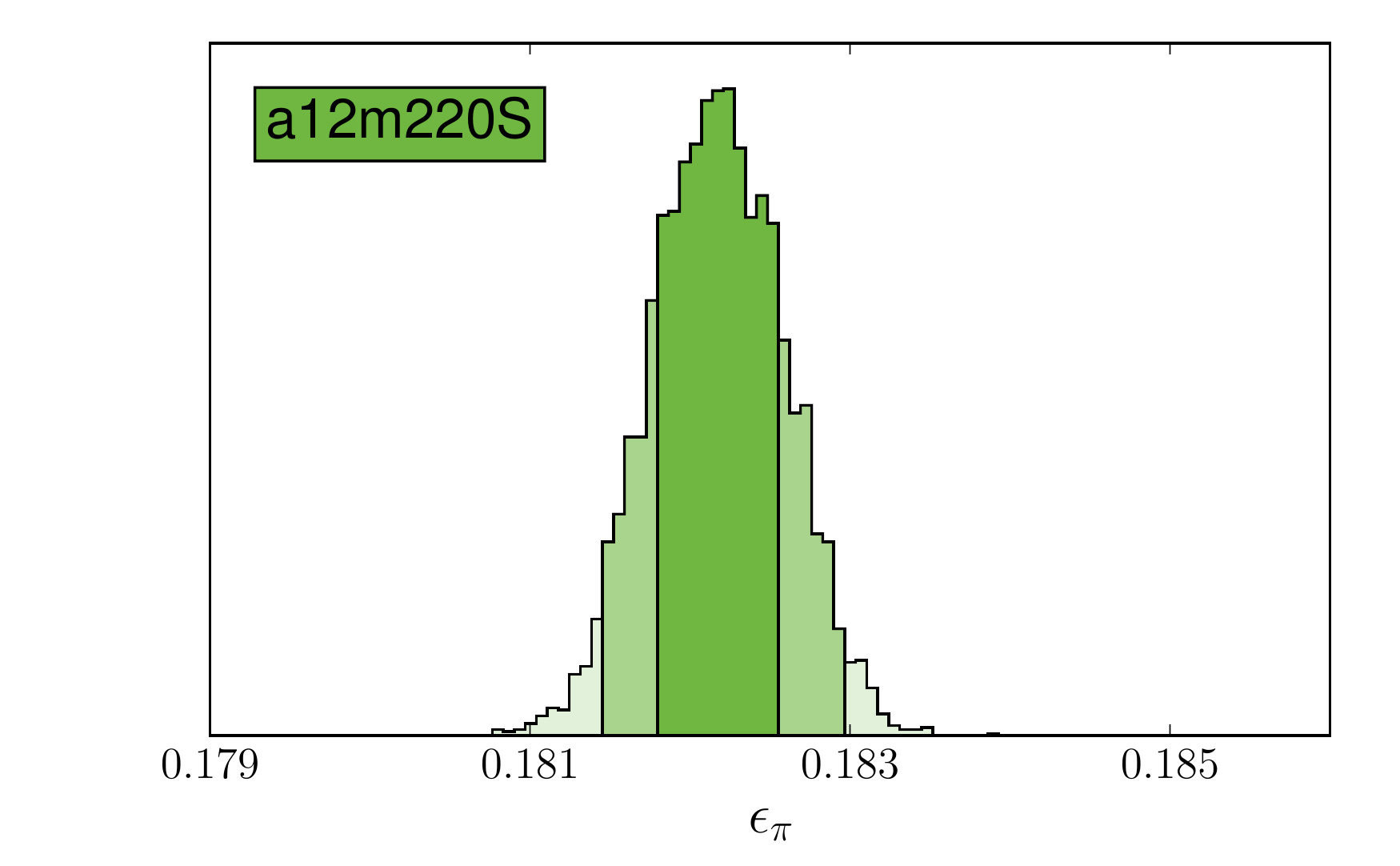}
\includegraphics[width=0.43\textwidth]{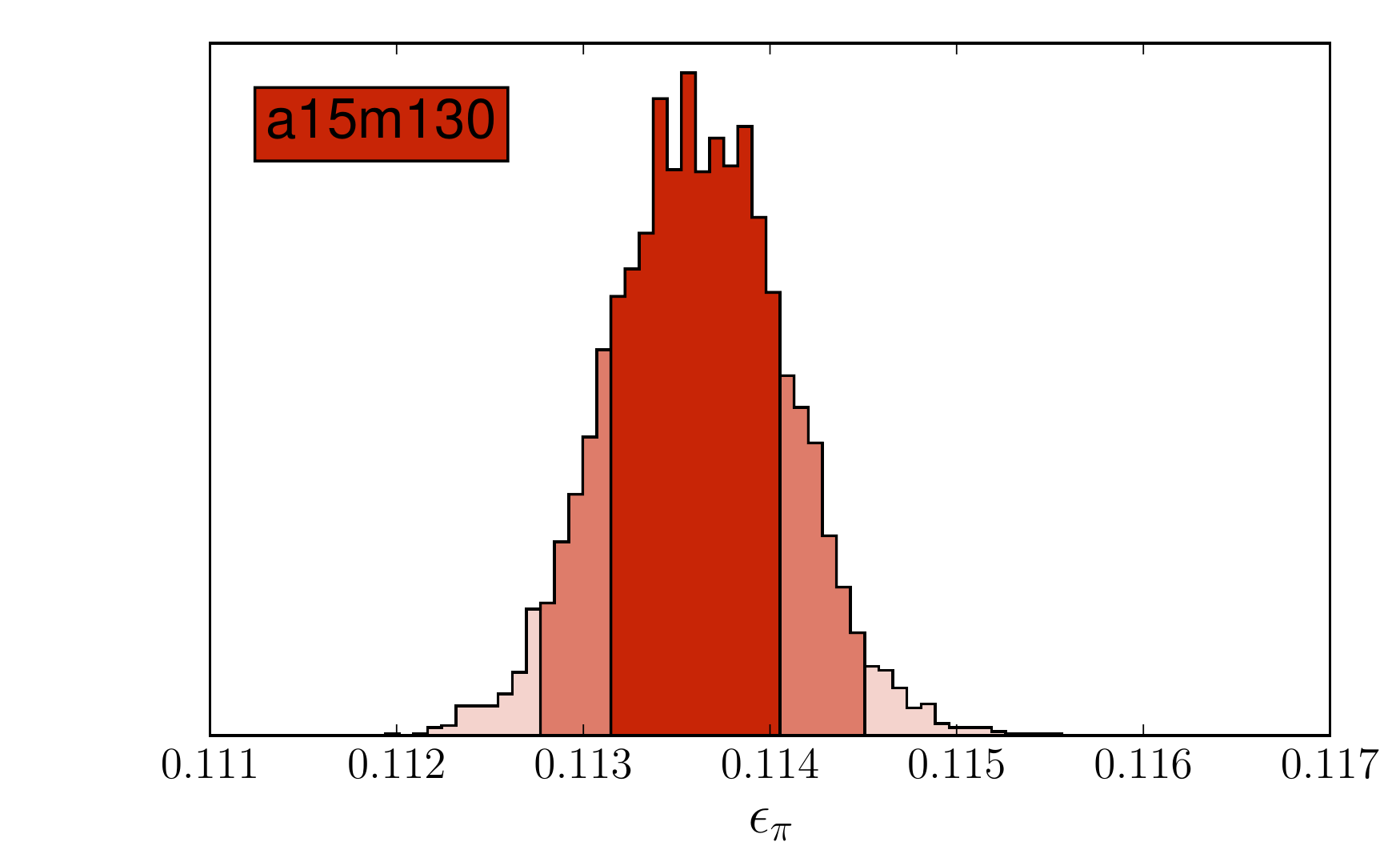}
\includegraphics[width=0.43\textwidth]{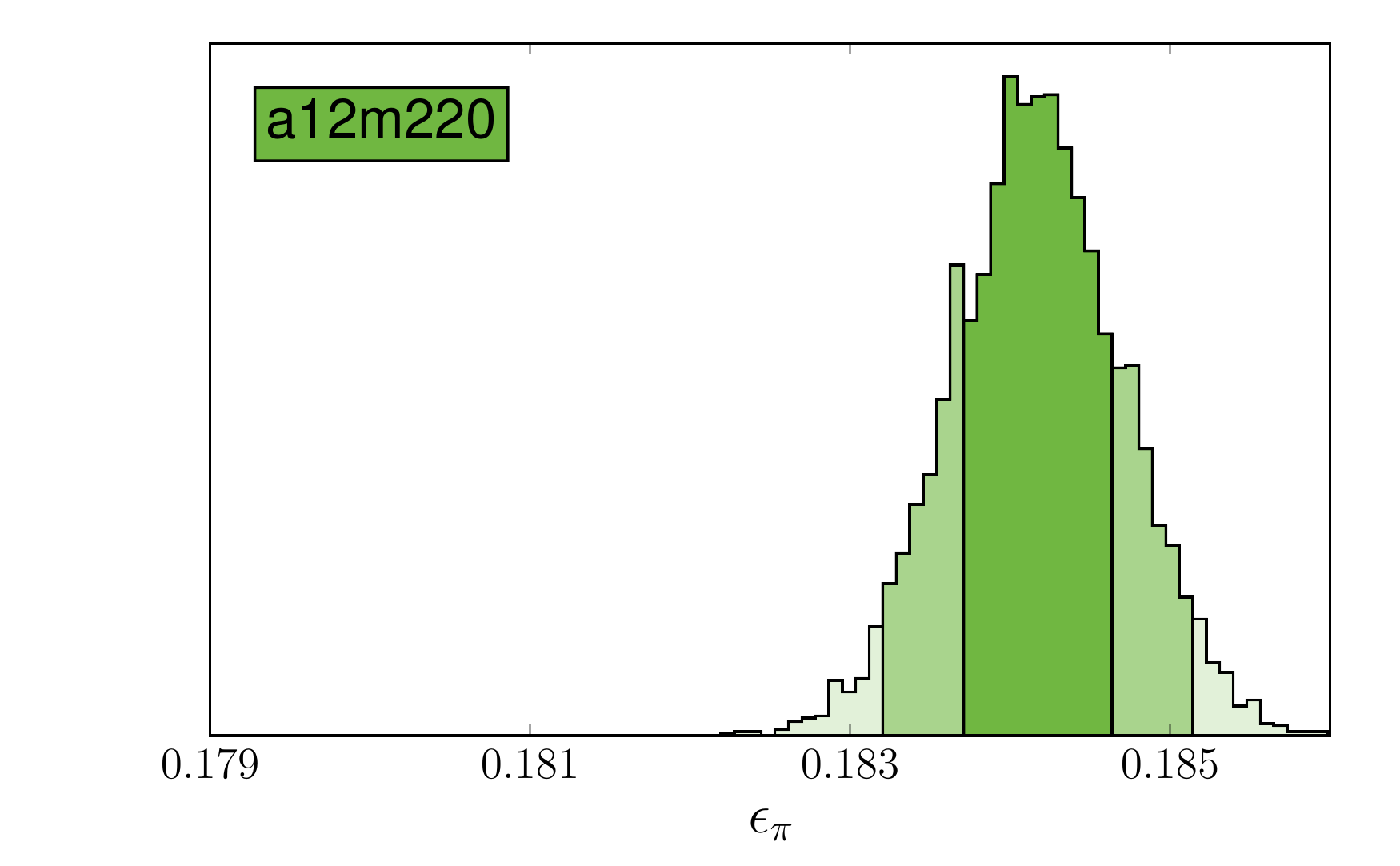}
\includegraphics[width=0.43\textwidth]{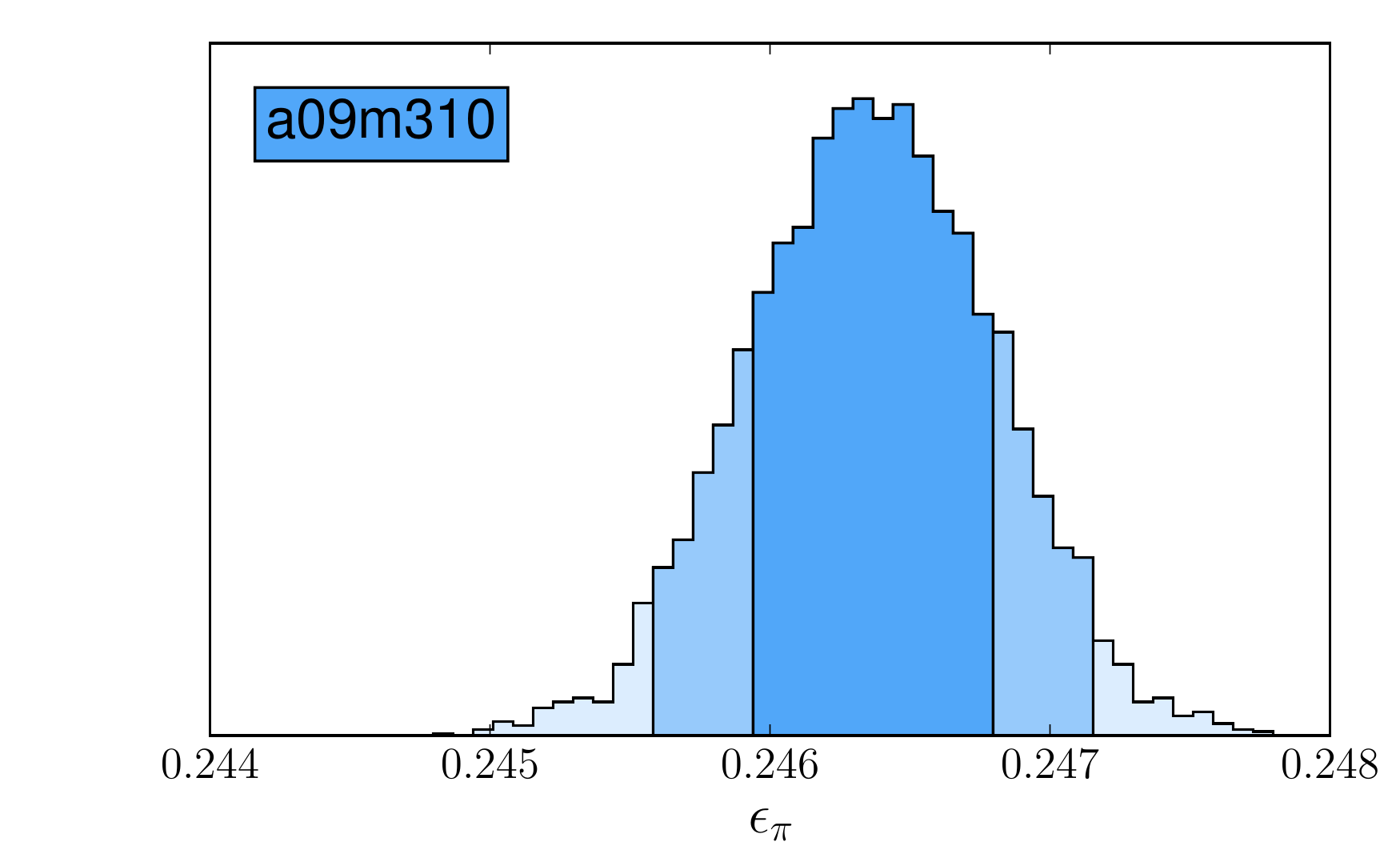}
\includegraphics[width=0.43\textwidth]{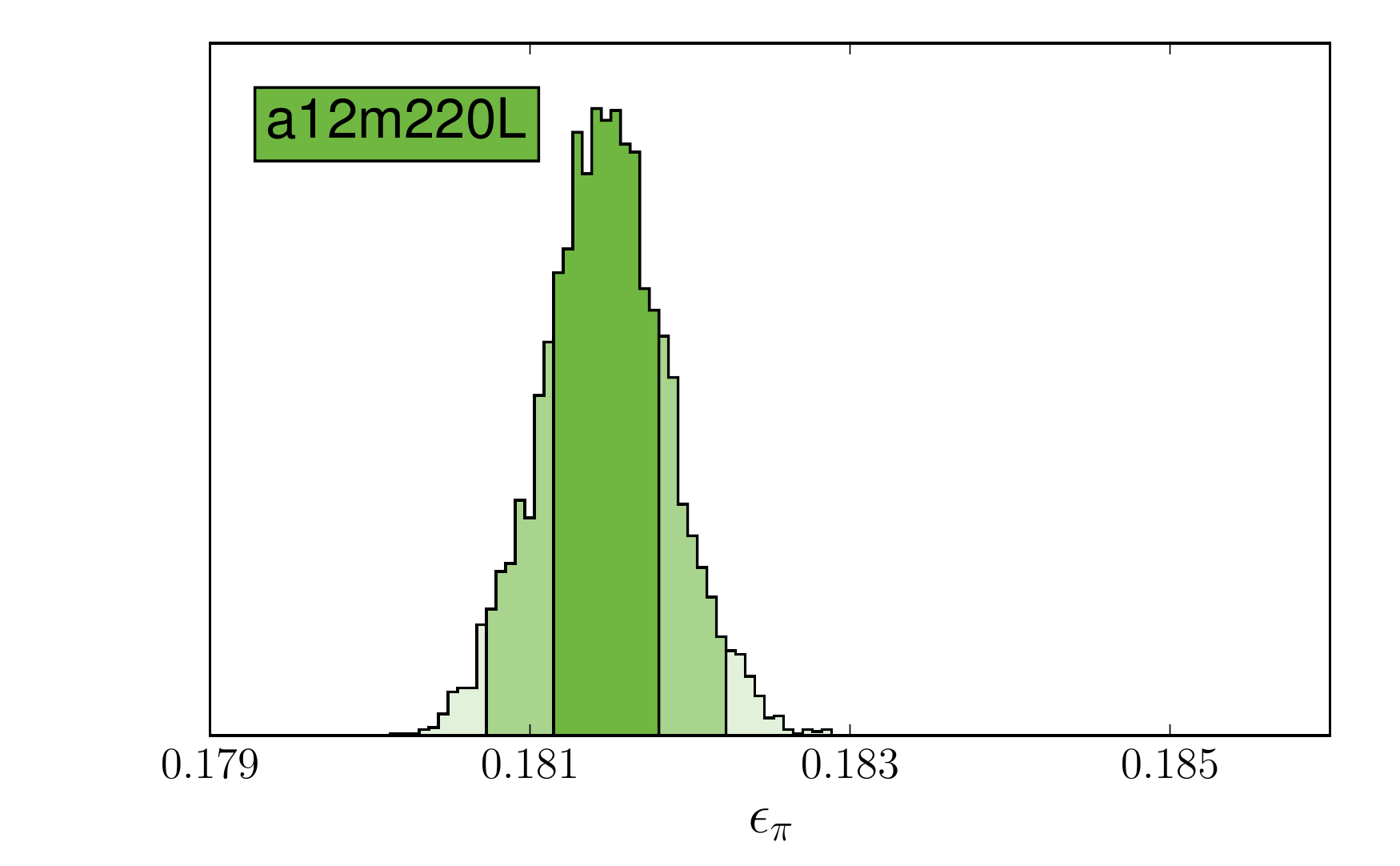}
\caption{Bootstrapped histograms from $\epsilon_\pi$. The abscissa is set to the same range for the three a12m220 ensembles.}
 \label{fig:Fpi_histograms}
\end{figure*}
\FloatBarrier

\end{document}